\documentclass[twocolumn]{aastex631}

  \newcommand {\nc} {\newcommand}
  \nc {\IR} [1]{\textcolor{red}{#1}}

\usepackage{amsmath}
\usepackage{footnote}
\usepackage{booktabs}
\usepackage{tabularx}

\begin{document}

\title{Impact of the latest $^{22}$Ne+$\alpha$ reaction rates \\on nucleosynthesis in massive stars and galactic chemical evolution}

\author{Emma Kotar}
\affil{National Nuclear Data Center, Brookhaven National Laboratory \\
Building 490, Upton, NY 11973-5000, USA}
\affil{Department of Physics and Astronomy, Wellseley College \\
106 Central Street,
Wellesley, MA 02481, USA}

\author{Shuya Ota}
\affil{National Nuclear Data Center, Brookhaven National Laboratory \\
Building 490, Upton, NY 11973-5000, USA}
\affil{Cyclotron Institute, Texas A\&M University \\ College Station, Texas 77843, USA}

\correspondingauthor{Emma Kotar}
\email{ek101@wellesley.edu}
\correspondingauthor{Shuya Ota}
\email{sota@bnl.gov}

\author{Allyson Dewey}
\affil{National Nuclear Data Center, Brookhaven National Laboratory \\
Building 490, Upton, NY 11973-5000, USA} 
\affil{Department of Physics and Astronomy, Rutgers University \\
57 US Highway 1, New Brunswick, NJ 08901-8554, USA}

\author{Joshua Millman}
\affil{National Nuclear Data Center, Brookhaven National Laboratory \\
Building 490, Upton, NY 11973-5000, USA} 
\affil{Department of Physics and Astronomy, Franklin and Marshall College \\
P.O. Box 3003, Lancaster, PA 17604-3003, USA}

\author{Lorenzo Roberti}
\affil{Istituto Nazionale di Fisica Nucleare - Laboratori Nazionali del Sud, Via Santa Sofia 62, Catania, I-95123, Italy}
\affil{Konkoly Observatory \\
Research Centre for Astronomy and Earth Sciences, HUN-REN, \\
Konkoly Thege M. út 15-17, 1121 Budapest, Hungary} 
\affil{CSFK, MTA Centre of Excellence, Budapest, Konkoly Thege Mikl\'os út 15-17, H-1121, Hungary}
\affil{Istituto Nazionale di Astrofisica – Osservatorio Astronomico di Roma, Via Frascati 33, Monte Porzio Catone, I-00040, Italy}

\author{Marco Pignatari}
\affil{Konkoly Observatory \\
Research Centre for Astronomy and Earth Sciences, HUN-REN, \\
Konkoly Thege M. út 15-17, 1121 Budapest, Hungary} 
\affil{CSFK, MTA Centre of Excellence, Budapest, Konkoly Thege Mikl\'os út 15-17, H-1121, Hungary}
\affil{University of Bayreuth, BGI, Universitätsstraße 30, 95447 Bayreuth, Germany}

\collaboration{6}{(NuGrid Collaboration) http://nugridstars.org}



\begin{abstract}
In massive stars (initial mass of $\gtrsim$ 9 \(M_\odot\)), the weak $s$ (slow neutron capture) process produces elements between Fe and Zr,
enriching the Galaxy with these elements through core-collapse supernova explosions. The weak s-process nucleosynthesis is driven by neutrons produced in the $^{22}$Ne($\alpha,n$)$^{25}$Mg reaction during convective He-core and C-shell burning. 
The yields of heavy elements thus depend on the $^{22}$Ne($\alpha,n$)$^{25}$Mg 
and the competitive $^{22}$Ne($\alpha,\gamma$)$^{26}$Mg reaction rates which are dominated by several narrow-resonance reactions. While the accuracy of these rates has been under debate for decades, recent experimental efforts, including ours, drastically reduced these uncertainties. In this work, we use 
a set of 280 massive star nucleosynthesis models calculated using different $^{22}$Ne($\alpha,n$)$^{25}$Mg 
and $^{22}$Ne($\alpha,\gamma$)$^{26}$Mg rates, and a galactic chemical evolution (GCE) study to probe their impact on the weak $s$-process elemental abundances in the Galaxy. The GCE was computed with the OMEGA+ code, using the new
sets of stellar yields with different $^{22}$Ne+$\alpha$ rates. From GCE, we find that these rates are causing up to 0.45 dex of variations in the [Cu/Fe], [Ga/Fe], and [Ge/Fe] ratios predicted at solar metallicity. The greatest impact on the stellar nucleosynthesis and GCE results derives from uncertainties in the ($\alpha$,$n$) strength ($\omega\gamma_{(\alpha,n)}$) of the E$_x$=11.32 MeV resonance. We show that the variations observed in the GCE calculations for the weak $s$-proess elements become negligibly smaller than dispersions found in observations once the $\omega\gamma_{(\alpha,n)}$ is accurately determined within the uncertaintiy of 10--20\% (typically reported experimental errors for the resonance) in future nuclear physics experiments.

\end{abstract}

\keywords{Massive stars(732) --- Nuclear astrophysics(1129) --- Stellar nucleosynthesis(1616) --- Galaxy chemical evolution(580)}


\section{Introduction} \label{sec:intro}
The slow neutron capture, namely the $s$-process, produces about half of the solar abundances for elements heavier than Fe \citep[][]{Kappeler2011, lugaro:23}. The other half of the heavy elements are mostly produced by the rapid neutron capture process \citep[r-process][]{Woosley1994,kajino:19,cowan:21,Siegel2022}, with much smaller contributions from the p-process \citep[][]{rauscher:13, pignatari:16}. The contribution of neutrino-winds ejecta in core-collapse supernovae (CCSN) \citep[][and references therein]{psaltis:22} and from the i-process \citep[e.g.,][]{Denissenkov2019, Choplin2019} is still a matter of debate.

The relevant stellar sources for the s-process elements are low- and intermediate-mass Asymptotic Giant Branch (AGB) stars and massive stars (initial mass $\gtrsim$ 9 $M_{\odot}$). AGB stars produce the bulk of the solar s-process abundances between Sr and Bi \citep[e.g.,][]{bisterzo:14,Prantzos2020}, while massive stars are responsible for most of the production between Fe and Sr \citep[weak $s$-process; e.g.,][]{2010ApJ...710.1557P}. Most of the neutrons powering the s-process in AGB stars are generated by the $^{13}$C$(\alpha,n)^{16}$O reaction in the radiative $^{13}$C-pocket, with a smaller contribution by the $^{22}$Ne$(\alpha,n)^{25}$Mg reaction during Thermal Pulse events \citep[e.g.,][]{Straniero2006, karakas:14, battino:16, lugaro:23}.
In massive stars, $^{22}$Ne$(\alpha,n)^{25}$Mg is the main neutron source powering the s-process, which is efficiently activated in the convective He-burning core and the convective C-burning shell \citep[e.g.,][]{raiteri:91, baraffe:92, Kappeler1994, rauscher:02, the:07, 2010ApJ...710.1557P}. The $^{13}$C$(\alpha,n)^{16}$O reaction may be activated in the C-burning core, but its products are typically further processed by more advanced stages and not ejected by the CCSN explosion \citep[][]{chieffi:98, Pignatari2013}. 
Thus, the most relevant s-process stellar sources have been well-established. On the other hand, several uncertainties are affecting the stellar yields of s-process elements, mainly for CCSN. For these stars, quantitative predictions of theoretical stellar yields  have been affected by uncertainties of both stellar models \citep[e.g.,][]{Kappeler1994, 2016ApJS..225...24P, sukhbold:16, limongi:18,roberti:24} and nuclear reaction rates \citep[][and references therein]{2010ApJ...710.1557P, nishimura:17, Ota2021, Wiescher2023, pignatari:23}. The $^{22}$Ne$(\alpha,n)^{25}$Mg cross sections are particularly complicated because the reactions involve many resonances in their Gamow window. Therefore, their large uncertainties still have a significant impact on s-process nucleosynthesis \citep[e.g., ][]{Wiescher2012, Adsley2021, Best2025}. 
Additionally, as the $^{22}$Ne$(\alpha,n)^{25}$Mg reaction is an endothermic reaction, the neutron emission channels only open above a certain energy \citep[$E_{CM}$=478 keV,][]{Wang2021} where $E_{CM}$ is a Center-of-Mass energy in $^{22}$Ne+$\alpha$ reaction). For resonances near the neutron threshold like this, $\gamma$-ray and neutron emission channels can compete. The n/$\gamma$ branching ratio of the resonance depends on the strength of its single-particle component \citep{Wiescher2023}. 
$^{22}$Ne in the massive stars can be consumed by $^{22}$Ne$(\alpha,\gamma)^{26}$Mg reactions, leading to less active $^{22}$Ne$(\alpha,n)^{25}$Mg reactions in the end \citep{Kappeler1994,Wiescher2012}. 
As such, $^{22}$Ne$(\alpha,\gamma)^{26}$Mg cross sections also need to be constrained to understand the s-process yields from the CCSN. 
 
Like other charged-particle reactions, measuring these cross sections directly at stellar temperature is extremely challenging because of the small cross sections due to the extremely energy-dependent Coulomb barriers. 
The development of underground accelerator facilities, where neutron background environments are dramatically improved compared to those above the ground, is rapidly progressing \citep{Best2016,Gao2022,Shahina2022}. 
Nevertheless, the cross sections at low energy still need to rely on indirect measurements. 
Although both $^{22}$Ne$(\alpha,n)^{25}$Mg and $^{22}$Ne$(\alpha,\gamma)^{26}$Mg cross sections have been studied by both direct \citep[e.g.,][]{Wolke1989,Harms1991,Jaeger2001,Hunt2019,Shahina2024} and indirect \citep[e.g.,][]{Giesen1993,Longland2009,deBoer2010,Massimi2012,Talwar2016,Massimi2017,Adsley2017,Adsley2018,Jayatissa2020,OTA2020135256,Ota2021,Chen2021} measurements for decades, many ambiguities remain. 

Unambiguously determining these rates will drastically reduce uncertainties of nucleosynthesis calculations in massive stars, which are essential for improving the robustness of CCSN yields. The impacts of different $^{22}$Ne+$\alpha$ reaction rates on stellar yields are frequently discussed, but limited to specific stellar conditions \citep[e.g.,][]{Longland2012,Talwar2016,Adsley2021,Ota2021}. Their direct impacts on galactic chemical evolution (GCE) have not yet been studied. 
In particular, we expect that the s-process in massive stars is the main producer of solar abundances of Cu, Ga, and Ge; its contributions to other elements are less certain due to various uncertainties, including those in the processing of s-process rich layers by the CCSN explosion \citep[][]{2010ApJ...710.1557P, Prantzos2020}.  

In the present work, we therefore perform simulations to study the GCE of Cu, Ga, and Ge and other elements (from Zn to Sr), particularly focusing on their sensitivity to the $^{22}$Ne + $\alpha$ reaction rates in massive stars. 
Different $^{22}$Ne + $\alpha$ rates, including a conventionally referenced rate \citep[][hereafter LO12]{Longland2012} and the rate we obtained from the experiment by \cite{OTA2020135256} (hereafter OT20), are used in our simulations. 
Since many uncertainties remain in the reaction rates, multiple reaction rates that adopt different $^{22}$Ne + $\alpha$ cross sections have been published \citep[e.g.,][]{Adsley2021,Wiescher2023}. OT20 produces the smallest weak s-process stellar yields among these rates, \citep{Ota2021} because the adopted rate for the ($\alpha,n$) reaction is the smallest and that for the ($\alpha,\gamma$) reaction is relatively large. In contrast, the LO12 rate produces the largest weak $s$-process yields among the four reaction rates mentioned here. 
Because the $^{22}$Ne + $\alpha$ reaction in $s$-process stars is dominated by resonance reactions, GCE sensitivity to $^{22}$Ne+$\alpha$ rates can be investigated by varying the strength of individual resonances. 
We estimate the impact of individual resonances on the GCE using this approach, comparing the GCE by LO12 with that by OT20. We aim to determine a desirable precision and upper limits for future measurements to constrain the strengths of the relevant $^{22}$Ne+$\alpha$ resonances. The same stellar set provided by \cite{Ritter2018} is used in our nucleosynthesis calculations for the GCE simulations. We thus focus on studying the impact of different reaction rates on the stellar abundances and GCE.

The paper is organized as follows.  
The present uncertainties in $^{22}$Ne$+$$\alpha$ reaction rates and the strengths of the relevant resonances
are discussed in Section \ref{sec:reaction_rates}. We describe the stellar evolution models used in our study and summarize the present knowledge of the production of light $s$-elements: Cu, Zn, Ga, and Ge in Section \ref{sec:abundances}. Presented in Section \ref{sec:yields} are the main
features of the stellar yields and GCE models used in this work. In Sections \ref{sec:results}, the calculated GCEs for Cu, Zn, Ga, and Ge from the pre-supernova (SN) explosions are compared with observations, discussing the sensitivity to the varied strengths of individual resonances and possible impacts by the SN explosions. Finally, summary and conclusions are discussed in Section \ref{sec:summary}.

\section{$^{22}$Ne+$\alpha$ reaction rates} \label{sec:reaction_rates}
Since the ground states of both $^{22}$Ne and $\alpha$ have spin-parity ($J^\pi$) of 0$^+$ and their reaction Q-value for populating the compound nucleus, $^{26}$Mg, 
is highly positive (Q = 10.614 MeV), $^{22}$Ne+$\alpha$ reaction primarily proceeds through resonant capture to natural-parity states in the highly excited $^{26}$Mg \citep[see review by][]{Wiescher2023}. 
While the compound nucleus's level density is high, only a limited number of resonances that have natural-parity and high $\alpha$-cluster strength are expected to contribute to the $^{22}$Ne+$\alpha$ reaction. 
The expectation has been verified by direct measurements, also confirming these resonances' widths ($\Gamma$) are narrow and do not overlap between each resonance. 
For a given resonance, the key properties that determine its contribution to the stellar rate are the $^{22}$Ne+$\alpha$ resonance energy (E$_x$ = $Q$ + E$_{CM}$) and the resonance strengths, $\omega\gamma_{(\alpha,n)}$ and $\omega\gamma_{(\alpha,\gamma)}$. Two (one possible) resonances, at $E_x$=11.32 and $\approx$11.17 MeV, have been intensively studied due to their importance in the weak s-process's temperature (0.3--0.35 GK). Although the weak s-process is reactivated at about 1 GK, the final abundances are less sensitive to the current uncertainties of the resonance strengths at the relevant energy. 

One of the two important resonances is well-identified at E$_{CM}$ = 0.702 MeV (E$_x$ = 11.32 MeV). Both the ($\alpha,n$) and ($\alpha,\gamma$) stellar rates are dominated by this resonance. 
It has been observed in both direct ($\alpha,n$) \citep{Harms1991,Drotleff1991,Giesen1993,Jaeger2001,Shahina2024} and ($\alpha,\gamma$) experiments \citep{Wolke1989,Hunt2019,Shahina2022}. 
Its $\omega\gamma_{(\alpha,\gamma)}$ is well established with these published measurements in excellent agreement with each other \citep{Best2025}. In contrast,
$\omega\gamma_{(\alpha,n)}$ extracted from the direct ($\alpha,n$) measurements are in poor agreement, making the $\omega\gamma_{(\alpha,n)}$ for the
E$_x$ = 11.32 MeV resonance the outstanding contributor to uncertainties
on the total $^{22}$Ne($\alpha,n$)$^{25}$Mg stellar rate. 

At lower temperatures ($<$0.25 GK), both reaction channels may be dominated by
one or more resonances between the neutron threshold ($S_n$=11.09 MeV) and E$_{CM}$ $\approx$ 0.635 MeV (E$_x$ $\approx$ 11.25 MeV). 
In this region, in particular the resonance at E$_x$ $\approx$ 11.17 MeV has long been controversial \citep[see e.g.,][]{Longland2012} since reported by \cite{Berman1969}. 
\cite{Jaeger2001} searched for this state in
their direct ($\alpha,n$) measurement, setting an upper limit on the $\omega\gamma_{(\alpha,n)}$ at E$_x$ $\approx$ 11.17 MeV, $<$60 neV. 
Recently, \cite{Shahina2022} also observed no strong transitions around this energy region in their ($\alpha,\gamma$) experiment, setting an upper limit of $\omega\gamma_{(\alpha,\gamma)}$ $<$150 neV on a candidate resonance at E$_x$ $\approx$ 11.17 MeV. Additionally, the $\omega\gamma_{(\alpha,n)}$ 
for the natural-parity resonances identified by \cite{Massimi2012}, the lowest-lying resonance above $S_n$, $E_x$ = 11.11 MeV, remain
poorly constrained ($\lesssim$60 neV) by direct measurements \citep{Jaeger2001}. 
Although this resonance has never been reported to have strong $\alpha$-cluster strength by indirect measurements, it may have the potential to completely dominate the ($\alpha,n$) rate below $\approx$0.3 GK due to its proximity to the Gamow peak \citep{Ota2021}.
While the $E_x$ = 11.32 MeV and 11.17 MeV resonances are critical for 0.2--0.35 GK, two resonances below the neutron threshold (therefore no ($\alpha,n$) cross sections), 11.08 and 10.95 MeV resonances, could contribute to consume $^{22}$Ne via ($\alpha,\gamma$) reactions at 0.1--0.2 GK \citep{OTA2020135256} before ($\alpha,n$) reaction is activated at a later (hotter) stage of the star \citep{Kappeler1994,Wiescher2023}.  

These situations are well-described \citep[in e.g.,][]{OTA2020135256,Adsley2021}. 
In this work, we will therefore study the sensitivity of the GCE to the strengths of these individual resonances ($\omega\gamma_{(\alpha,n)}$ and $\omega\gamma_{(\alpha,\gamma)}$), including those above 11.32 MeV, by varying their strengths. 

\section{Stellar models and production of the elements copper, zinc, gallium and germanium} \label{sec:abundances}

\subsection{Stellar models and CCSN explosions}
For the nucleosynthesis calculations presented in this work, we used the 20 massive star models from the stellar set by \cite{Ritter2018} \footnote{Initial mass $M_{ZAMS}$=12, 15, 20, and 25 $M_{\odot}$ (where ZAMS means Zero Age Main Sequence) and metallicity $Z$ = 0.0001, 0.001, 0.006, 0.01, and 0.02 calculated with the MESA code, rev. 3372 \citep[][]{2011ApJS..192....3P}}. The CCSN models are calculated with a semi-analytical approach based on the Sedov blast wave (SBW) solution for the determination of the peak velocity of the shock throughout the stellar structure \citep{Sedov1946}, while the remnant mass is determined using the delayed explosion prescription from \cite{Fryer2012}. 
For further detail, readers are referred to \cite{2016ApJS..225...24P}, \cite{Ritter2018} and \cite{Roberti2024}.

For $Z$=0.01 and 0.02, the initial abundances adopted are solar-scaled \citep[][]{Grevesse1993}.  At $Z$ = 0.006 and below, $\alpha$-enhanced abundances were used, where the isotopes of $^{12}$C, $^{16}$O, $^{20}$Ne, $^{24}$Mg, $^{28}$Si, $^{32}$S, $^{36}$Ar, $^{40}$Ca, and $^{48}$Ti are enhanced compared to a solar scaled composition \citep[see Table 1, in][]{Ritter2018}, based on stellar observations. Different $\alpha$-enhancements are applied to different species, varying between [Ca/Fe]=+0.222 (where the enhancement is applied to $^{40}$Ca, and the remaining Ca isotopes are solar scaled), and [O/Fe]=+0.886 (the enhancement is only applied to $^{16}$O). Although the solar abundances by \cite{Grevesse1993} have been updated several times \citep[more recently, see][]{Magg2022}, these variations would not affect the results of the present study.

The Multi-zone Post-Processing Nucleosynthesis in Parallel code (MPPNP) developed by the NuGrid collaboration is used for the calculations of the complete nucleosynthesis \citep[e.g.,][]{2016ApJS..225...24P} as used in \cite{Ritter2018}. The MPPNP post-processes the production and destruction of all relevant isotopes during stellar evolution and during CCSN explosions, allowing us to derive stellar isotopic yields. In this work, the nucleosynthesis simulations are performed for the aforementioned stellar models, where only the $^{22}$Ne+$\alpha$ rates are modified. The stellar structure and CCSN models can be used unchanged for different $^{22}$Ne+$\alpha$ rates, as the reactions do not impact energy generation and therefore do not affect stellar evolution properties. Both the pre-SN and post-SN nucleosynthesis were calculated using version V2.2 of the JINA Reaclib database if reaction rates from experiments were not available. For instance, for the $^{22}$Ne($\alpha$,n)$^{25}$Mg rate and the $^{22}$Ne($\alpha$,$\gamma$)$^{26}$Mg rate, \cite{Ritter2018} adopted \cite{Jaeger2001} and \cite{Angulo1999}, respectively. 
Experimental neutron capture reaction rates were taken, when available, from the KADoNIS compilation \citep{Dillmann2006}. We refer to \cite{2016ApJS..225...24P} and \cite{Ritter2018} for more details of all the reaction rates used for the post-processing calculations.

\subsection{Productions of Cu, Zn, Ga, and Ge}
\label{subsec:lights_production}
While the bulk of the s-process products are expected to be built in pre-SN hydrostatic evolution, the shock wave and explosive nucleosynthesis during CCSN explosions can affect the ejected s-process yields \citep[e.g.,][]{rauscher:02}. Therefore, explosive nucleosynthesis should be taken into account to derive the complete stellar yields. In the following part of this sub-section, we will explore the productions of the elements Cu, Zn, Ga, and Ge, which are located just above the iron-group element Ni, and are produced by the s-process in massive stars \citep[e.g.,][]{raiteri:91, raiteri:91a, rauscher:02, the:07, 2010ApJ...710.1557P}. For the remainder of this work, we will focus our analysis and discussions on these elements. As mentioned, it is established that most of the solar Cu, Ga, and Ge are produced by the s-process in massive stars, while the origin of Zn remains a puzzle \citep[e.g.,][]{2010ApJ...710.1557P, Prantzos2020, kobayashi:20}. Thus, it is beneficial to discuss the production processes of these elements in this subsection to better understand their productions in the different layers of the CCSN ejectas, which will lead to better understanding of the impact of the $^{22}$Ne+$\alpha$ rates discussed in the later sections. 

\subsubsection{Copper}
\begin{figure*}
    \centering
    \includegraphics[scale=0.41]
{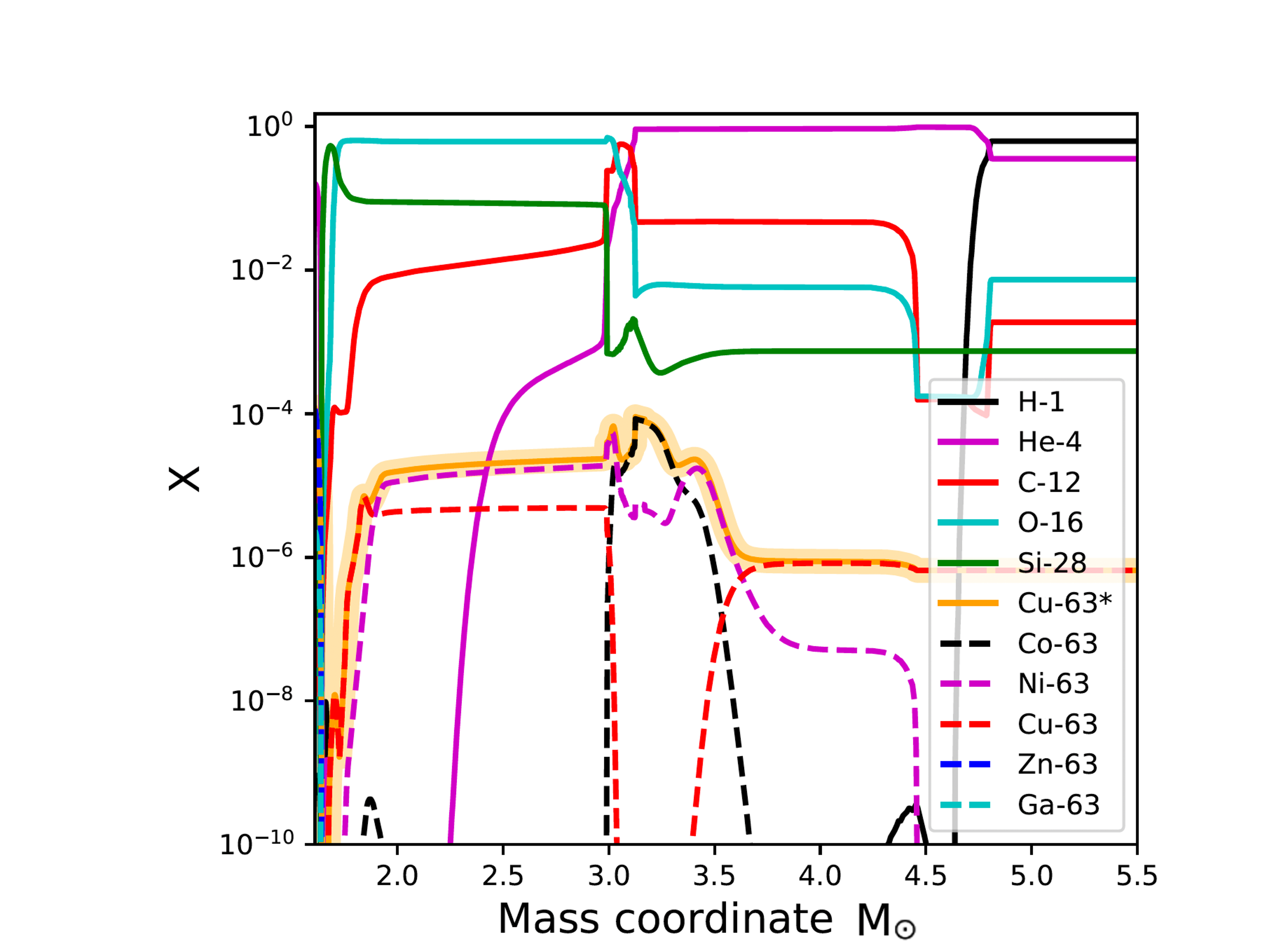}
\includegraphics[scale=0.41]
{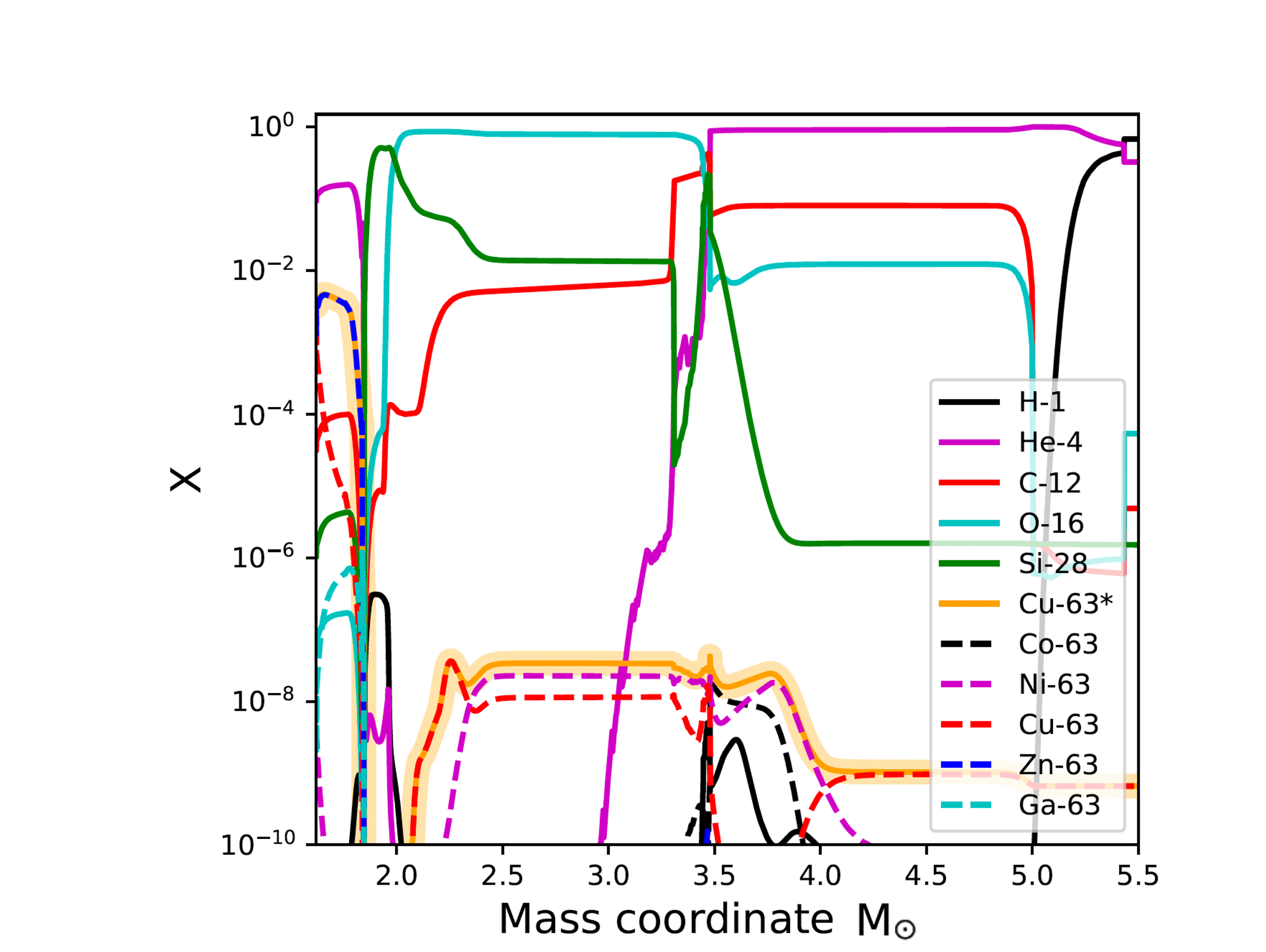}
    \caption{Abundance profiles for relevant isotopes that help identify the different parts of the CCSN ejecta ($^1$H, $^4$He, $^{12}$C, $^{16}$O and $^{28}$Si), $^{63}$Cu ejected and after considering the radiogenic contributions ($^{63}$Cu*), are shown with respect to mass coordinate (mass fraction $X$ in y-axis) for the models of M$_{ZAMS}$=15M$_{\odot}$ at $Z$=0.02 (left panel) and $Z$=0.0001 (right panel). The relevant radiogenic parent species of $^{63}$Cu are also presented and plotted with dashed lines.}
    \label{fig:15_cu63}
\end{figure*}

In the solar system, Cu is made up of 69.15\% of $^{63}$Cu and 30.85\% of $^{65}$Cu. In Figure \ref{fig:15_cu63}, we show the abundance profiles for $^{63}$Cu in the CCSN ejecta of the 15M$_{\odot}$ at $Z$=0.02 and $Z$=0.0001 models calculated by \cite{Ritter2018}. 
Note that here $X$ denotes the abundance in mass fraction of a given isotope $i$ for each mass coordinate in the model\footnote{In a given differential mass zone $\Delta m$, at the mass coordinate $m$, $X_i(m) = \Delta m_i(m) / \Delta m(m)$.}.
Although the two models have the same initial mass, the model at lower $Z$ shows a more relevant primary\footnote{i.e., not directly affected from the initial composition of the star, in contrast to secondary.} explosive nucleosynthesis component, within an $\alpha$-rich region in the deepest 0.2-0.3 M$_{\odot}$ region of the ejecta, right above the newly formed proto-neutron star (PNS), where the $\alpha$-process takes place \citep[e.g.,][]{woosley:92, 2016ApJS..225...24P}. Therefore, we will discuss the primary production of Cu (and other elements discussed in this subsection) in CCSN using the $Z$=0.0001 model. 

In the model at $Z$=0.02, $^{63}$Cu shows a mild s-process production in the material ejected from the former convective He shell at mass coordinates between 3.5 M$_{\odot}$ and 4.5 M$_{\odot}$. $^{63}$Cu at $\approx$3.1 M$_{\odot}$--3.5 M$_{\odot}$ is produced by n-process \citep[][and references therein]{meyer:00, pignatari:18} mostly as radiogenic\footnote{By radiogenic production, we mean that the given stable isotope, in this case $^{63}$Cu, is not ejected directly as itself but in the form of radioactive isotopes that will decay later to $^{63}$Cu, based on their decay half-life.} $^{63}$Ni and $^{63}$Co, triggered by the $^{22}$Ne($\alpha$,n)$^{25}$Mg activation during explosive He burning. The bulk of the s-process production is visible at about 1.8 M$_{\odot}$--3 M$_{\odot}$ as ashes of the pre-explosive C burning. Traces of more extreme explosive components are present in the most internal ejected layers, but are only barely noticeable in this model. 

In the model at $Z$=0.0001, the $^{63}$Cu production discussed above is about three orders of magnitude smaller, due to the smaller $Z$ of the progenitor and the well-expected reduction of both the neutron source $^{22}$Ne (as it is secondary) and the iron seeds \citep[e.g.,][]{baraffe:92, raiteri:92, pignatari:08,roberti:24}. Instead, a strong primary production is obtained in the $\alpha$-rich freezout material at $\approx$1.6 M$_{\odot}$--1.8 M$_{\odot}$. $^{63}$Cu show abundances up to a few per mill in mass fractions, and is mostly produced from the radiogenic contribution from $^{63}$Zn, with much smaller direct production as $^{63}$Cu and the radiogenic contribution from $^{63}$Ga. 

\begin{figure*}
    \centering
    \includegraphics[scale=0.41]
{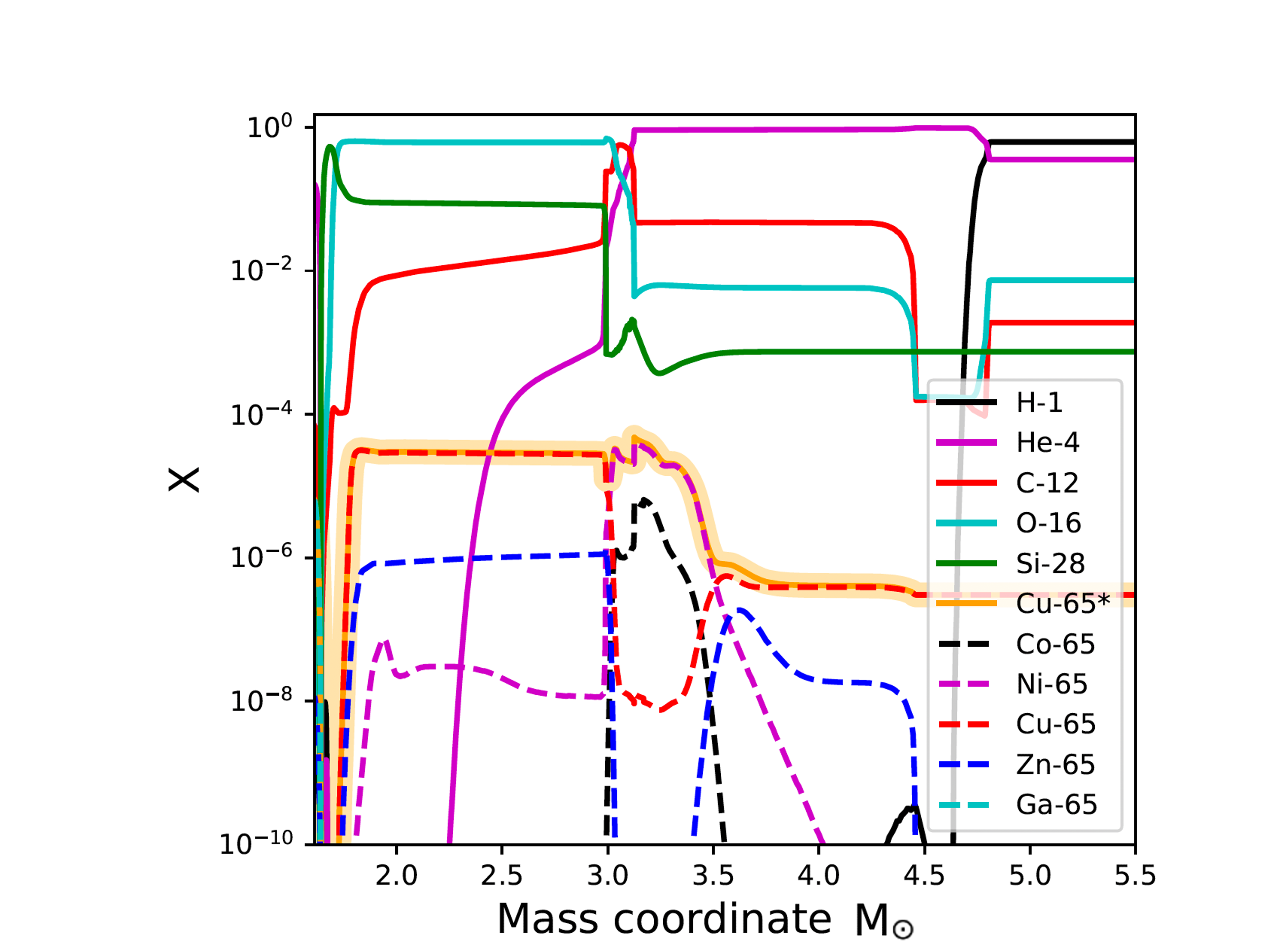}
    \includegraphics[scale=0.41]
{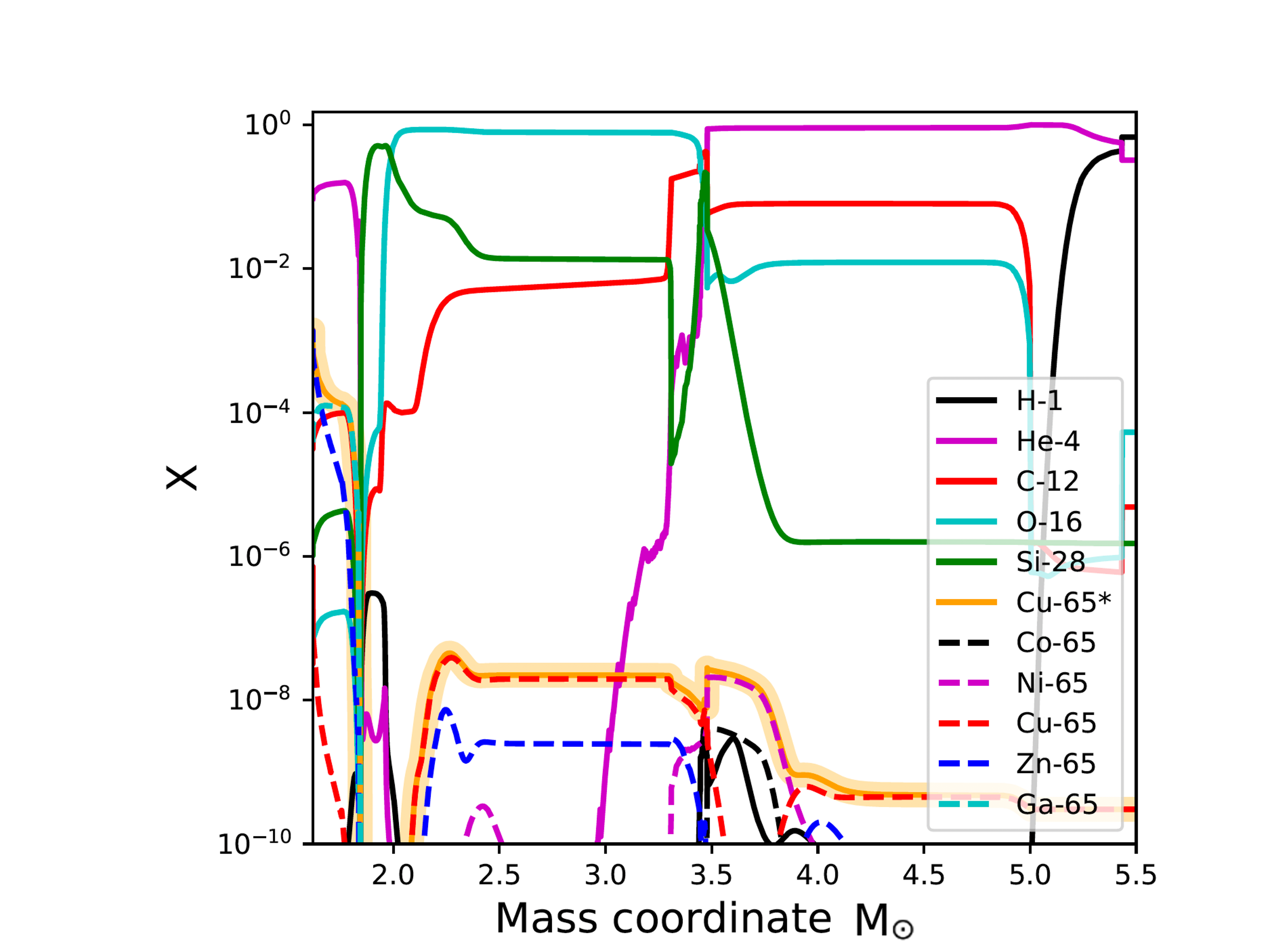}
    \caption{The same as in Figure \ref{fig:15_cu63}, but for $^{65}$Cu. }
    \label{fig:15_cu65}
\end{figure*}

In Figure \ref{fig:15_cu65}, the same production pattern as Figure \ref{fig:15_cu63} is observed for $^{65}$Cu, which are ejected in the same CCSN layers as $^{63}$Cu. The s-process makes $^{65}$Cu mostly directly, with a marginal radiogenic contribution from $^{65}$Zn (which is made from neutron-capture on $^{64}$Zn). The n-process during the explosion instead produces $^{65}$Cu mostly as $^{65}$Ni, with a small contribution from $^{65}$Co. $^{65}$Cu also has a strong primary component above the PNS in the model at low $Z$, specifically as radiogenic contribution from $^{65}$Zn and $^{65}$Ga (Figure \ref{fig:15_cu65}, right panel). 

\subsubsection{Zinc}

\begin{figure*}
    \centering
    \includegraphics[scale=0.41]
    {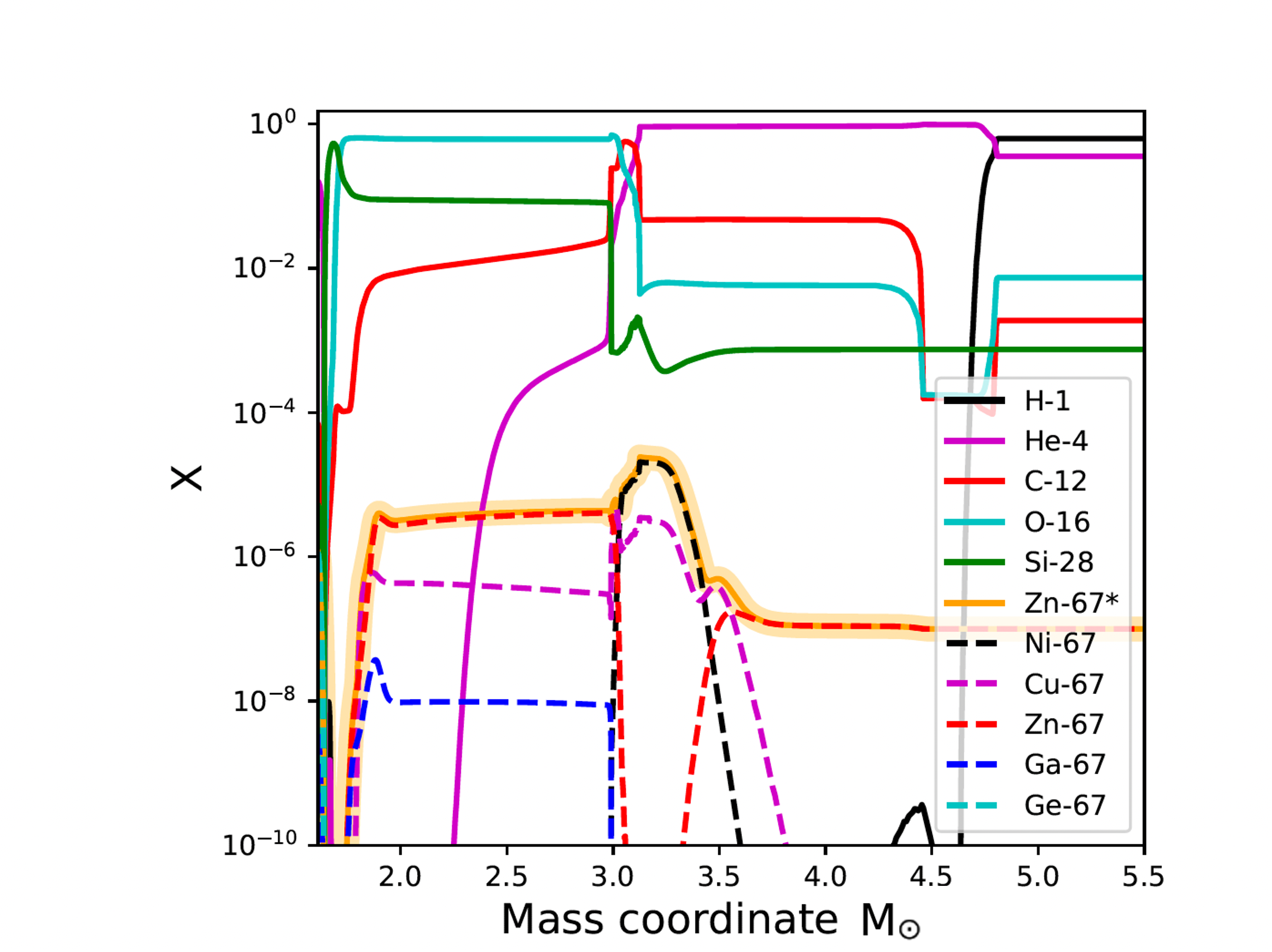}
    \includegraphics[scale=0.41]
    {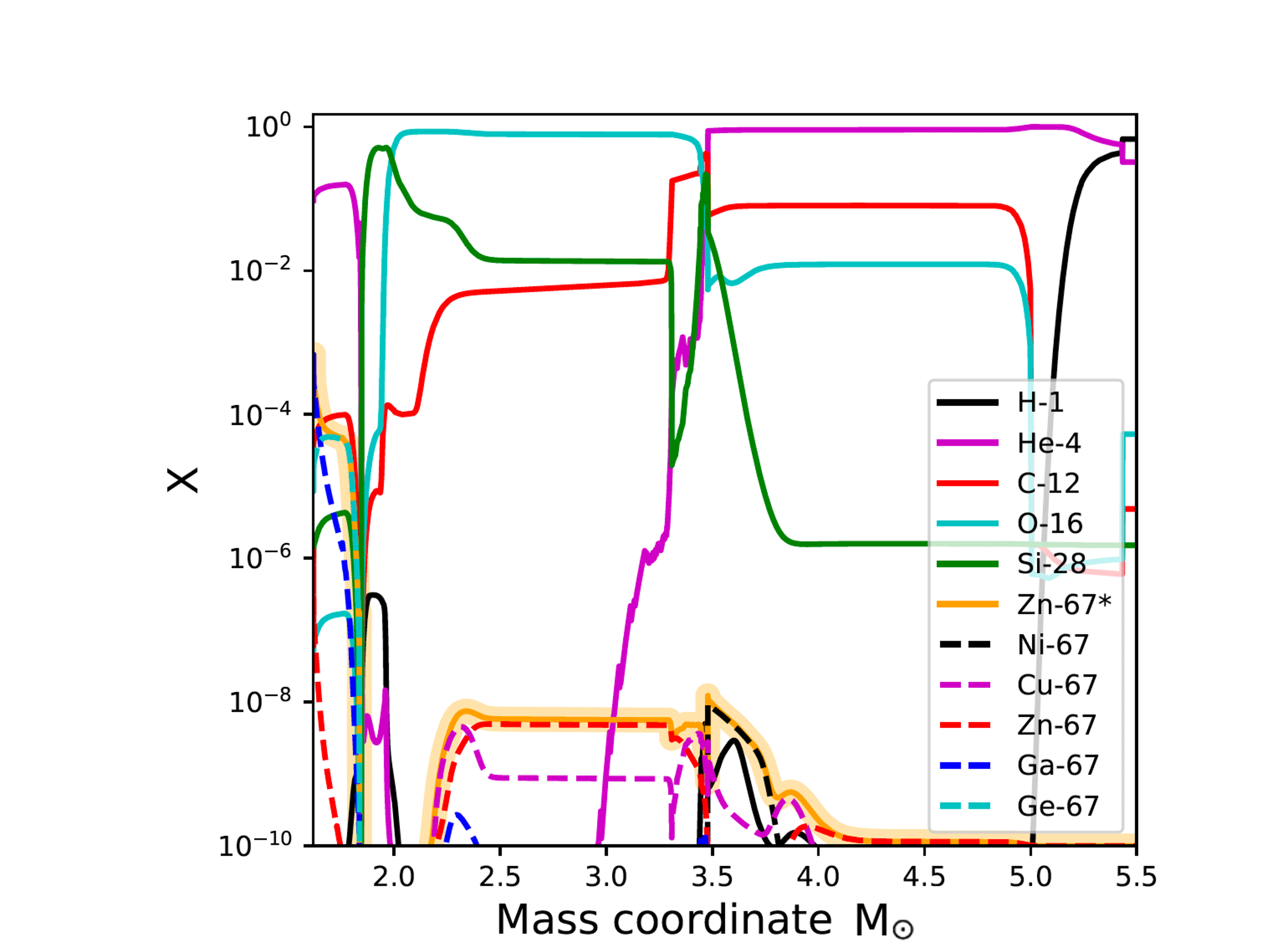}
    \caption{The same as in Figure \ref{fig:15_cu63}, but for $^{67}$Zn. }
    \label{fig:15_zn67}
\end{figure*}

The solar Zn composition is made up of 49.17\% of $^{64}$Zn, 27.73\% of $^{66}$Zn, 4.04\% of $^{67}$Zn, 18.45\% of $^{68}$Zn and 0.61\% of $^{70}$Zn. The nucleosynthesis of Zn in massive stars has been studied extensively for both CCSN and the most energetic hypernovae \citep[][and references therein]{kobayashi:20}, due to the puzzling high abundance of Zn with respect to Fe observed in the early Galaxy. This is mostly due to the primary production of $^{64}$Zn (and partly $^{66}$Zn). In our models, an effective s-process production is obtained for all Zn isotopes except for $^{64}$Zn. Since the s-process marginally produces Zn, here we focus in detail on $^{67}$Zn (Figure \ref{fig:15_zn67}) and $^{68}$Zn (Figure \ref{fig:15_zn68}). Analogous figures for $^{64}$Zn, $^{66}$Zn and $^{70}$Zn are available in Appendix. 

\begin{figure*}
    \centering
    \includegraphics[scale=0.41]
    {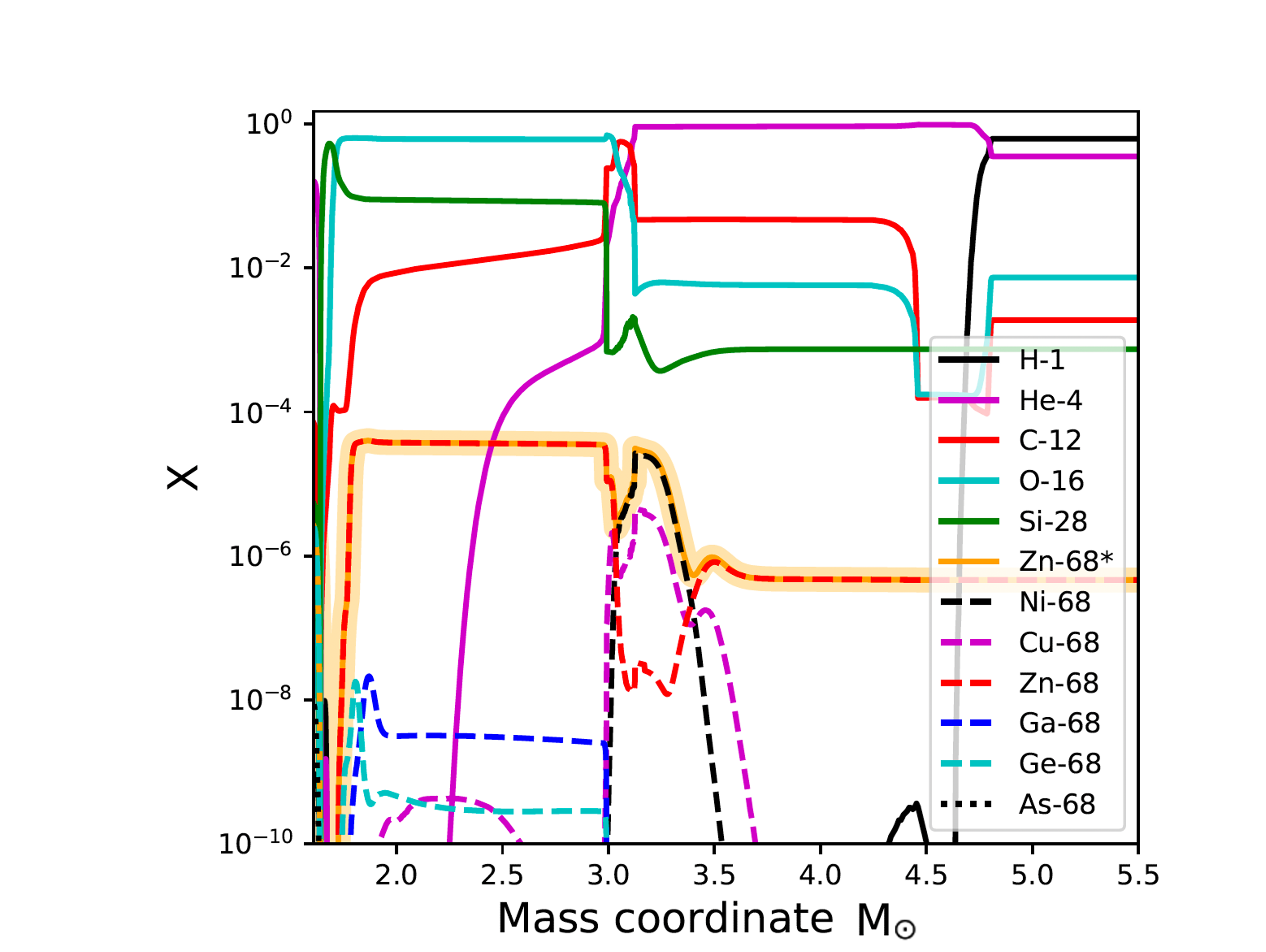}
    \includegraphics[scale=0.41]
    {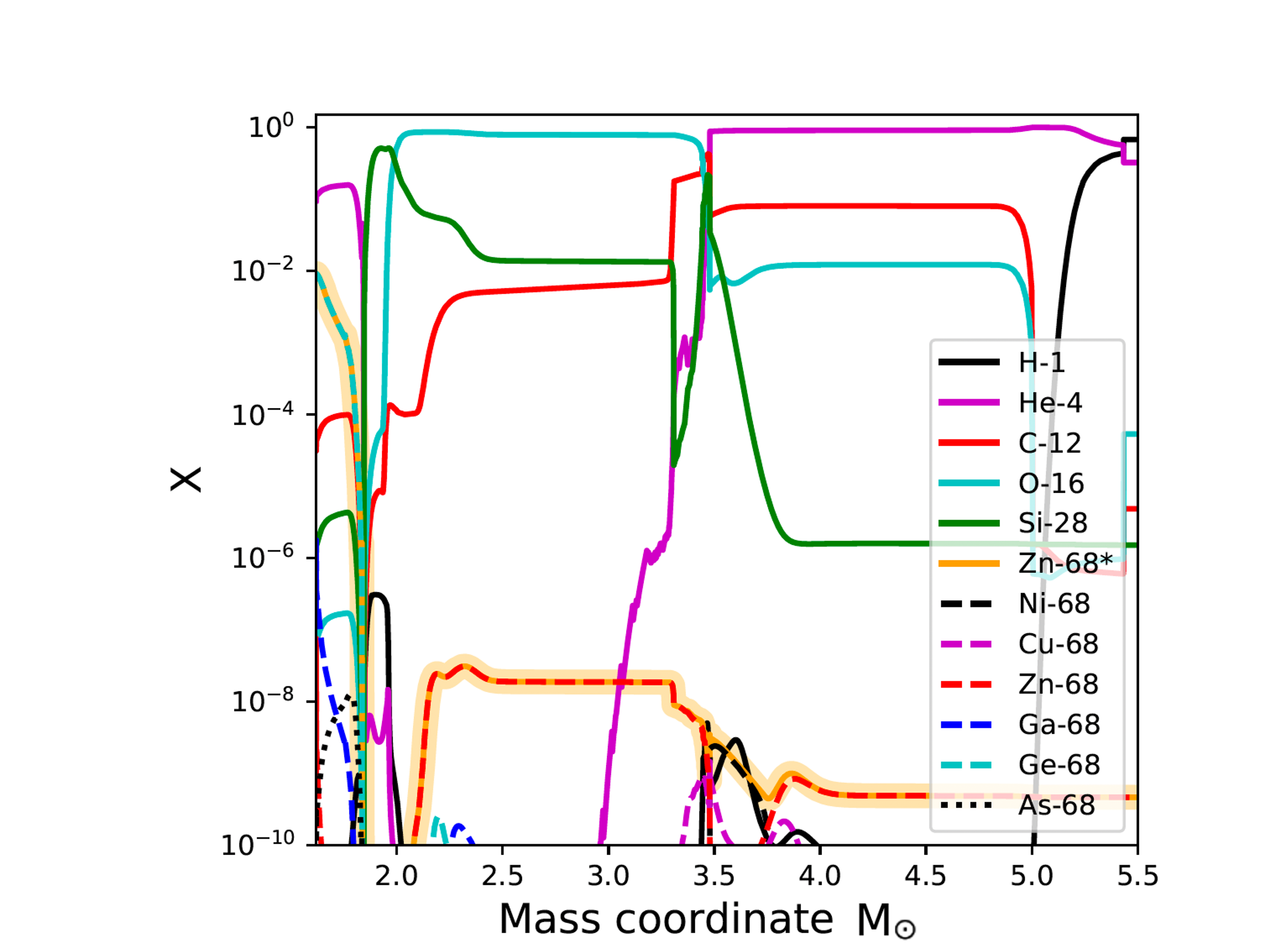}
    \caption{The same as in Figure \ref{fig:15_cu63}, but for $^{68}$Zn. }
    \label{fig:15_zn68}
\end{figure*}

In Figure \ref{fig:15_zn67} left panel ($Z$=0.02), $^{67}$Zn shows a less efficient s-process production compared to $^{66}$Zn
and $^{68}$Zn. This is a well-known effect in the s-process, where the neutron capture rate of $^{67}$Zn (with an odd number of neutrons) is much larger than the nearby species with an even neutron number. Interestingly, we can see in this model a strong n-process radiogenic production of $^{67}$Zn mostly as $^{67}$Ni and $^{67}$Cu, which in the integrated stellar yields is comparable to the s-process contribution. In the model at $Z$=0.0001 (Figure \ref{fig:15_zn67} right panel), there is a strong $\alpha$-process radiogenic production as $^{67}$Ge and $^{67}$Ga. The nucleosythesis of $^{68}$Zn shown in Figure \ref{fig:15_zn68} is very similar to the Cu isotopes. At low $Z$, the primary $^{68}$Zn production is dominated from the radiogenic contribution from $^{68}$Ge. 

\subsubsection{Gallium}

In the solar system, Ga is made of $^{69}$Ga (60.108\%) and $^{71}$Ga (39.892\%). The nucleosynthesis of Ga in CCSN shows remarkable similarities to Cu. Due to the difficulty in observing Ga in stars, a good understanding of their nucleosynthesis production could allow us to use Cu as an observational diagnostic for Ga \citep[][]{2010ApJ...710.1557P}. 

\begin{figure*}
    \centering
    \includegraphics[scale=0.41]
    {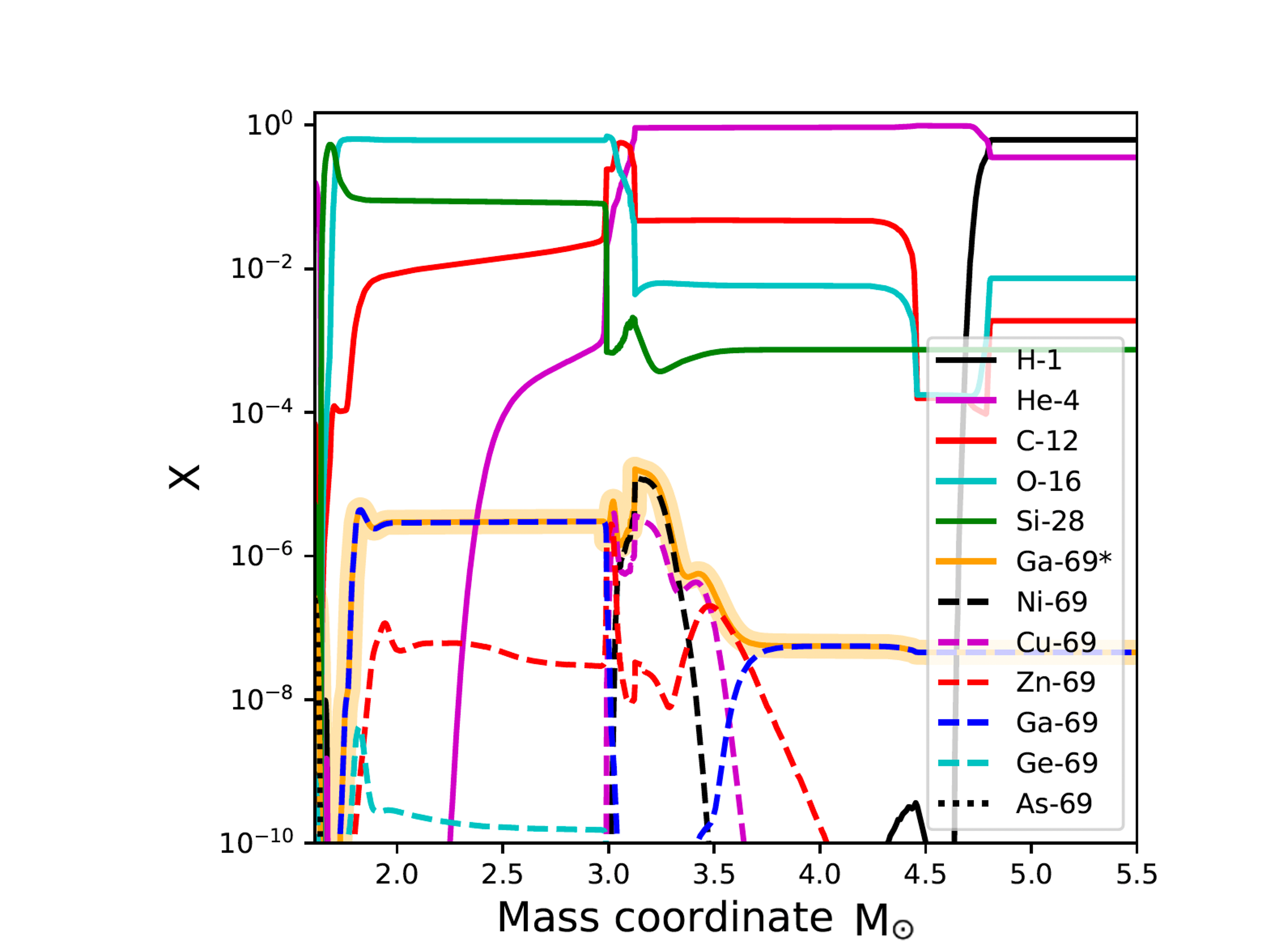}
    \includegraphics[scale=0.41]
    {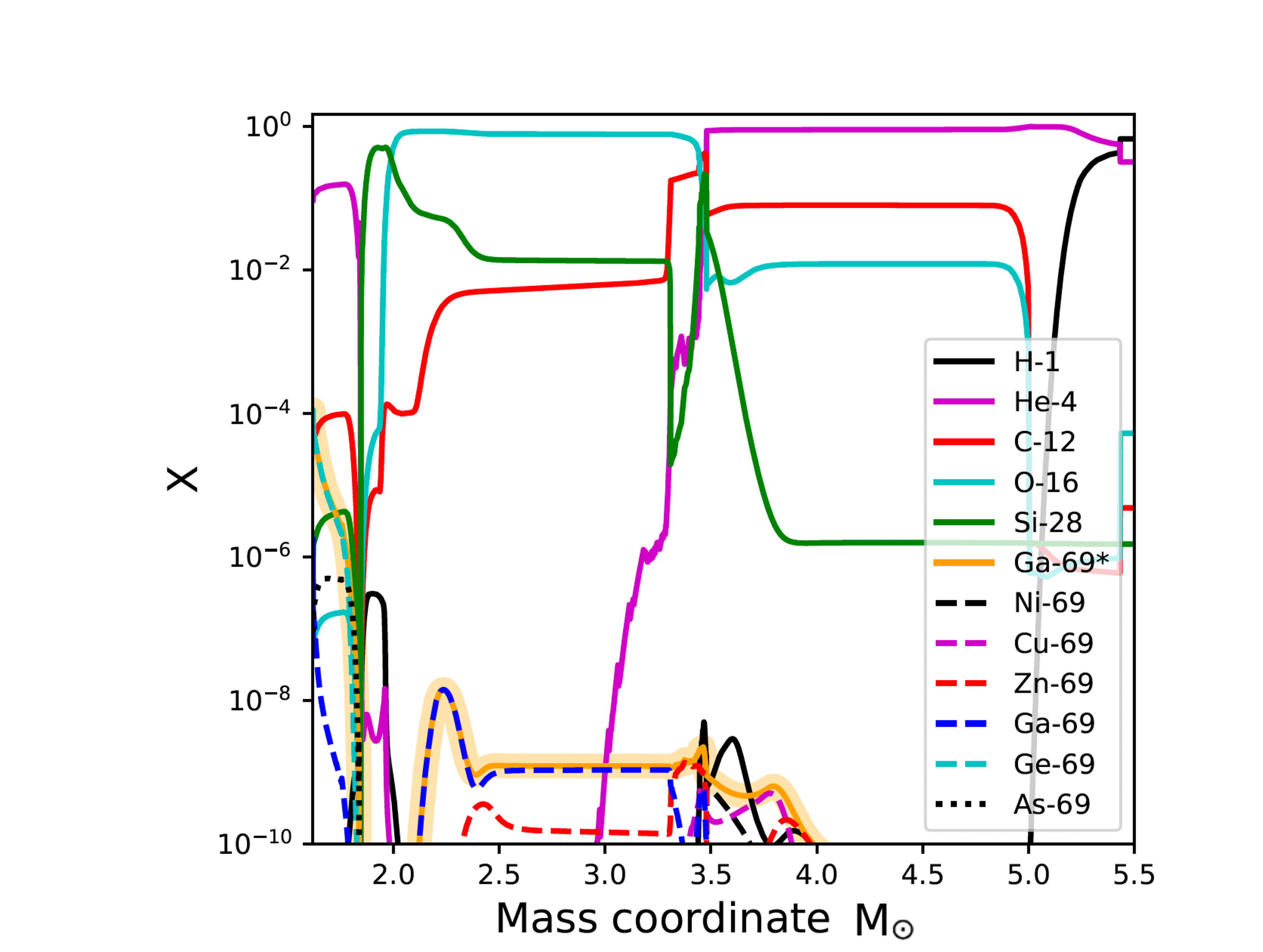}
    \caption{The same as in Figure \ref{fig:15_cu63}, but for $^{69}$Ga. }
    \label{fig:15_ga69}
\end{figure*}

In Figure \ref{fig:15_ga69}, $^{69}$Ga is produced directly by the s-process. In the left panel, the n-process component is built by the radiogenic contribution of $^{69}$Ni, $^{69}$Cu and $^{69}$Zn. In the right panel, the $\alpha$-process component is made of radiogenic $^{69}$Ge and small amounts of $^{69}$As. 

\begin{figure*}
    \centering
    \includegraphics[scale=0.41]
    {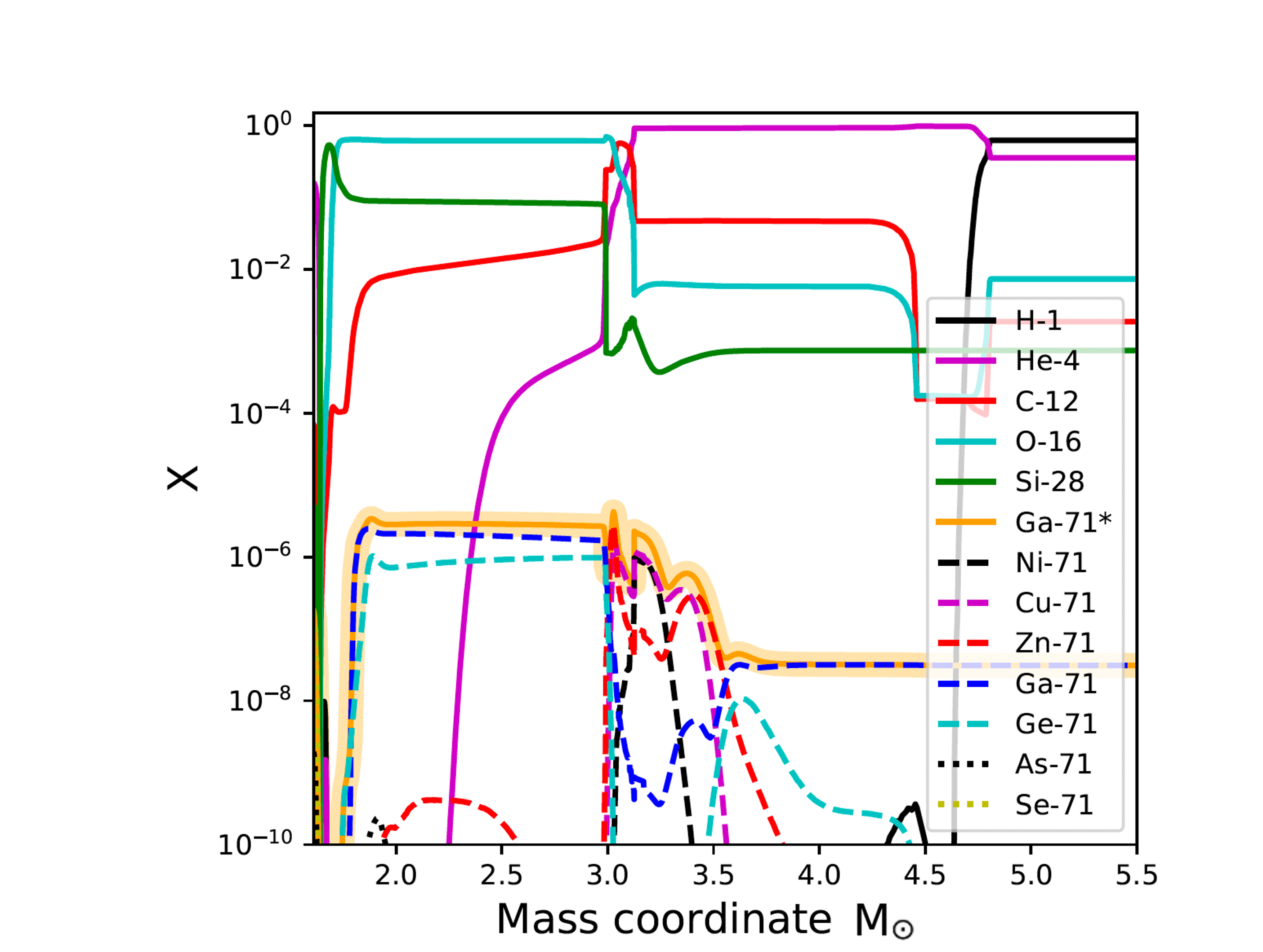}
    \includegraphics[scale=0.41]
    {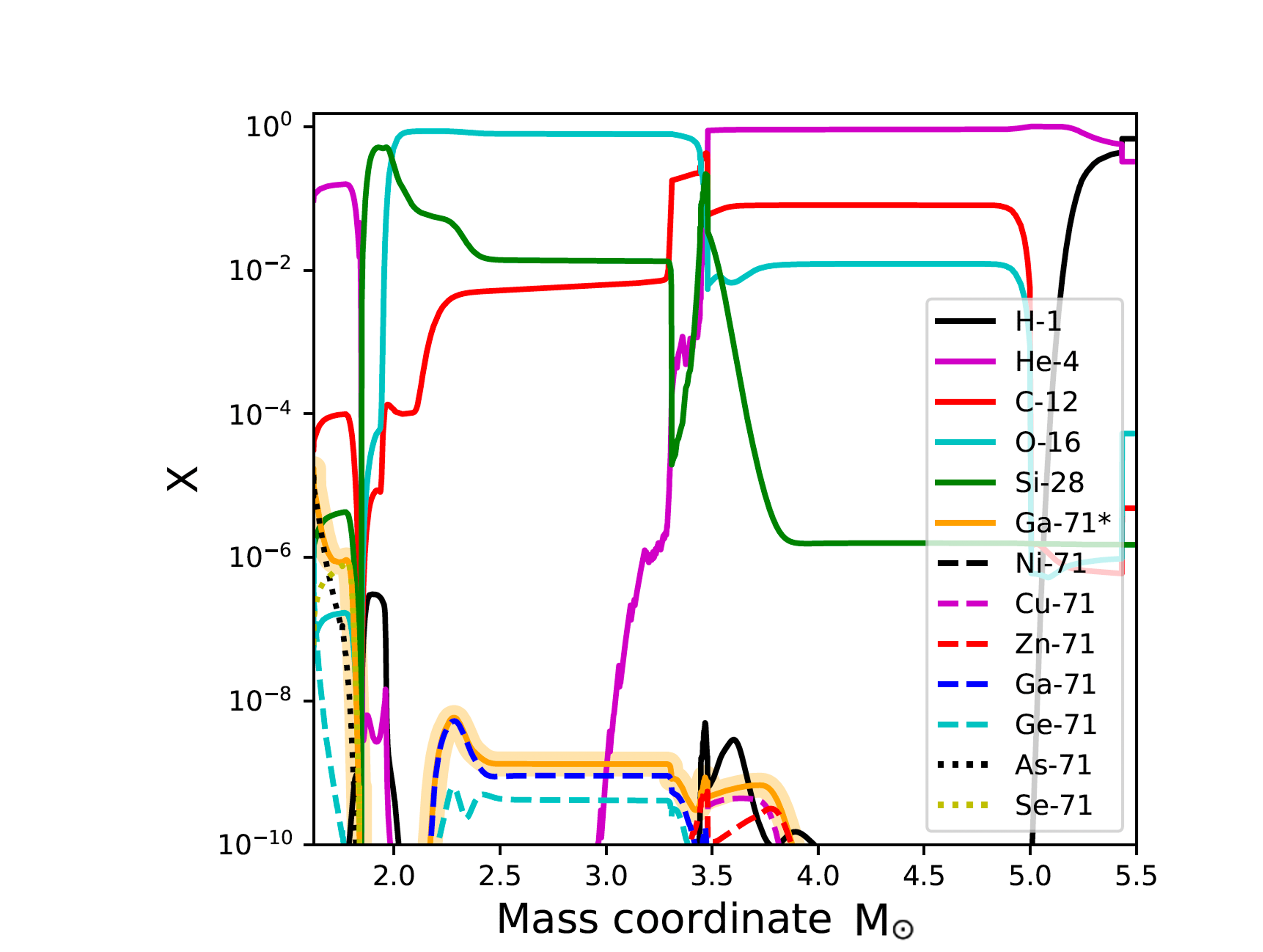}
    \caption{The same as in Figure \ref{fig:15_cu63}, but for $^{71}$Ga. }
    \label{fig:15_ga71}
\end{figure*}

\subsubsection{Germanium}

\begin{figure*}
    \centering
    \includegraphics[scale=0.41]
    {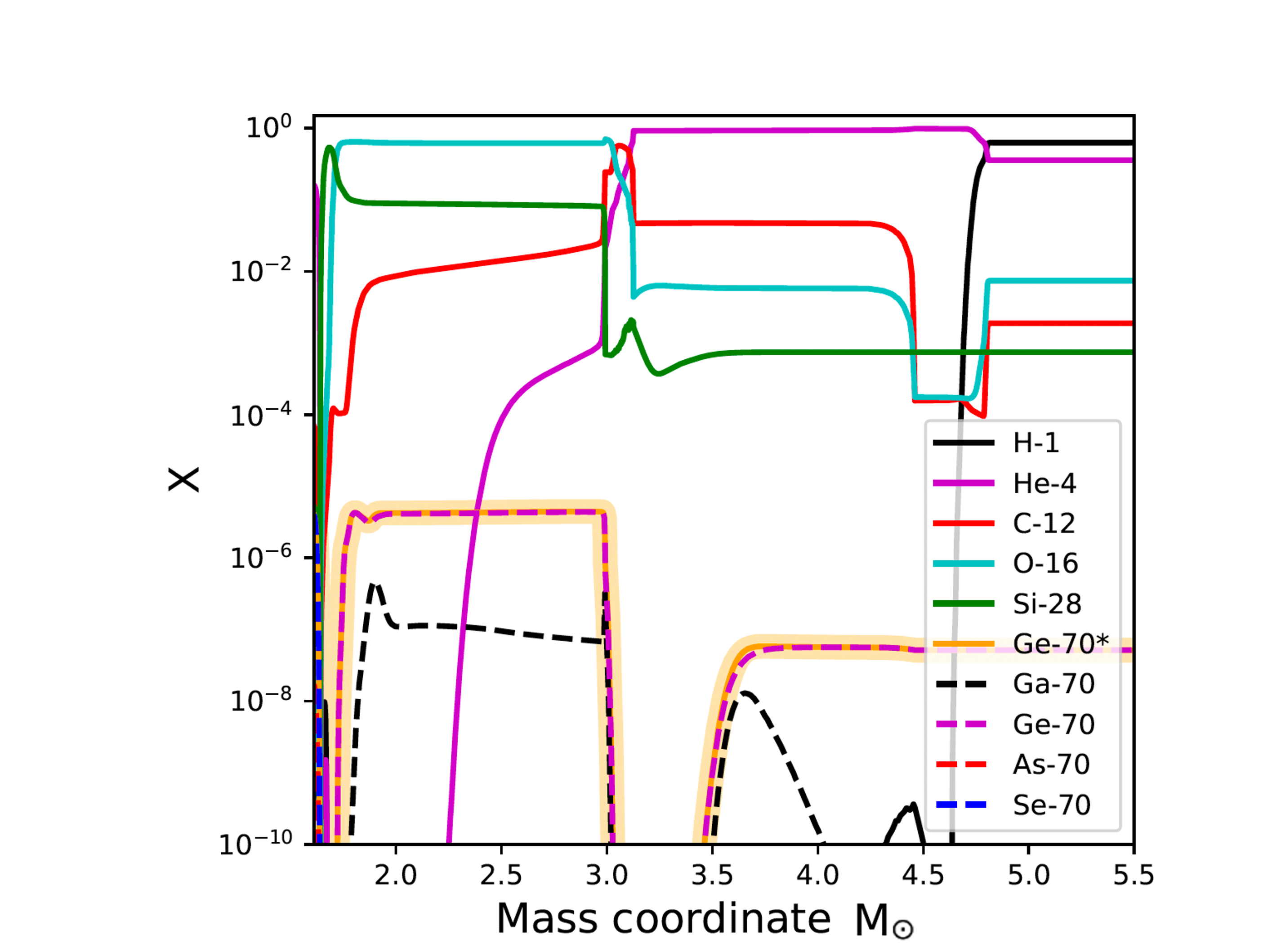}
    \includegraphics[scale=0.41]
    {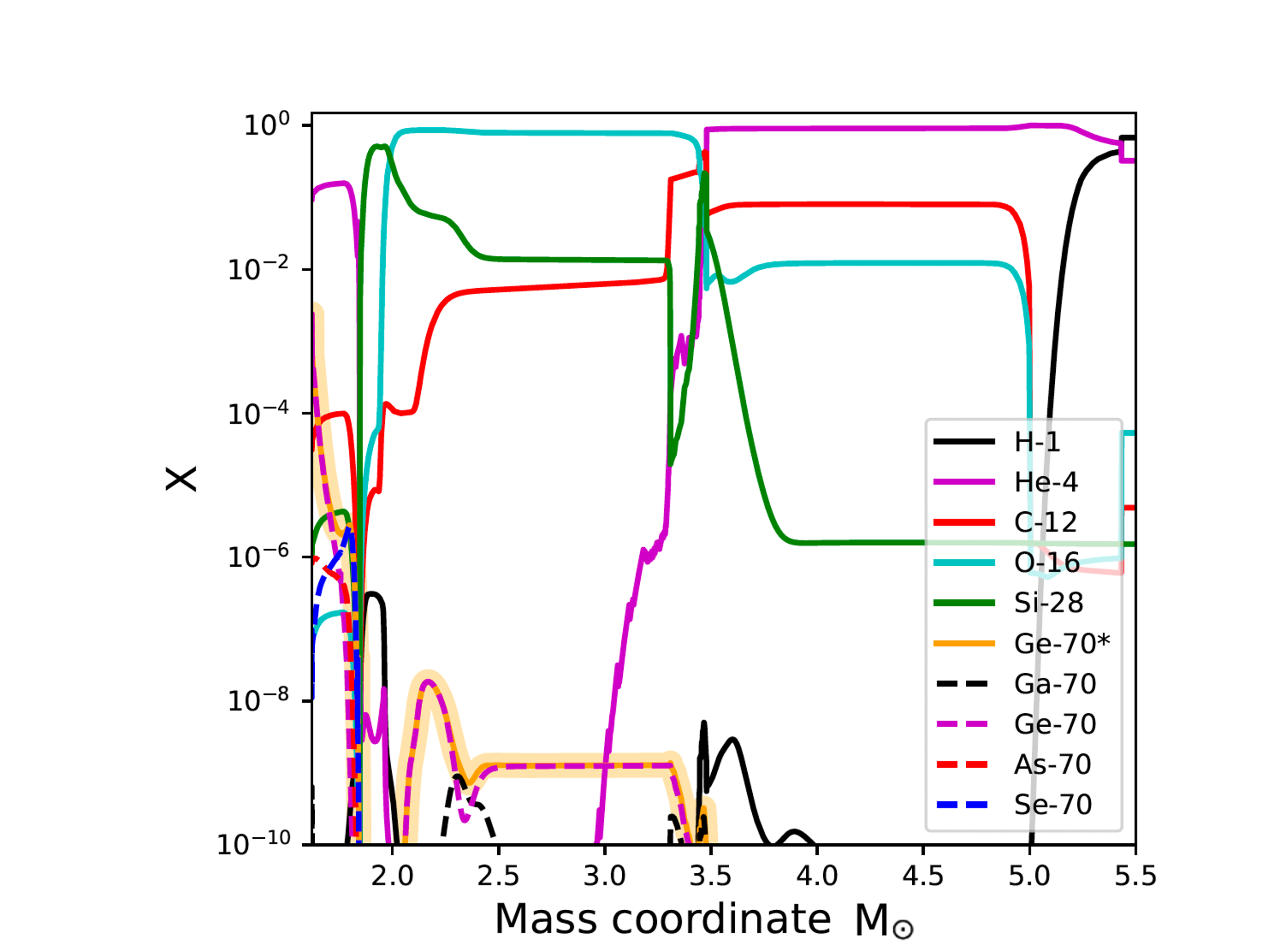}
    \caption{The same as in Figure \ref{fig:15_cu63}, but for $^{70}$Ge. }
    \label{fig:15_ge70}
\end{figure*}

The stable Ge isotopes $^{70}$Ge, $^{72}$Ge, $^{73}$Ge, $^{74}$Ge and $^{76}$Ge account for 20.52\%, 27.45\%, 7.76\%, 36.52\% and 7.75\% of the solar Ge, respectively. $^{70}$Ge is defined as an s-only isotope, meaning that its production is due to the s-process only, as the r-process contribution is shielded by the stable isobar $^{70}$Zn \citep[e.g.,][]{Kappeler2011}. In Figure \ref{fig:15_ge70}, left panel, we confirm this also for the n-process, where there is no direct or radiogenic contribution feeding $^{70}$Ge abundances. On the other hand, such a definition may not be entirely accurate, since it is possible to have a primary $^{70}$Ge production by the $\alpha$-process, mostly directly making the isotope and with some small radiogenic contribution from $^{70}$Se and $^{70}$As. A small primary component for Ge is indeed observed in metal-poor stars \citep[][]{cowan:05, 2010ApJ...710.1557P}. 

\begin{figure*}
    \centering
    \includegraphics[scale=0.41]
    {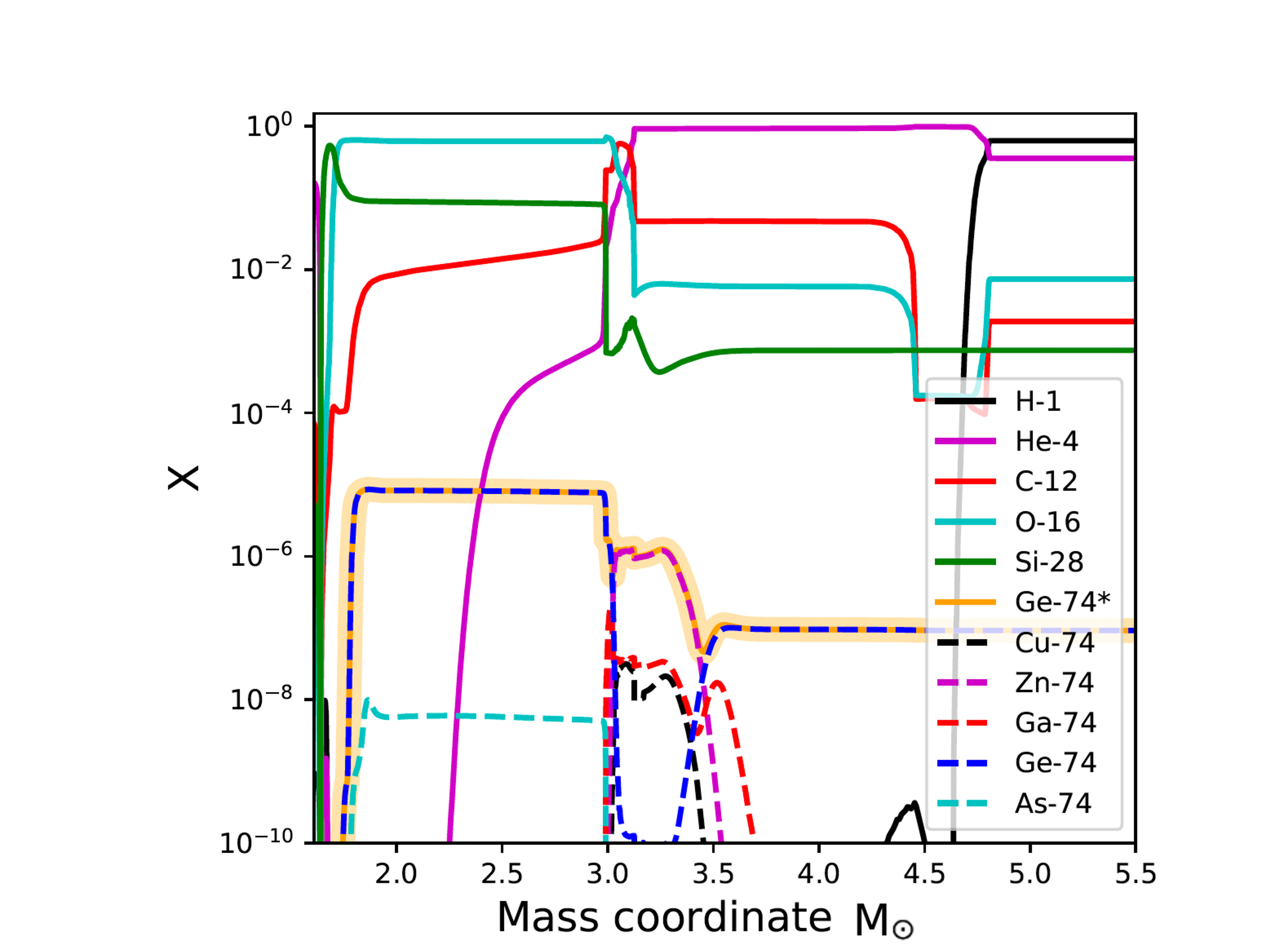}
    \includegraphics[scale=0.41]
    {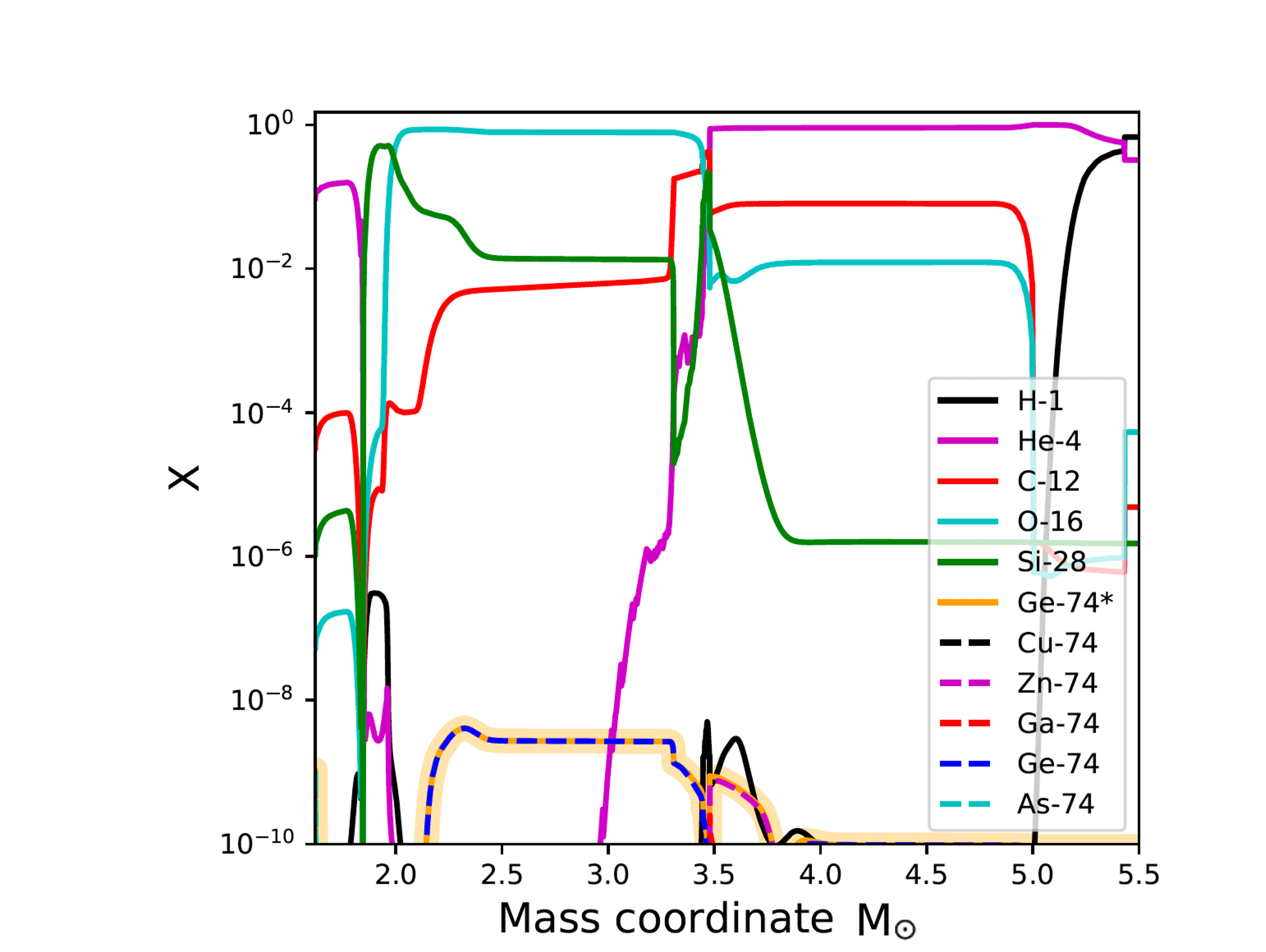}
    \caption{The same as in Figure \ref{fig:15_cu63}, but for $^{74}$Ge. }
    \label{fig:15_ge74}
\end{figure*}

Unlike $^{70}$Ge, in Figure \ref{fig:15_ge74}, left panel, $^{74}$Ge can also be ejected with a small n-process component due to the radiogenic contribution of $^{74}$Zn. However, the direct s-process production will dominate the production of this isotope. In Figure \ref{fig:15_ge74}, right panel, $^{74}$Ge does not show relevant primary production via the $\alpha$-process, due to the shielding from the stable isobar $^{74}$Se. 

Analogous figures are available in the Appendix for $^{72}$Ge, $^{73}$Ge and $^{76}$Ge.


\section{Stellar yields and GCE calculations} \label{sec:yields}

\subsection{Nucleosynthesis and stellar yield table - testing the impact of the $^{22}$Ne+$\alpha$ rates}
\label{subsec:ne22an_on_ccsn_profiles}
As mentioned, we performed the post-processing nucleosynthesis simulations for the 20 massive star progenitor models 
in the stellar set considered, at the given initial mass ($M_{ZAMS}$ = 12, 15, 20, and 25 $M_{\odot}$) and $Z$ ($Z$ = 0.0001, 0.001, 0.006, 0.01, and 0.02). For each progenitor, we performed three calculations using the upper bound, recommended, and lower bound for each of the LO12 and OT20 rates. The upper bound of each reaction rate was probed by using the upper limit of their $^{22}$Ne($\alpha$,n)$^{25}$Mg rate and the lower limit of their $^{22}$Ne($\alpha$,$\gamma$)$^{26}$Mg rate. The recommended calculation was made using the recommended rates for both reactions. For the lower bound, we instead used the lower limit of their $^{22}$Ne($\alpha$,n)$^{25}$Mg rate and the upper limit of their $^{22}$Ne($\alpha$,$\gamma$)$^{26}$Mg rate. 
A total of 120 (20 stellar models $\times$ 6 reaction rates) complete nucleosynthesis simulations were thus performed for LO12 and OT20, including the CCSN nucleosynthesis, to compare with the original yields set by \cite{Ritter2018}. 

Additional pre-SN nucleosynthesis calculations are performed for six variants of the OT20 recommended rate, and for the more recently evaluated rates by \cite{Adsley2021} and \cite{Wiescher2023} to better interpret the differences observed between the LO12 and OT20 rates. Thus, an additional 160 nucleosynthesis simulations were performed.

\begin{figure*}
    \centering
    \includegraphics[width=0.88\textwidth]{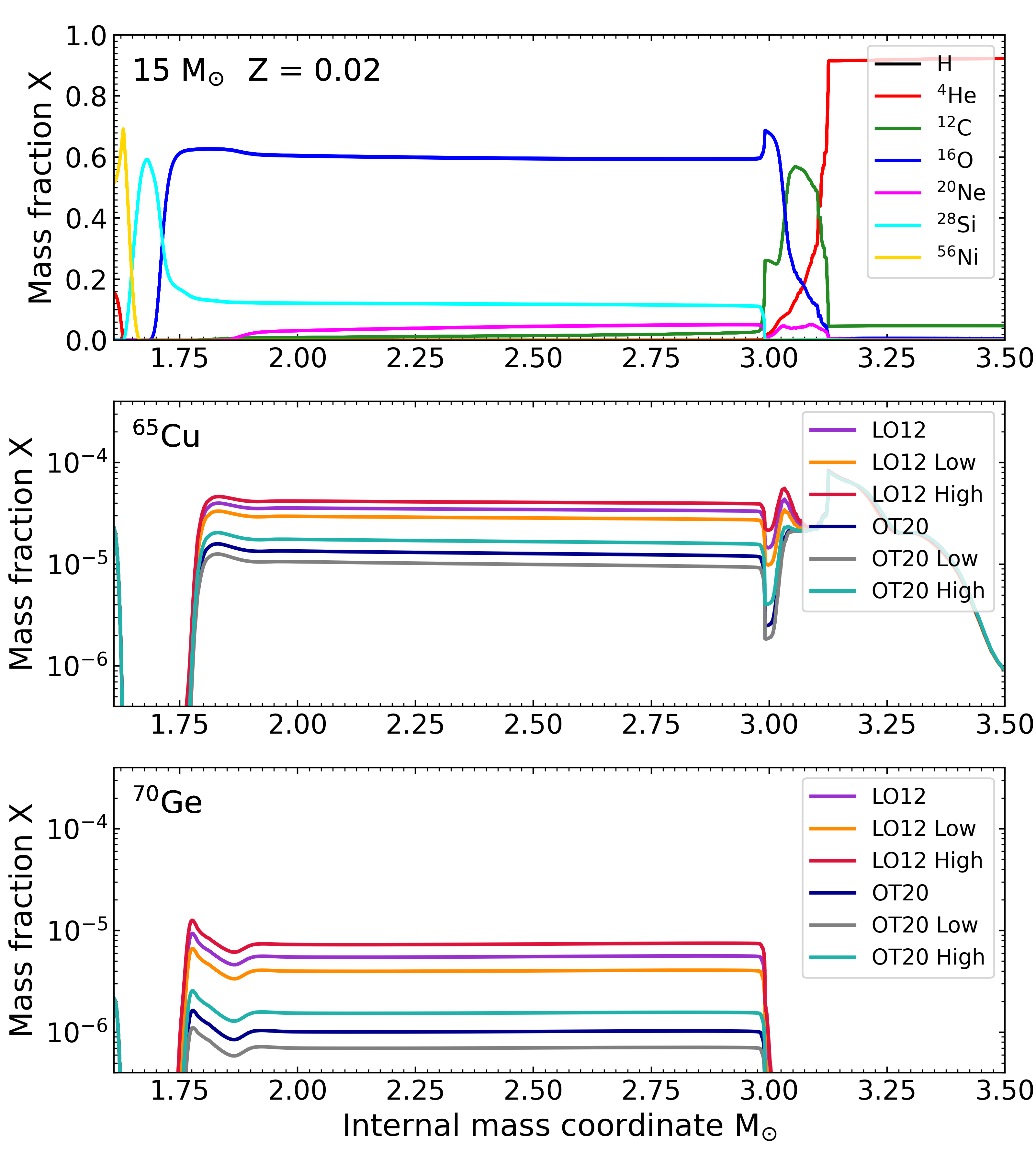}
    \caption{Post-SN stellar abundance profiles for major fuels ($\rm ^{1}H$, $\rm ^4 He$, $\rm ^{12} C$, $\rm ^{20} Ne$, $\rm ^{16} O$, $\rm ^{28} Si$, and $\rm ^{56} Ni$, upper panel), $^{65}$Cu (central panel) and $^{70}$Ge (lower panel) in the 15 $M_{\odot}$ model at $Z$=0.02. 
    }
    \label{fig:plots_with_postSN_Inside}
\end{figure*}

Figure \ref{fig:plots_with_postSN_Inside} shows the post-SN abundance profiles of $^{65}$Cu and $^{70}$Ge 
for the 15 $M_{\odot}$ at $Z$=0.02 model produced in the present work (similar to Figure \ref{fig:15_cu65} and \ref{fig:15_ge70} left panels). 
As expected from the discussion in subsection \ref{subsec:lights_production}, large variations caused by the different $^{22}$Ne+$\alpha$ rates are observed mainly between 1.8 $M_{\odot}$ and 3 $M_{\odot}$ (where the $s$-process is active), for both $^{65}$Cu and $^{70}$Ge. The largest variation is seen between the high bound setup of LO12 rates and the low bound setup of OT20. 
Such variations are mostly generated in the convective He-burning core in the progenitor star, carried in the following shell C-burning regions and finally ejected by the CCSN explosion. Regarding the explosive processes discussed earlier, the n-process is not largely affected by $^{22}$Ne+$\alpha$ uncertainties, since it operates at temperatures around 1 GK \citep[e.g.,][]{pignatari:18} and variations between different rates are much smaller than those induced by under He-burning conditions (see e.g., $^{65}$Cu in Figure \ref{fig:plots_with_postSN_Inside}). 

The $\alpha$ process is also not affected by the $^{22}$Ne+$\alpha$ rates. 
On the other hand, uncertainties in the explosion model may significantly affect the appearance of this component, and the use of a more sophisticated CCSN explosion simulation is mandatory to study in detail the nucleosynthesis nearby the PNS \citep[see, e.g.,][and references therein]{curtis:19,boccioli:23,BR24}. We plan to address this challenge in future work.

In the following section, we focus on the impact of the pre-SN yields on the GCE, aiming to isolate the effects of uncertainties in the $^{22}$Ne+$\alpha$ reaction rate. This approach allows us to disentangle these effects from those related to CCSN modeling, which will affect the production of Zn and the other elements discussed in the present work more at low $Z$. 

\subsection{GCE Simulations}

We calculated stellar yields from all the 280 nucleosynthesis simulations described in Section \ref{subsec:ne22an_on_ccsn_profiles} and used them as inputs for GCE calculations. Stellar wind yields, i.e., the time-integrated mass lost by stellar winds throughout the lifetime of the AGB and massive stars, are included in the total yields used for GCE calculations, although their contribution to the elements considered in the present work is negligible.

For AGB stars (with initial masses 1-7$M_{\odot}$), we used the original yields from \cite{Ritter2018}, while for Type Ia SN (SNIa), we used the fast deflagration W7 yields from \cite{Iwamoto1999}. We did not perform additional simulations on these astrophysical sources because the productions of Cu, Zn, Ga, and Ge are dominated by massive stars. Likewise, we did not consider the contribution from popIII stars or Hypernovae. These sources could potentially contribute to the abundances of these elements at very low $Z$ \citep[e.g.,][]{kobayashi:20}. For the present work, we will focus on the GCE of stars with [Fe/H]$>$-2. 
Stellar yields are interpolated below and above the transition mass of 8 $M_{\odot}$ for the AGBs and massive stars, respectively \citep[see][for detail]{Ritter2018syg}.

We used a two-zone GCE code, OMEGA+ \citep{2018APJ_Cote}. The first (inner) zone consists of the star-forming region in the galaxy modeled by OMEGA \citep{2017APJ_Cote} and the second (outer) zone represents 
the circumgalactic medium (CGM). The mass fraction of a given isotope ($X$) in the gas (with its mass denoted by $M_{gas}(t)$) within the inner zone at a given time $t$ will be obtained within OMEGA+ as follows \citep{Cote2016,2017APJ_Cote}. The temporal evolution of the gas within the inner zone ($\dot{M}_{gas}$) is calculated with the following equation:
\begin{equation}
    \dot{M}_{gas}(t) = \dot{M}_{inflow}(t) + \dot{M}_{ej}(t) - \dot{M}_{*}(t) - \dot{M}_{outflow}(t)
\end{equation}

where {$\dot{M}_{inflow}(t)$} represents the gas inflow from the CGM at $t$ after the Big Bang ($t$=0), {$\dot{M}_{ej}(t)$} the stellar ejecta, {$\dot{M}_{*}(t)$} the gas used for star formation, and {$\dot{M}_{outflow}(t)$} is the gas outflow into the CGM. Similarly, the evolution of gas within the CGM ($\dot{M}_{CGM, gas}$) is calculated as:
\begin{equation}
\begin{split}
    \dot{M}_{CGM, gas}(t) = \dot{M}_{CGM, in}(t) + \dot{M}_{outflow}(t) - \\\dot{M}_{CGM, out}(t) - \dot{M}_{inflow}(t)
\end{split}
\end{equation}
where $\dot{M}_{CGM, in}(t)$ and $\dot{M}_{CGM, out}(t)$ are the gas coming in or out from external intergalactic medium (IGM), respectively. 
In the present work, 
$\dot{M}_{CGM, in}(t)$ and $\dot{M}_{CGM, out}(t)$ 
are set to 0. We tested different constant values for them following the prescription in \cite{2017APJ_Cote} (and references therein; $\dot{M}_{CGM, in}(t)$) and within the range listed in Table 1 of \cite{2018APJ_Cote} ($\dot{M}_{CGM, out}(t)$) and observed negligible impacts on our GCE results. 

\begin{figure*}
    \centering
    \includegraphics[scale=0.55]
    {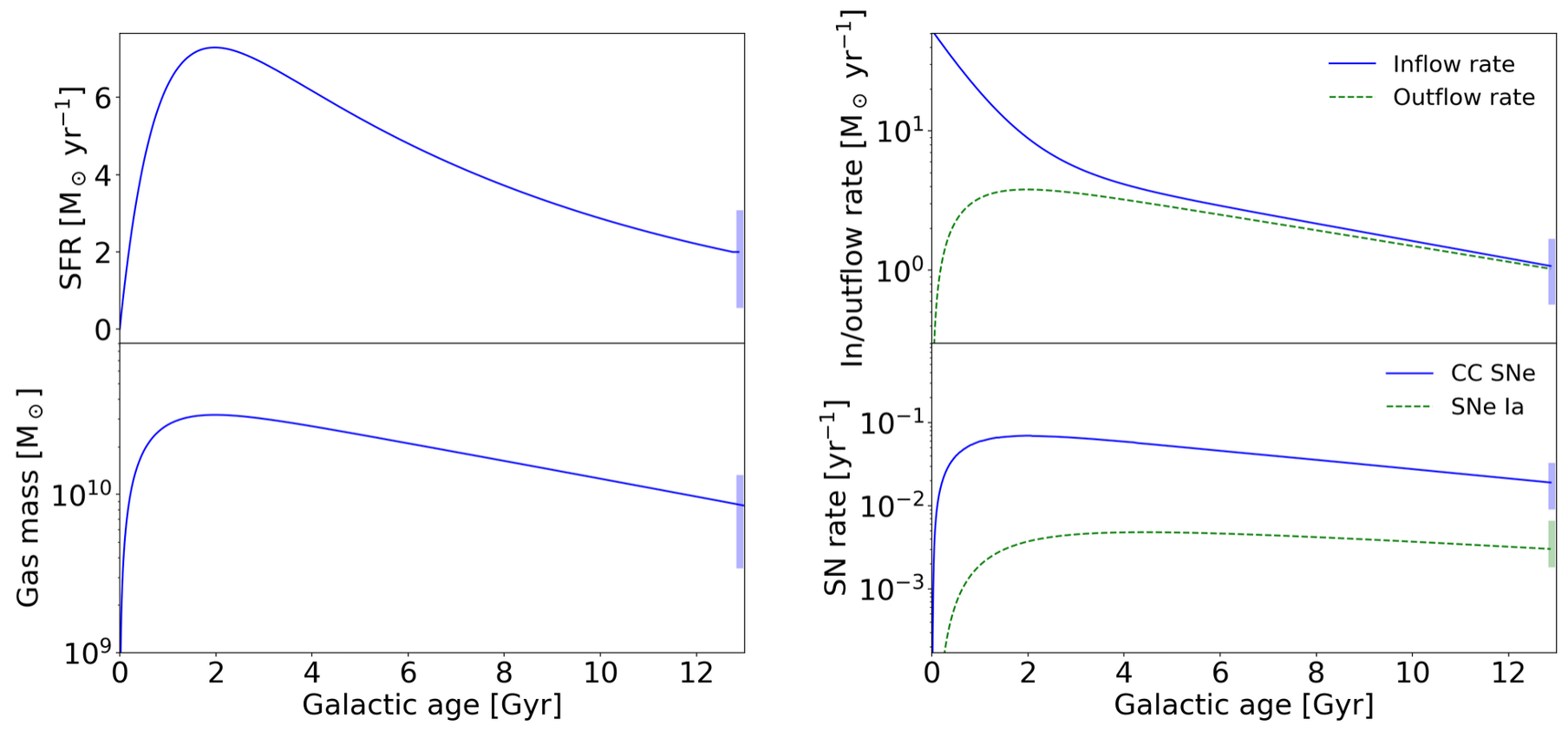}
    \caption{The stellar formation rate, gas mass, inflow and outflow rates, and supernova (CCSN and SNIa) rates as a function of the galactic age are plotted for the OMEGA+ model. The observational constraints \citep[from][]{kubryk:15} are plotted as the solid bars on the right side ($t\approx13$ Gyr). 
    }
    \label{fig:Diagonstic}
\end{figure*}

At each time-step $t_j$, OMEGA creates a simple stellar population (SSP) using SYGMA \citep{Ritter2018syg} to model the stellar ejecta. The SSP is a group of stars created at the same time with the same initial metallicity ($Z_{ZAMS}$, zero-age main sequence), where the initial mass function (IMF) of \cite{10.1046/j.1365-8711.2001.04022.x} defines the mass distribution of the AGB and massive stars. 
The stellar ejecta, $\dot{M}_{ej}$, at a certain time $t$ is calculated by taking the sum of the SSP's ejecta $\dot{M}_{ej}^j$ from $j$=1 to the total number of SSPs that have formed by time $t$. Thus,
\begin{equation}
    \dot{M}_{ej}(t) = \sum_j \dot{M}_{ej}^j(M_j, Z_j, t-t_j) 
\end{equation}
where $\dot{M}_{ej}^j$ depends on the initial mass population, $M_j$, the initial metallicity, $Z_j$ (=$Z_{ZAMS}$), and the age of the population, $t-t_j$.

For other variables ($\dot{M}_{inflow}$, $\dot{M}_{*}$, and $\dot{M}_{outflow}$), we followed a prescription provided by \cite{2019APJ_Cote}, in which model parameters were tuned to best reproduce observational constraints for the Milky Way disk \citep[][]{kubryk:15}. 

We used a two-infall model of gas inflow by \cite{1997ApJ...477..765C}, which accounts for two episodes of exponential gas inflow. This model is described with the following equation:
\begin{equation}
    \dot{M}_{inflow}(t) = A_1e^{\frac{-t}{\tau_1}} + A_2e^{\frac{t_{max}-t}{\tau_2}}
\end{equation}
where the time of maximum infall is set by the $t_{max}$, the continuation time of the infall is set by $\tau_1$ and $\tau_2$ for the first and second episode, respectively, and the normalization constants $A_1$ and $A_2$ represent the strength of each infall. 

\begin{table}[!htbp]
\begin{center}
    \caption{The values of the free parameters used in the two-infall model and the SFR from \cite{10.1093/mnras/stac3180}.}
    \label{table:1}
\begin{tabular}{c c}
 \hline
 Parameter & Value \\ [0.5ex] 
 \hline
 \hline
 $A_1$ $(M_{\odot}/yr)$ & 46 \\ 
 $A_2$ $(M_{\odot}/yr)$ & 5.9 \\
 $t_{max}$ (Gyr) & 1.0 \\
 $\tau_1$ (Gyr) & 0.8 \\
 $\tau_2$ (Gyr)  & 7.0\\
 $\epsilon_*$ & 0.23 \\
 $\tau_*$ (Gyr) & 1.0 \\ 
 $\eta$ & 0.52 \\ [1ex] 
 \hline
 \end{tabular}
\end{center}
\end{table}

The stellar formation rate (SFR) of the inner zone, $\dot{M}_{*}$, is set with the following equation:
\begin{equation}
    \dot{M}_{*}(t) = \frac{\epsilon_*}{\tau_*}M_{gas}(t)
\end{equation}
where {$\epsilon_*$} is the stellar formation efficiency and {$\tau_*$} is the star formation time scale. The outflow of the inner galaxy, $\dot{M}_{outflow}$, is proportional to the SFR, where the scale factor is set using a mass loading factor, $\eta$:
\begin{equation}
    \dot{M}_{outflow}(t) = \eta \dot{M}_{*}(t)
\end{equation}
The values of the parameters used above are taken from \cite{2019APJ_Cote} and \cite{10.1093/mnras/stac3180}, and they are listed in Table \ref{table:1}.

In Figure \ref{fig:Diagonstic}, 
we plotted some important galactic properties: $\dot{M}_{*}$ (SFR), ${M}_{gas}$ (the gas mass of the galaxy), $\dot{M}_{inflow}$, $\dot{M}_{outflow}$, the CCSN rate, and SNIa rate as a function of the galactic age. 
All these diagnostics fall within the observational limits, reproducing the results from \cite{2018APJ_Cote}. 
Note that the two infall-episodes are not as clearly distinguishable in these plots (e.g., $\dot{M}_{inflow}$) as those of \cite{1997ApJ...477..765C}. The bimodality derived from the two episodes is, however, more clearly visible when using different parameter values as shown in \cite{2019APJ_Cote}. Furthermore, since we do not include the threshold in stellar formation process as \cite{1997ApJ...477..765C} adopted, the two episodes are more smoothly connected to each other, making the bimodality less apparent. 


It should also be noted that, as discussed in \cite{Cote2016} and \cite{2019APJ_Cote}, choice of different $\dot{M}_{inflow}$ and $\dot{M}_{outflow}$ has non-negligible impact on GCE. For instance, when using sets of the parameter values to reproduce the high and low limits of the observational constraints \citep[see Fig. \ref{fig:Diagonstic} and ][]{2019APJ_Cote}, variations of $^{+0.2}_{-0.1}$ dex for Cu and Zn and $^{+0.3}_{-0.2}$ dex for Ga and Ge are found in the GCE results below [Fe/H]$=$--0.5, respectively. However, we confirmed that the relative differences observed in GCE using different reaction rates, which are the main interest of the present work, are negligibly influenced by the choice of the parameter sets. No GCE variations are indeed observed near solar $Z$.


Since the values of $\eta$, the mass-loading constant that controls $\dot{M}_{outflow}$, are not largely varied in the above parameter sets (0.52, 0.50, and 0.45 for the best-fit, low, and high limits, respectively; defined in Table 1 of \cite{2019APJ_Cote}), we further tested different values for $\eta$ to specifically see the role of $\dot{M}_{outflow}(t)$ in the GCE. The resulting differences in the GCE curves are negligibly small except for the final $Z$ (or [Fe/H]) where the GCE reaches at the end of the simulation ($t$=13 Gyrs). About 0.1 dex higher (lower) [Fe/H] is reached with a lower (higher) value of $\eta$=0 (1.0) because the enriched heavy elements in the Galaxy are exchanged for less enriched medium in the CGM. Note that we stop the simulations at $t$=13 Gyrs. The same simulation time was assumed e.g., in \cite{Kobayashi2006}, and using the more recently reported Milky Way galaxy's age ($\approx$13.6 Gyrs) only leads to small differences in our GCE results.

\begin{figure}
    \centering
    \includegraphics[scale=0.45]{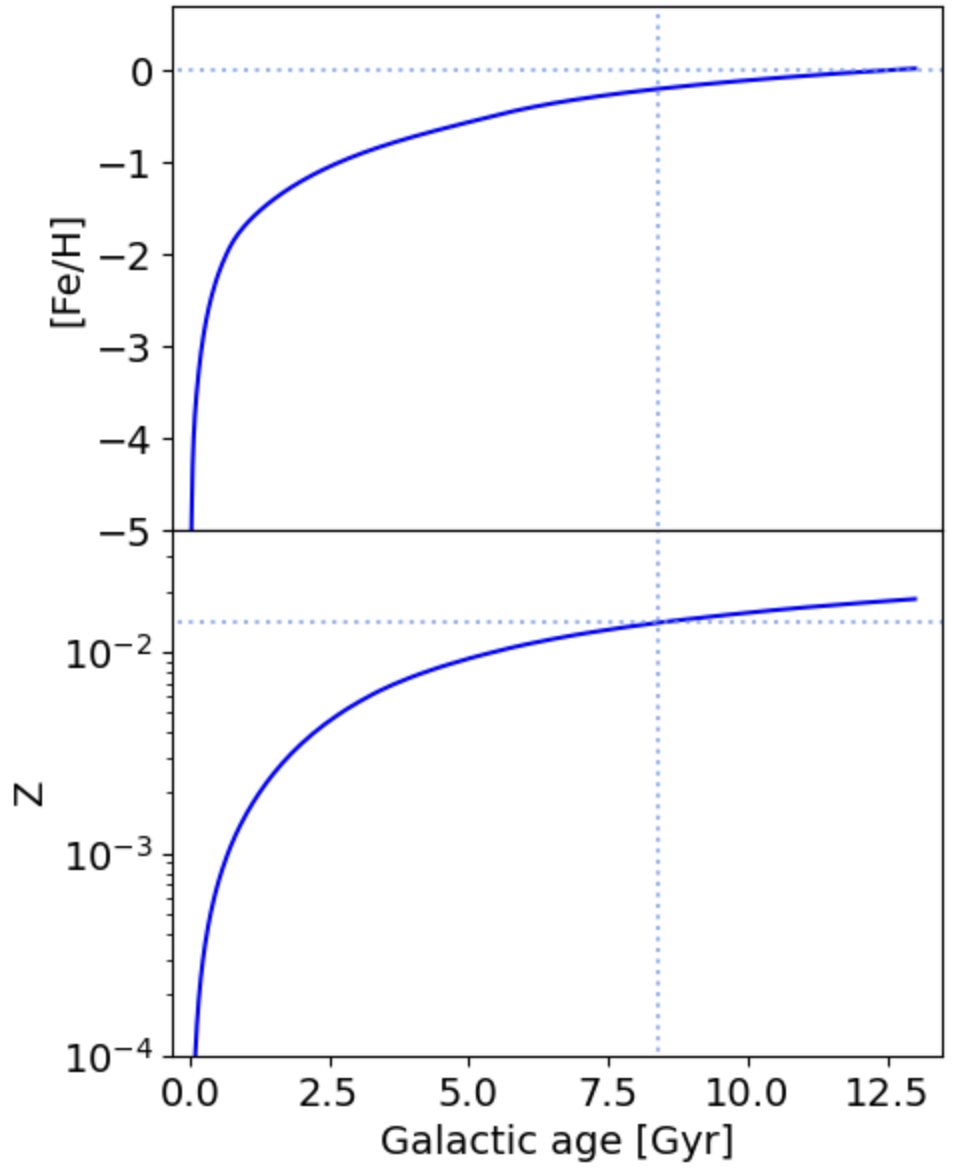}
    \caption{Evolution of the gas metalicity ([Fe/H] (top) and $Z$ (bottom)) predicted by the Milky Way Galaxy model used in the present work. The vertical and horizontal dotted lines represent the time at which the Sun is assumed to form and the solar metalicity \citep[]{Asplund2009}, respectively. See \cite{2019APJ_Cote} for detail as well.}
    \label{fig:diagnostic2}
\end{figure}

As additional benchmarking, we plotted the temporal evolution of [Fe/H] and $Z$ in Figure \ref{fig:diagnostic2}. In the figure, $Z$ reaches the solar value \citep[$Z$=0.014 from][]{Asplund2009} by the time the solar system was formed ($\approx$8.5 Gyrs) and continues toward moderate supersolar enrichment ($Z$=0.02). [Fe/H] however is slightly underestimated ($\approx$--0.2 dex) at $t$$\approx$8.5 Gyrs, possibly because we used $\alpha$-enhanced initial abundances for stellar yield calculations at $Z$$\le$0.006. 
Since the initial abundances of Fe are relatively small for the corresponding $Z$ (see above), it could lead to a smaller amount of Fe ejecta from massive stars and SNIa \citep[see e.g., ][]{Ritter2018syg}. We thus conclude that our GCE simulations are reasonably calibrated and uncertainties derived from the calibration process do not impact the conclusions of this work. 

Further details of the GCE code and the adopted GCE setup can be found in \cite{2018APJ_Cote}.

\subsection{Observational data}

The simulated GCE was compared with observational data collected from STELLAB: a code with a database of spectra observations from individual stars within the Milky Way \citep[see][]{2017APJ_Cote}. Most of STELLAB’s data are from the galactic disc, halo, and globular clusters within the Milky Way. In this work, we use the available stellar observations from the Milky Way disk to compare with our GCE simulations. 

To supplement the STELLAB data for Ga and Ge, we imported stellar abundance data from JINAbase \citep{2018ApJS..238...36A} 
and added observations from \cite{Ivans2003}, and from \cite{cowan:05}, \cite{Sneden1998}, and \cite{Placco2015} for Ga and Ge, respectively. 

\section{GCE Results} \label{sec:results}
\subsection{Impact of the $^{22}$Ne+$\alpha$ rate uncertainties on the GCE of Cu, Zn, Ga and Ge: simulations using preSN yields}
\label{subsec: gce_preSN}

\begin{figure*}
    \centering
    \includegraphics[scale=0.70]
    {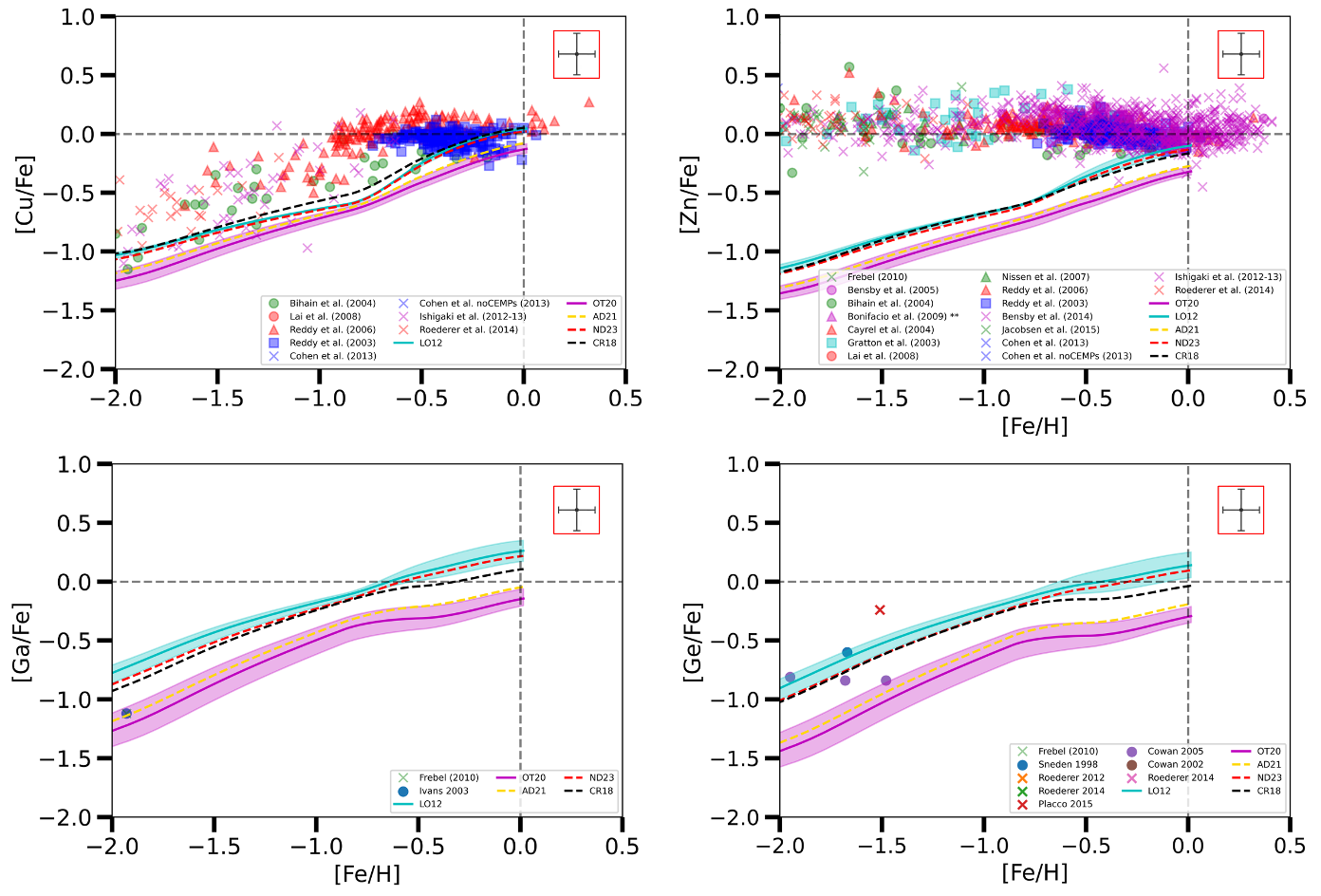}
    \caption {
    The results of GCE calculations using pre-SN yields for the selected elements (Cu, Zn, Ga, and Ge) in comparison with observational data. The error bars in the red box show the typical observational uncertainties ($\sigma$; $\pm$0.1 dex for x and $\pm$0.2 dex in y). 
    The uncertainties from the LO12 and OT20 rates are represented by the band (see text for details). The calculated results using AD21, ND23, and CR18 rates are also presented. The crossing point of the dashed lines represents the solar metalicity ([X/Fe] = [Fe/H] = 0). The observational data is from \cite{Bihain2004,Lai2008,Reddy2003,Reddy2006,Cohen2013,Ishigaki2013,Roederer2014,Frebel2010,Bensby2005,Bonifacio2009,Cayrel2004,Gratton2003,Bensby2014,Jacobson2015,Ivans2003,Sneden1998,cowan:05,Placco2015}.
    }
    \label{fig:plots}
\end{figure*}

In Figure \ref{fig:plots}, we show the pre-SN GCE curves for Cu, Zn, Ga, and Ge calculated from the massive star yields using different $^{22}$Ne+$\alpha$ reaction rates, in comparison with observational data. 
Abundance evolution for each element is shown as 
[X/Fe] ratios\footnote{We remind that [X/Y] = $log_{10}({X}/{Y}) - log_{10}({X}/{Y})_{_{\odot}}$ (mass fractions are used for $X$ and $Y$)}. The solar values are from \cite{Asplund2009}. 
Only the gas component (i.e., isotopes not trapped within stars) is considered. 

The GCE curves shown in Fig. \ref{fig:plots} are produced using the yields from the nucleosynthesis calculated with LO12 and OT20 rates including their uncertainties. Additional curves are calculated with other $^{22}$Ne+$\alpha$ rates, in particular the recommended rates from \cite{Adsley2021} (Adsley-TAMU, hereafter AD21), from \cite{Wiescher2023} (Notre-Dame (ND), hereafter ND23), and the original \cite{Ritter2018} configuration (hereafter CR18). Although not shown in the figure, we confirmed that the upper and lower limits of the GCE curves using the yields from ND23 and AD21 rates lie within the curves bounded by the LO12 and OT20 rates. 

\begin{table*}[htbp]
\begin{center}
    \caption{The differences (in unit of dex) of elemental abundances ([X/Fe]) in the GCE calculated using LO12 and OT20 rates. These are all evaluated by [X/Fe]$_{LO12}$--[X/Fe]$_{OT20}$ (dex) at [Fe/H]=0. The values in parenthesis denote the uncertainties taking into account those from LO12 and OT20. 
    }
    \label{table:2}
\begin{tabular}{c c c c c c c c c c c}
\hline
\hline
 Element & Cu & Zn & Ga & Ge & As & Se & Br & Kr & Rb & Sr\\ [0.5ex] 
 $Z_{el}$ & 29 & 30 & 31 & 32 & 33 & 34 & 35 & 36 & 37 & 38 \\
 \hline
 LO12-OT20 & 0.18 (6) & 0.22 (7) & 0.41 (12) & 0.44 (14) & 0.53 (17) & 0.37 (15) & 0.38 (14) & 0.37 (19) & 0.41 (20) & 0.41 (26) \\ 
 \hline
\end{tabular}
\end{center}
\end{table*}

A wealth of observational data is available for Cu and Zn; therefore, these elements have been studied in detail in previous GCE research 
\citep[e.g.,][]{Matteucci1993, Timmes1995, 1997ApJ...477..765C, Mishenia2002, Bisterzo2005, Kobayashi2006, mishenina:17, kobayashi:20, lach:20}. The availability of observational data for Ga and Ge is instead scarce, with only a handful of metal-poor stars measured in the $Z$ range considered for this work ([Fe/H]$>-2$). 

The [Cu/Fe] and [Zn/Fe] variations due to the OT20 and LO12 uncertainties are less than $\pm$0.05 dex, while they increase up to $\pm$$\approx$0.1 dex for [Ga/Fe] and [Ge/Fe] (see Figure \ref{fig:plots}). 
However, the OT20 and LO12 distributions do not overlap in the figure, with [Cu/Fe] and [Zn/Fe] variations between the two rates (LO12 upper bound and OT20 lower bound) up to 0.4 dex at solar metallicity, and up to about 0.7 dex for [Ge/Fe] and [Ga/Fe]. 

The [X/Fe] values relative to the GCE calculation performed with yields using other rates typically evolve between the OT20 and LO12 limits. Except for Zn, all simulations are consistent with the solar values within the observational uncertainties. The inconsistency found in Zn is not surprising because of the strong primary Zn component that is not accounted for in these simulations \citep[][]{2010ApJ...710.1557P}. 
As mentioned earlier, the source of the primary Zn production is still a matter of debate \citep[][]{Bisterzo2005}, with possible large contributions from hypernovae in the early Galaxy \citep[][and references therein]{kobayashi:20}. Thus, the Zn discrepancy shown in Figure \ref{fig:plots} is not related to the current sensitivity study of the $^{22}$Ne$+$$\alpha$ reaction rates.  

While the solar abundance of Cu is better reproduced by the LO12 or the CR18 rates, those of Ga and Ge seem to be better reproduced by the CR18 rate, lying between the OT20 and LO12 rates. 
Future stellar observations of Ge and Ga for $Z$ typical of the Milky Way disk would be needed to derive stronger constraints.
For Cu, the GCE models tend to underproduce [Cu/Fe] for $Z$ lower than solar. This could be justified by the missing primary Cu component in the pre-SN yields. 

The variation of [X/Fe] obtained using the OT20 and LO12 recommended rates is summarized in Table \ref{table:2}. Here we report the variation for all elements between Cu and Sr. In general, 0.2-0.5 dex variation was found for [X/Fe]. However, the impact on the GCE of As and other heavier elements should be re-evaluated in the future 
since the massive star's $s$-process contribution for As and other heavier elements is expected to be smaller compared to AGB stars and/or explosive nucleosynthesis \citep[][]{Kappeler2011}. 

\subsection{Resonance Sensitivity}
\label{subsec:sensitivity}
In this section, we investigate the difference in the LO12 and OT20 rates in more detail. We study the impact on the GCE of the differences in the two rates, focusing on the contributions from individual resonances. 
The $\omega\gamma$ of relevant resonances to the weak $s$-process identified as having large uncertainties by 
\cite{Ota2021} and by the recent review article \cite{Wiescher2023} are summarized in Table \ref{table:3}. 
It should be noted that the adopted strengths and upper limits (denoted by "$<$") have substantially different meanings in stellar reaction rates as described below. The upper limits are not calculated using the specified $\omega\gamma$. 

While the adopted strength was used in the stellar reaction calculations, the upper limit was used to determine the median and probability distribution assuming the Porter-Thomas distribution \citep{Longland2010,Longland2012} in the present work. As such, the actual (e.g., median) strength used in stellar rate calculations is much smaller than the upper limit used as it is in e.g., \cite{Angulo1999,Kappeler1994,Massimi2017}. 
For example, $\omega\gamma_{(\alpha,\gamma)}$ for the $E_x$=11.17 MeV resonance in LO12 has a median of $\approx$10$^{-8}$ eV ($\ll$ 4.30$\times$10$^{-7}$ eV upper limit). 
Thus, the resonance's impact on the stellar yield is negligibly small using either the LO12 or OT20 ($<$5.42$\times$10$^{-11}$ eV) rate. 

\begin{table*}[!htbp]
\begin{center}
    \caption{Varied $^{22}$Ne+$\alpha$ resonance strengths $\omega\gamma_{(\alpha,x)}$ (in unit of eV; $x$=$n$ or $\gamma$) to test the GCE sensitivity. $E_x$ and $E_{CM}$ are from \cite{Wiescher2023}. $\omega\gamma$ (case $n$) denotes the $\omega\gamma$ used for a specific resonance in Case $n$'s calculations while $\omega\gamma$ for all other resonances are kept the same as $\omega\gamma$(OT20) (see text for more details). For $E_x$$>$11.32 MeV, no specific resonances were used but the reaction rate is changed (thus denoted as "--").}
    \label{table:3}
\begin{tabular}{c c c c c c c}
\hline\hline
 $E_x$ (MeV) & 10.95 $(\alpha,\gamma)$ & 11.09 $(\alpha,\gamma)$ & 11.11 $(\alpha,$n$)$ & 11.17 $(\alpha,\gamma)$ & 11.32 $(\alpha,$n$)$ & $>$11.32 $(\alpha,$n$)$  \\ [0.5ex] 
 $E_{CM}$ (MeV)  &0.336 & 0.471 & 0.498 & 0.551 & 0.702 & $>$0.702 \\ [0.5ex] 
 \hline
  $\omega\gamma$ (LO12) & $<$8.70$\times$10$^{-15}$ & 0 & $<$5.80$\times$10$^{-8}$ & $<$4.30$\times$10$^{-7}$ & 1.40$\times$10$^{-4}$ (30) & -- \\ 
  $\omega\gamma$ (OT20) & 9.00$\times$10$^{-14}$ (270) & 2.85$\times$10$^{-10}$ (157) & $<$1.10$\times$10$^{-9}$ & $<$5.42$\times$10$^{-11}$ & 4.21$\times$10$^{-5}$ (69) & -- \\ 
  $\omega\gamma$ (Case 1) &  &  &  &  & 1.40$\times$10$^{-4}$ &  \\   
  $\omega\gamma$ (Case 2) &  &  &  & 6.60$\times$10$^{-7}$ &  &  \\
  $\omega\gamma$ (Case 3) &  &  & 1.35$\times$10$^{-8}$ &  &  &  \\  
  $\omega\gamma$ (Case 4) &  & 0 &  &  &  &  \\ 
  $\omega\gamma$ (Case 5) & 8.70$\times$10$^{-15}$ &  &  &  &  &  \\ 
  $\omega\gamma$ (Case 6) & &  &  &  &  & $\times$0.5 \\ 
 \hline
\end{tabular}
\end{center}
\end{table*}

As mentioned in Section \ref{sec:yields}, we performed GCE calculations for six different variants of the OT20 recommended rate (Case 1-6, as summarized in Table \ref{table:3}). 

In Case 1, the LO12 $\omega\gamma_{(\alpha,n)}$ rate was used for the $E_x$=11.32 MeV resonance. 

In Case 2, the $\omega\gamma_{(\alpha,\gamma)}$ for the $E_x$=11.17 MeV resonance was increased to 660 neV, which is obtained by \cite{Talwar2016} from their indirect measurement. Note that the existence of the resonance was invalidated by the latest direct measurement experiment by \cite{Shahina2022}, and an upper limit of 150 neV was determined. Thus, the median strength should be much smaller ($<$10 neV). Still, we use 660 neV to probe the maximum impact of this resonance. 

In Case 3, we increased the $\omega\gamma_{(\alpha,n)}$ for the $E_x$=11.11 MeV resonance to 13.5 neV (approximately the value adopted by \cite{Massimi2017}; see \cite{Ota2021}). It should be noted that the resonance is dominated by ($\alpha$,n) channel and the $\omega\gamma_{(\alpha,n)}$ is constrained to the same level as the $E_x$=11.17 MeV resonance ($\lesssim$60 neV) by direct measurements \citep{Jaeger2001}. 
The upper limit used by LO12 (58 neV) reflects this value. 
We studied this resonance only for its potential contributions, as theorized 
by \cite{Massimi2017}. 
However, since this is the lowest possible resonance for ($\alpha,n$) channel, the weak $s$-process could be activated at lower temperature ($<$0.3 GK) if the resonance strength is high.

In Case 4, the $\omega\gamma_{(\alpha,\gamma)}$ for the $E_x$=11.09 MeV resonance was set to 0 as LO12. 

In Case 5, the same $\omega\gamma_{(\alpha,\gamma)}$ for LO12 was used for the $E_x$=10.95 MeV resonance (but not as an upper limit). 

In Case 6, we decreased the OT20's ($\alpha$,$n$) reaction rate at $\geq$ 1 GK by a factor of 2. Since a number of resonances (primarily $E_x$$\approx$11.8 MeV and above) contribute to the $^{22}$Ne($\alpha$,$n$)$^{25}$Mg reaction, the relevance of the $^{22}$Ne($\alpha$,$\gamma$)$^{26}$Mg becomes negligible.  
In this temperature range, we estimated the summed impact of all the high energy resonances by simply applying a correction factor to the reaction rate. 

\begin{figure*}
    \centering
    \includegraphics[scale=0.7]
    {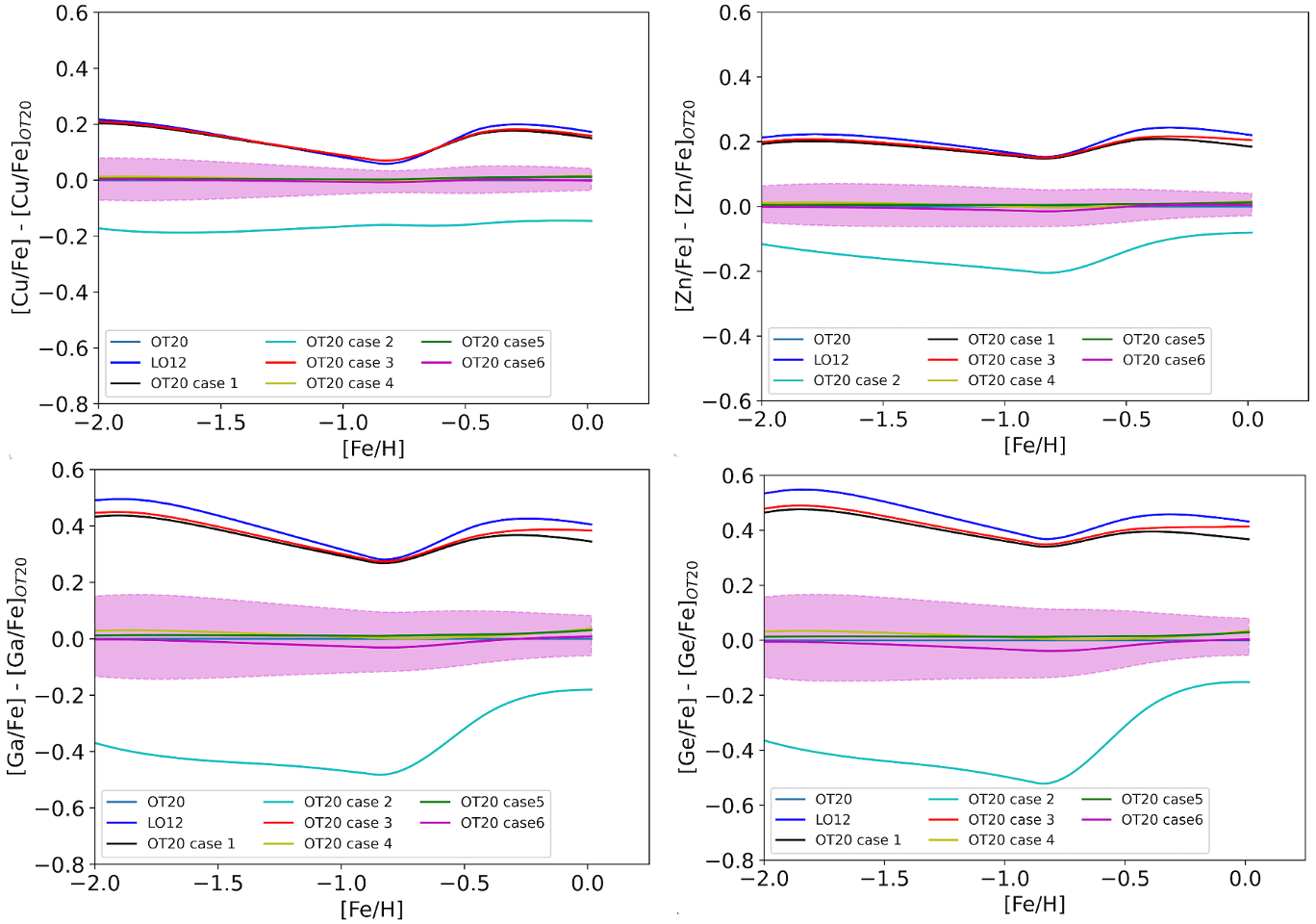}

    \caption{Sensitivity of GCE to varied resonance strengths (see Table \ref{table:3}). The top row is [Cu/Fe] and [Zn/Fe], and the bottom row is [Ga/Fe] and [Ge/Fe], respectively. The results from each condition are normalized to the OT20 values. The magenta band shows uncertainties derived from the OT20 rate. 
    }
    \label{fig:plots_with_cases}
\end{figure*}

Results of the GCE calculations are shown in Figure \ref{fig:plots_with_cases}, which are normalized with respect to the OT20 rate GCE calculation. Apart from the coordinates around [Fe/H]$\approx$-1, the differences from the OT20 curve are nearly independent of the metallicity. 
The change in GCE observed by each case is summarized in Table \ref{table:4}. For all elements considered, the differences in the GCE by the LO12 and OT20 rates are due mostly ($>$ 90\%) to the differences of the $\omega\gamma_{(\alpha,n)}$ at $E_x$=11.32 MeV resonance (Case 1). 

Given the difference in $\omega\gamma_{(\alpha,n)}$ at $E_x$=11.32 MeV is only a factor of 3.3, it is clear that this $\omega\gamma_{(\alpha,n)}$ is the most impactful on s-process nucleosynthesis in massive stars among the remaining uncertain $\omega\gamma$ in the $^{22}$Ne + $\alpha$ reaction. The recent direct measurement by \cite{Shahina2024} reported the $\omega\gamma_{(\alpha,n)}$ to be 100 (22) $\mu$eV, which varies from OT20 by a factor of 2.4. Its impact on the GCE is therefore estimated to be $\approx$0.15 dex (Cu and Zn) and $\approx$0.3 dex (Ga and Ge), which would result in elemental abundances distributed between OT20 and LO12. 
Upcoming verification of their results by future measurements in underground facilities can considerably reduce the uncertainties in the GCE \citep{Masha2022,Best2025}. 
If $\omega\gamma_{(\alpha,n)}$ is unambiguously constrained with a precision of better than 10--20\% (or e.g., $\Delta(\omega\gamma_{(\alpha,n)})$=10-20 $\mu$eV; which are typical or slightly better experimental uncertainties reported for this resonance), it will reduce the uncertainties from this resonance on the GCE simulations to $<$$\pm$0.05 dex for Cu and Zn, and $<$$\pm$0.1 dex for Ga and Ge, respectively.

As mentioned above, while the $E_x$=11.17 MeV (Case 2) and 11.11 MeV (Case 3) resonances vary the GCE as much as the $E_x$=11.32 MeV resonance, the $\omega\gamma$ used in this exercise does not correspond to the $\omega\gamma$ used in LO12. 
However, the GCE variations observed in Case 2 and 3 indicate the upper limits that future nuclear physics experiments should probe, without assuming any probability distributions (e.g., Porter-Thomas). Since such assumptions are ambiguous depending on the probability distribution, it is more instructive to determine the upper limit such that the resonance is negligible if its upper limit is below a certain value, especially when the limits may be reachable by the experiment. 

For the $\omega\gamma_{(\alpha,\gamma)}$ of the $E_x$=11.17 MeV resonance (Case 2), the GCE varies by 0.2--0.4 dex when increasing the $\omega\gamma_{(\alpha,\gamma)}$ from 0 to 660 neV. 
Assuming a simple linearity, we could safely say that the GCE uncertainties will be $<$0.05 dex for all the elements considered if the upper limit was reduced 10 times from 660 neV. 
If future experiments set an upper limit of $\approx$60 neV on this resonance (the current limit: 150 neV by \cite{Shahina2022}), rather small GCE uncertainties ($<$0.05 dex) would be achievable. 
If the $\omega\gamma_{(\alpha,n)}$ for the $E_x$=11.11 MeV resonance (Case 3) was similarly constrained by reducing the upper limit (13.5 neV) 10 times, the GCE uncertainties would be $<$0.05 dex for all the elements considered. 
Therefore, a $\approx$1 neV upper limit would be enough for this resonance.
Again, no experimental evidence has been reported that the $E_x$=11.11 MeV resonance has a large $\alpha$-cluster strength. 
In addition, the direct measurements show no signature of a strong resonance near the corresponding energy. Nevertheless, it will be beneficial to constrain the upper limit of this $\omega\gamma_{(\alpha,n)}$ to $\approx$ 1 neV in future direct measurements due to its potential impact and experimental feasibility. 
At underground facilities such measurements can be possible \citep{Rapagnani2022}. 
It should also be mentioned that these upper limits may slightly change as the neutron productions compete among different resonances using limited amounts of $^{22}$Ne. For instance, if the dominant $E_x$=11.32 MeV resonance has a higher $\omega\gamma_{(\alpha,n)}$ value, the impact of the lower energy resonances could be reduced. Thus, even higher upper limits on the resonances may be acceptable.

Case 4 and 5 do not affect the GCE results, indicating $\omega\gamma_{(\alpha,\gamma)}$ for these resonances have already been well constrained. Case 6 has a negligible impact on the GCE of the elements shown in Figure \ref{fig:plots_with_cases}. 
However, the production of neutron-rich isotopes close to branching points in the C-burning shell could be affected, due to neutron density variations. 

 \begin{table*}[!htbp]
\begin{center}
    \caption{GCE uncertainties ($\Delta$[X/Fe] in unit of dex) resulted from varied resonance strengths $\omega\gamma_{(\alpha,x)}$ (see Table \ref{table:3}). They are evaluated at [Fe/H]$=$0. Default $\Delta$[X/Fe] corresponds to the case using LO12 and OT20 (taken from Table \ref{table:2}).} 
    \label{table:4}
\begin{tabular}{c c c c c c c c}
\hline\hline
 Case & 1 & 2 & 3 & 4 & 5 & 6 & Default \\ 
 \hline
 $\Delta$[Cu/Fe] & +0.15(5) & --0.15(5) & +0.14(5) & +0.01(5) & +0.01(5) & $\approx$0.00(5) & 0.18(6) \\ 
 $\Delta$[Zn/Fe] & +0.20(4) & --0.09(4) & +0.18(4) & +0.01(4) & +0.01(4) & +0.01(4) & 0.22(7) \\ 
 $\Delta$[Ga/Fe] & +0.38(9) & --0.18(9) & +0.33(9) & +0.03(9) & +0.03(9) & $\approx$0.00(9) & 0.41(12) \\ 
 $\Delta$[Ge/Fe] & +0.40(8) & --0.16(4) & +0.36(4) & +0.03(4) & +0.03(4) & $\approx$0.00(4) & 0.44(14) \\ 
 \hline
\end{tabular}
\end{center}
\end{table*}

\subsection{Post-SN GCE}

The purpose of the present work is to study the impact of uncertainties of the $^{22}$Ne+$\alpha$ rates on the s-process in massive stars; we have explored this topic in great detail in the previous sections. 
As we have seen in Section \ref{subsec:ne22an_on_ccsn_profiles}, the $\alpha$-process component found in some CCSN models by \cite{Ritter2018} is not affected by the $^{22}$Ne+$\alpha$ rates, and depends instead on the CCSN explosion setup. It is possible to roughly estimate the primary CCSN production of Cu to 6-16\%, based on observations of metal-poor stars \citep[][]{2010ApJ...710.1557P}. A similar contribution could be estimated for Ge (around 20\%) and Ga, but the limited number of spectroscopic observations for these elements limits the ability to derive relevant constraints.  

\begin{figure*}
    \centering
        
\includegraphics[scale=0.7]{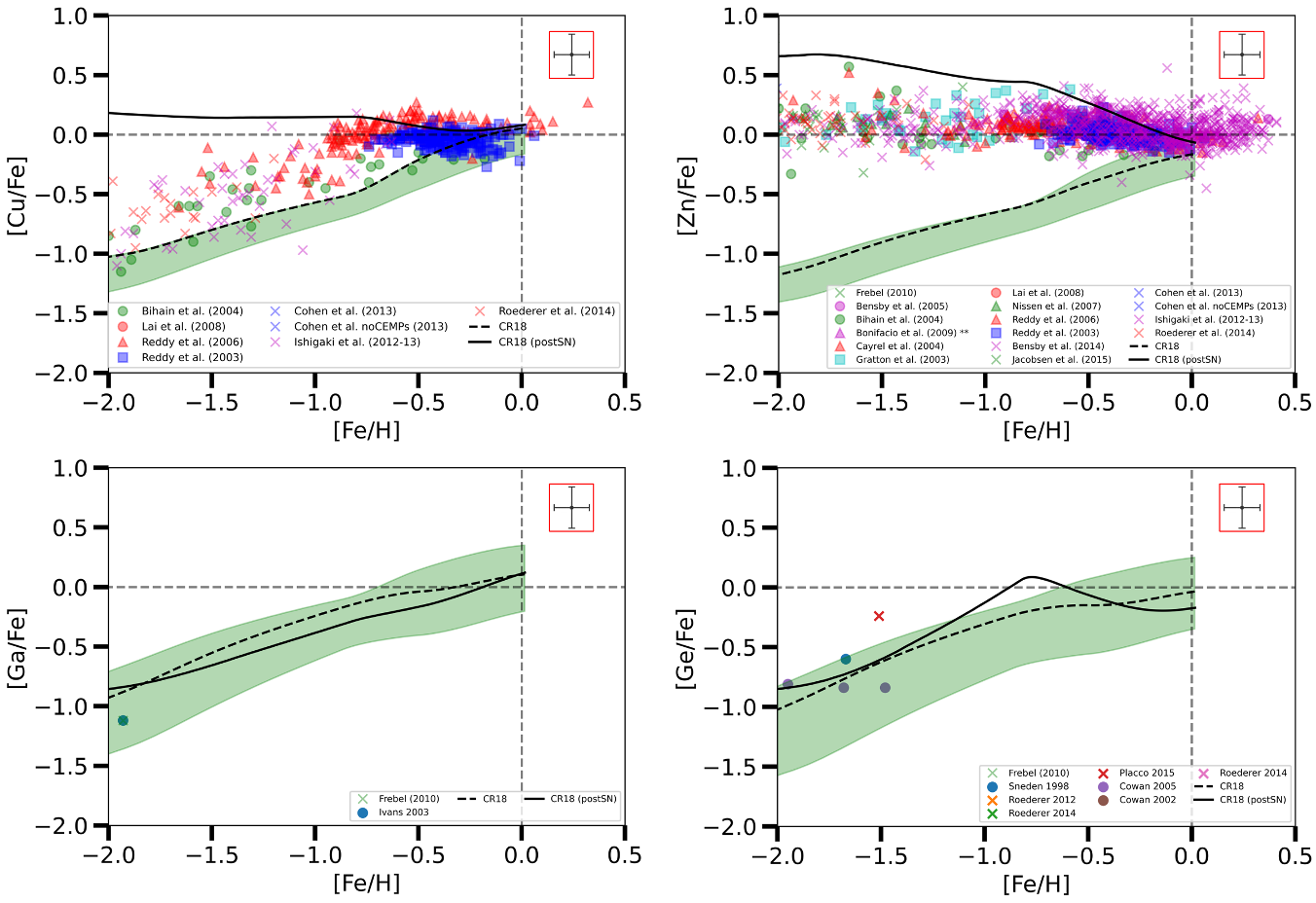}           
    \caption{The post-SN results of GCE calculations for Cu, Zn, Ga, and Ge in comparison with observational data. The green bands show the pre-SN GCE variations using the upper limit by LO12 and the lower limit by OT20, thus the possible maximum range from the current $^{22}$Ne$+$$\alpha$ reaction rate uncertainties. The pre-SN GCE using the CR18 rate is shown as well. The error bars in the red box shows the typical observational uncertainties ($\sigma$; $\pm$0.1 dex for x and $\pm$0.2 dex for y). See Figure \ref{fig:plots} for the observation data.}
    \label{fig:plots_with_postSN}
\end{figure*}

Using the same GCE framework adopted for the previous simulations, we have calculated a GCE model using the CCSN yield table provided by \cite{Ritter2018}. 
Figure \ref{fig:plots_with_postSN} shows the GCE curves of Cu, Zn, Ga, and Ge. As a reference, the pre-SN GCE scatter between the LO12 upper limit and OT20 lower limit (see section \ref{subsec: gce_preSN}) is shown as green bands, with the pre-SN GCE using the CR18 rate. Within the $Z$ range covered in the figure,  the GCE model using CCSN yields for Ga and Ge does not appear to diverge significantly from the GCE models adopting the pre-SN yields, with the few observations available consistent with the results. 

Simulations of Cu and Zn show instead some tension with observations \citep[e.g.,][]{mishenina:17}. The [Cu/Fe] trend is consistent with the solar abundances, but it is quite flat for different $Z$. Observations show instead solar ratios between $Z$$_{\odot}$ and [Fe/H] $\approx$ -0.5, but then they start to decrease for lower $Z$. Using the same phenomenological considerations in \cite{2010ApJ...710.1557P}, this would imply that in the GCE model adopting \cite{Ritter2018} CCSN yields, about 30-50\% of the solar Cu would be made by primary CCSN components, assuming that 1/3 or 1/2 of the solar Fe is provided by massive stars. Therefore, in \cite{Ritter2018} standard CCSN yields the primary Cu production is overestimated by a factor between 2 and 8 with respect to observations, and this is directly noticeable in the CCSN abundance profiles from 12 and 15 $M_{\odot}$ stars at $Z$$\leq$0.006 (see e.g., Figures \ref{fig:15_cu63} and \ref{fig:15_cu65} for the 15 M$_{\odot}$ model at Z=0.0001). An overestimation compared to observations is also obtained for the [Zn/Fe] ratios. This is interesting, since Zn has been historically under-produced by GCE simulations, leading to scenarios where a boosted Zn contribution from hypernovae has been claimed at low $Z$ \citep[][and references therein]{kobayashi:20}. Our GCE simulations have no hypernovae, but we still see the huge potential impact of the $\alpha$-process on Zn, from the most internal 0.1-0.2 M$_{\odot}$ in the CCSN ejecta.  

More careful investigation is in progress, using observations of these elements (Cu, Zn, Ga, and Ge) in metal-poor stars, to find new constrains for the deepest CCSN ejecta. However, this will not affect the s-process production of these elements. 

\section{Summary and future work} \label{sec:summary}
While $^{22}$Ne($\alpha$,$n$)$^{25}$Mg and $^{22}$Ne($\alpha$,$\gamma$)$^{26}$Mg are among the most important reactions for s-process nucleosynthesis, their reaction rates still have significant uncertainties. \cite{Jayatissa2020,OTA2020135256} and \cite{Ota2021} used indirect measurements to report weaker strengths of relevant resonances, such as the $E_x$=11.32 MeV for the ($\alpha$,n) channel and 11.17 MeV for the ($\alpha$,$\gamma$) channel. 
The most recent direct measurement by \cite{Shahina2024} obtained a more moderate value of $\omega\gamma_{(\alpha,n)}$ for the $E_x$=11.32 MeV resonance which is more consistent with other experiments within their experimental uncertainties (22\%). Their measurements of the $\omega\gamma_{(\alpha,\gamma)}$ for the $E_x$=11.17 MeV are consistent with our measurements, implying that this resonance probably does not contribute significantly to $s$-process nucleosynthesis. 
As described in \cite{Wiescher2023}, there are remaining uncertainties in $\omega\gamma$ for other resonances below and above the energies of these two resonances. 

The uncertainties from each of these resonances affect abundance yields of theoretical stellar models and the GCE of the s-process elements. This work provides the first detailed and quantitative estimation of the impact of different $^{22}$Ne+$\alpha$ rates on the s-process in massive stars and on the GCE of its main nucleosynthesis products. 
In particular, we focus on the production of the element group Cu, Zn, Ga, and Ge with Z = 29--32. For stellar nucleosynthesis simulations, we used the post-processing nucleosynthesis code, MPPNP, developed by the NuGrid Collaboration. For GCE simulations, we used the GCE code, OMEGA+.
For the analysis, we considered 14 different $^{22}$Ne+$\alpha$ reaction rate combinations. For each reaction rate configuration, the nucleosynthesis in 20 massive star models (four initial stellar masses, 12, 15, 20, and 25 M$_{\odot}$, and five metalicities, Z=0.02, 0.01, 0.006, 0.001, and 0.0001) were calculated, for a total of 280 nucleosynthesis models. 

In comparison with observations, the simulated GCE reproduce their solar abundances except for Zn. Our simulations underproduce Cu at low $Z$ partly because of the uncertainties of the GCE parameters used in the present work. The stellar models that are used can also impact the obtained final yields. For instance, \cite{pumo:10} discusses that convective overshooting during the He core burning influences how much $^{22}$Ne is available and the duration of neutron-capture conditions. Their work demonstrates that using different values for the overshooting parameter $f$ (10$^{-5}$--0.035; c.f., 0.022 used \citep[in the present models][]{Ritter2018}), the weak $s$ abundances could decrease/increase by a factor of 2--3 in 15--25 M$_{\odot}$ stars. Furthermore, fast rotating massive stars may potentially increase the $s$ process yields compared to the non-rotating stars used in the present work, especially at low $Z$ \citep[][]{pignatari:08,frischknecht:16,roberti:24}. As well as ($n,\gamma$) cross sections, uncertainties of the other nuclear cross sections relevant to neutron economy in the $s$ process stellar environment such as neutron poison (e.g., $^{16}$O($n,\gamma$)$^{17}$O) could also help explain the inconsistency with the observations. The ratio of $^{17}$O($\alpha,\gamma$)/$^{17}$O($\alpha,n$) cross sections influences the available neutrons for the weak $s$ process. As shown in \cite{best:13}, the ratio's current uncertainties could vary the abundances of weak $s$ elements by an order of magnitude. Therefore, many uncertainties in the physics input remain that hinder comparisons between the GCE simulations and observational data. Nonetheless, we consider the relative differences in GCE by different $^{22}$Ne$+$$\alpha$ reaction rates reported in the present work are still valid.

We obtain that using our recent $^{22}$Ne($\alpha$,$n$)$^{25}$Mg and $^{22}$Ne($\alpha$,$\gamma$)$^{26}$Mg rates, the elements Cu, Zn, Ga, and Ge result in up to a factor of 1.5-3 ($\approx$0.2-0.45 dex) reductions in their galactic mass fractions compared to GCE models using the rates by \cite{Longland2012}. Such reductions are almost exclusively explained by the decrease of $\omega\gamma_{(\alpha,n)}$ at the $E_x$=11.32 MeV resonance. It confirms that this resonance is undoubtedly the current dominant source of the $^{22}$Ne+$\alpha$ rate's uncertainties for the s-process in massive stars. 

We also investigated the impact on GCE predictions to relevant lower (10.95, 11.08, and 11.11 MeV) and higher energy resonances ($>$11.32 MeV), which could contribute to the s-process in massive stars during convective He-core burning and during convective C-shell burning, respectively. 
Apart from the $E_x$=11.11 MeV resonance, these do not impact the GCE when reasonable $\omega\gamma$ or upper limits are assumed. 

As such, it is critical to unambiguously constrain the $\omega\gamma_{(\alpha,n)}$ of the $E_x$=11.32 MeV resonance with a precision better than 10--20\% in future experiments. 
Thanks to the worldwide ongoing/planned experimental efforts at underground nuclear physics facilities such as LUNA, this should be feasible within the next few years. Lowering the upper limits on the $\omega\gamma_{(\alpha,\gamma)}$ of the $E_x$=11.17 MeV and $\omega\gamma_{(\alpha,n)}$ of the $E_x$=11.11 MeV to negligible levels ($\approx$60 neV and $\approx$1 neV, respectively) will also be possible \citep{Rapagnani2022}. These experiments will have 10-100 times lower background environment compared to the work by \cite{Jaeger2001}. 
We have demonstrated that well-constrained $^{22}$Ne$+$$\alpha$ reaction rates will result in much smaller uncertainties than the observational data. This will provide a crucial milestone to better constrain stellar evolution models of massive stars and the chemical enrichment history of our Galaxy. 

Finally, we discussed the GCE of Cu, Zn, Ga, and Ge using CCSN yields. The primary explosive components observed in metal-poor stars are found in our nucleosynthesis simulations. Such nucleosynthesis products are located in the deepest 0.1-0.2 M$_{\odot}$ ejected, and they depend on the CCSN explosion setup. We show that they are not affected by the $^{22}$Ne$+$$\alpha$ reaction rates, or by their current uncertainties. 

We emphasize the importance of including complete CCSN yields to compare stellar and GCE simulations with observational results more consistently, as they complement the pre-SN nucleosynthesis production. In future work, 
we indeed expect to use the s-process elements discussed here 
to constrain CCSN models by comparisons of GCE and stellar results with galactic archaeology observations, applying similar calibration methods adopted in the past using Ni \citep[][]{limongi:18} or Zn \citep[][]{Nomoto2013} as reference elements. In turn, we will apply the newly-constrained CCSN models to study GCE for all the elements.

\begin{acknowledgments}
EK, SO, AD, and JM acknowledge support from the Office of Nuclear Physics, Office of Science of the U.S. Department of Energy under Contract No.DE-AC02-98CH10886 with Brookhaven Science Associates, LLC. 
Financial support for this work was provided by the US Department of Energy, award Nos.\ DE-FG02-93ER40773 and DE-SC0018980, the US National Nuclear Security Administration, award No.\ DE-NA0003841. 
EK, AD, and JM were supported in part by
the U.S. Department of Energy, Office of Science, and Office of Workforce Development for Teachers and Scientists (WDTS) under the Science Undergraduate Laboratory Internships (SULI) Program, and in part by the BNL
Supplemental Undergraduate Research Program (SURP). MP acknowledge the support to NuGrid from the "Lendulet-2023" Program of the Hungarian Academy of Sciences (LP2023-10, Hungary), the ERC Consolidator Grant funding scheme (Project RADIOSTAR, G.A. n. 724560, Hungary), the ERC Synergy Grant Programme (Geoastronomy, grant agreement number 101166936, Germany), the ChETEC COST Action (CA16117), supported by the European Cooperation in Science and Technology, and the IReNA network supported by NSF AccelNet (Grant No. OISE-1927130). MP also thanks the support from NKFI via K-project 138031 (Hungary). LR acknowledges the support from the ChETEC-INFRA -- Transnational Access Projects 22102724-ST and 23103142-ST and the PRIN URKA Grant Number \verb |prin_2022rjlwhn|. We acknowledges support from the ChETEC-INFRA project funded by the European Union’s Horizon 2020 Research and Innovation programme (Grant Agreement No 101008324). We thanks access to {\sc viper}, the University of Hull HPC Facility. This research has used the Astrohub online virtual research environment (https://astrohub.uvic.ca), developed and operated by the Computational Stellar Astrophysics group (http://csa.phys.uvic.ca) at the University of Victoria and hosted on the Computed Canada Arbutus Cloud at the University of Victoria. SO was supported in part by the IReNA visiting fellowship. This work benefited from
interactions and workshops co-organized by The Center for Nuclear astrophysics Across Messengers (CeNAM) which is supported by the U.S. Department of Energy, Office of Science, Office of Nuclear Physics, under Award Number DE-SC0023128.
We thank Prof. Michael Wiescher and Dr. Richard deBoer (Univ. Notre Dame) for providing us with their reaction rates. We also thank Dr. K. A. Womack (Univ. Hull) for providing us with OMEGA+ parameters.

\end{acknowledgments}

\software{MPPNP \citep{2016ApJS..225...24P},  
          OMEGA+ \citep{2018APJ_Cote}, NuPyCEE \citep{2017APJ_Cote}.
}

\appendix
\label{sec:appendix}

\begin{figure*}
    \centering
    \includegraphics[scale=0.41]
    {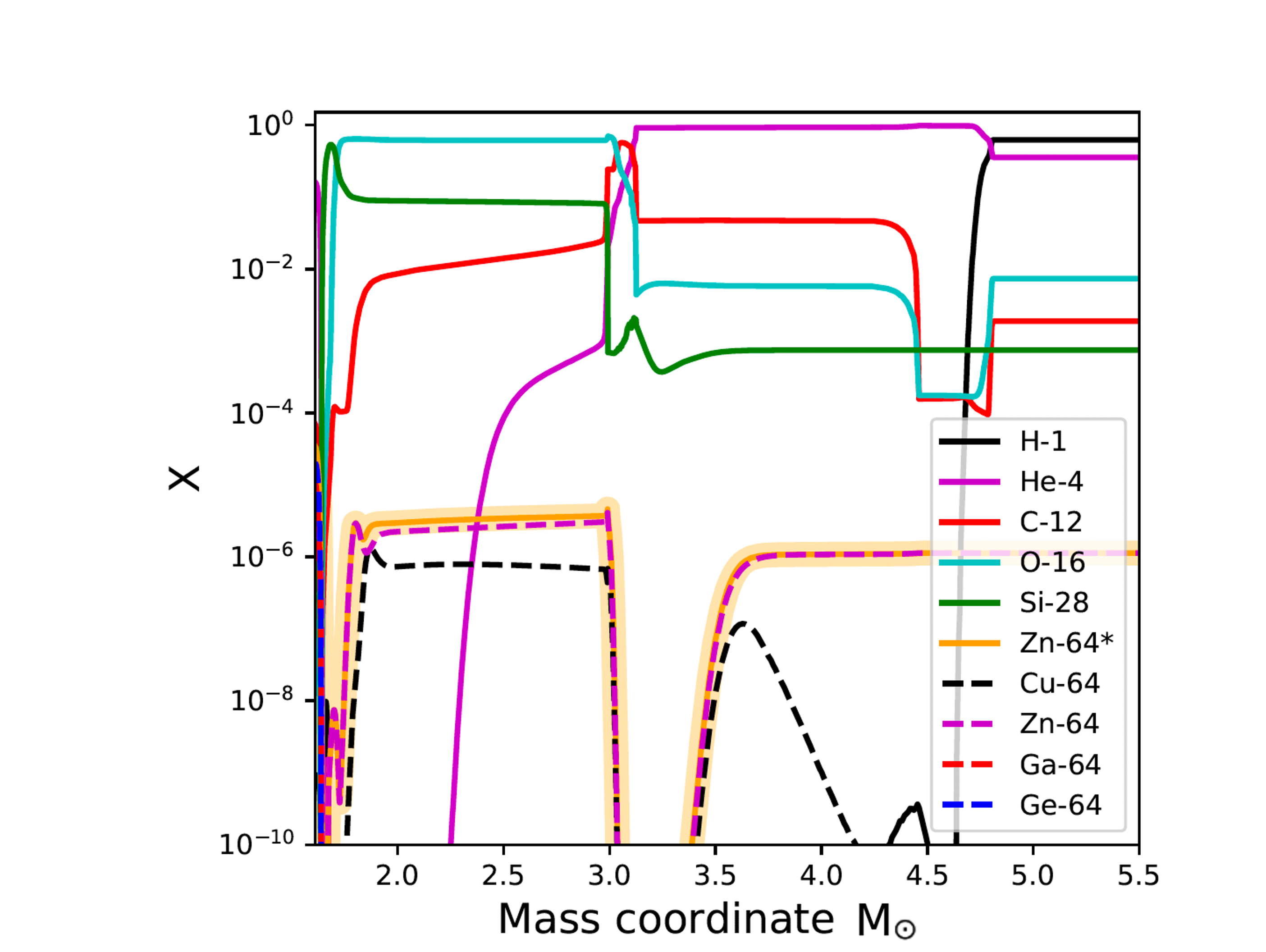}
    \includegraphics[scale=0.41]
    {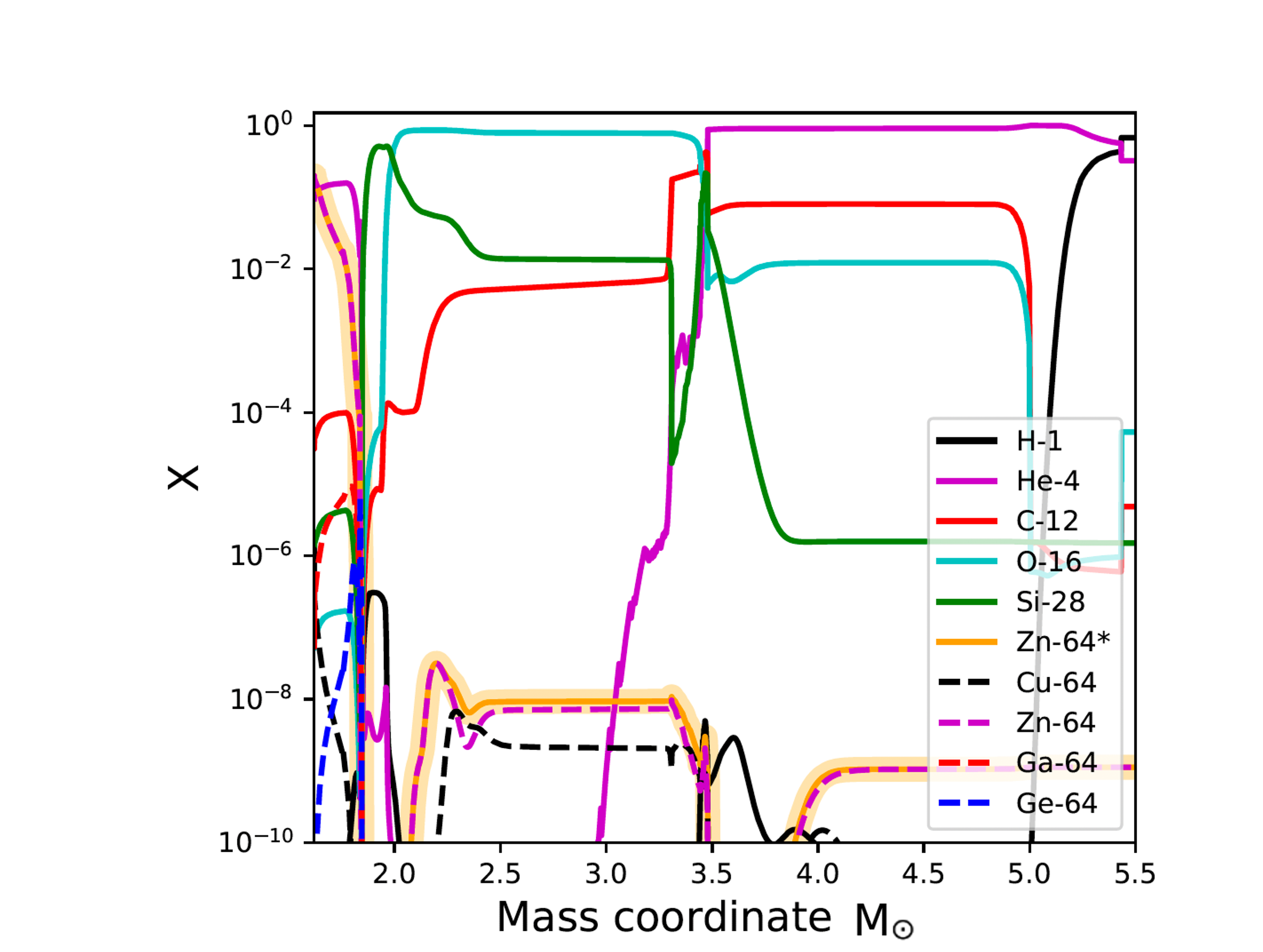}
    \caption{The same as in Figure \ref{fig:15_cu63}, but for $^{64}$Zn. }
    \label{fig:15_zn64}
\end{figure*}
\begin{figure*}
    \centering
    \includegraphics[scale=0.41]
    {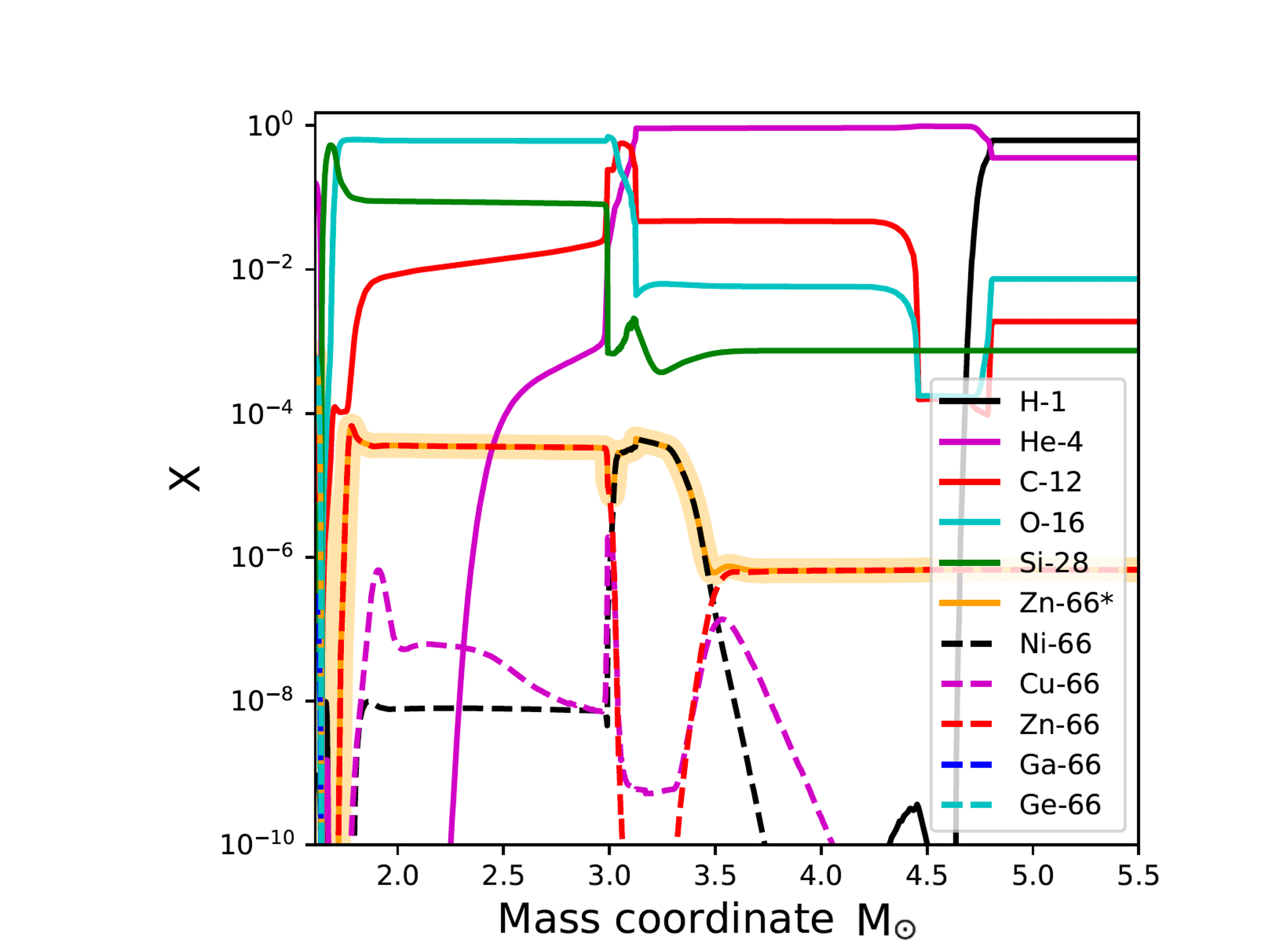}
    \includegraphics[scale=0.41]
    {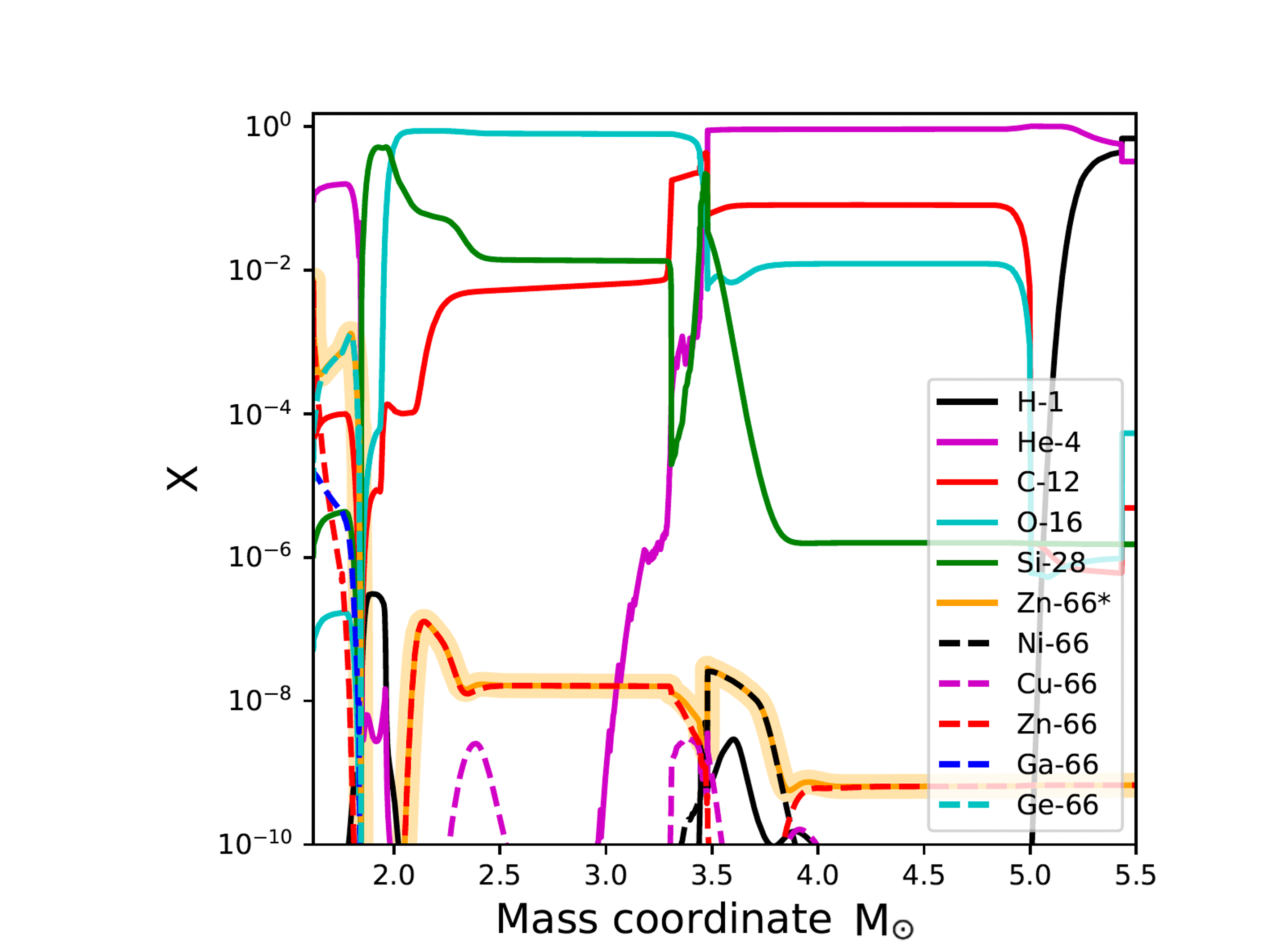}
    \caption{The same as in Figure \ref{fig:15_cu63}, but for $^{66}$Zn. }
    \label{fig:15_zn66}
\end{figure*}\begin{figure*}
    \centering
    \includegraphics[scale=0.41]
    {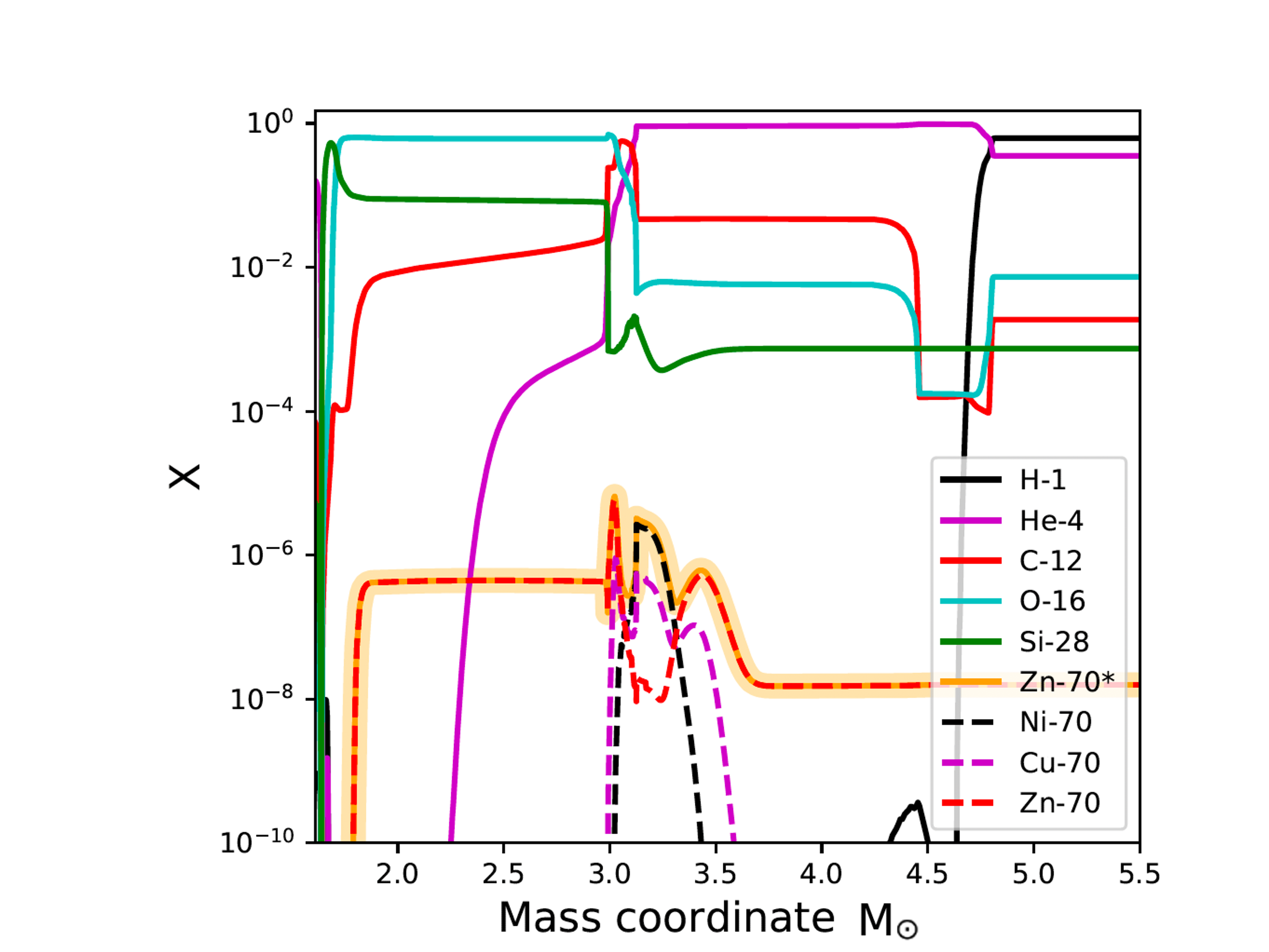}
    \includegraphics[scale=0.41]
    {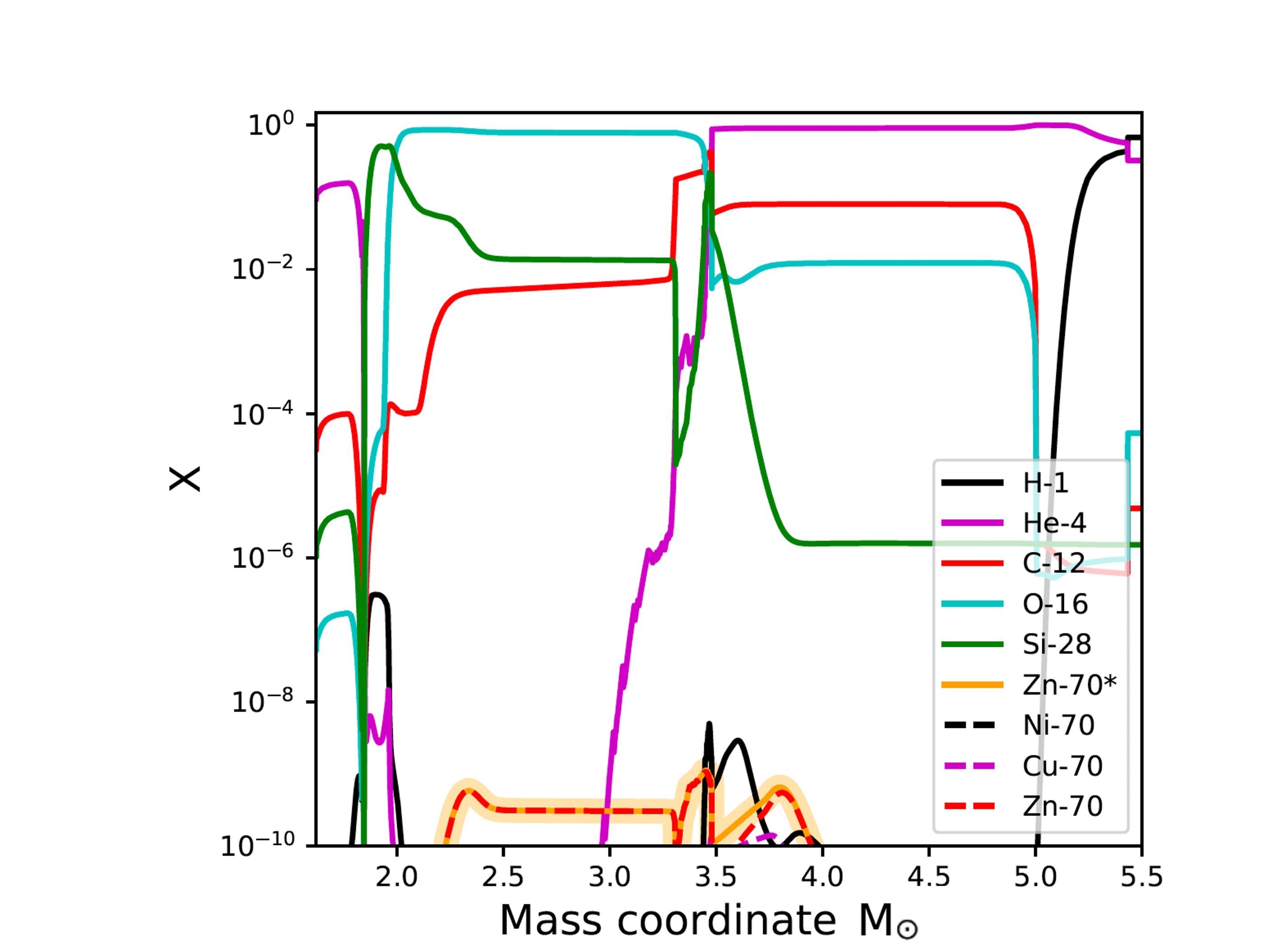}
    \caption{The same as in Figure \ref{fig:15_cu63}, but for $^{70}$Zn. }
    \label{fig:15_zn70}
\end{figure*}

\begin{figure*}
    \centering
    \includegraphics[scale=0.41]
    {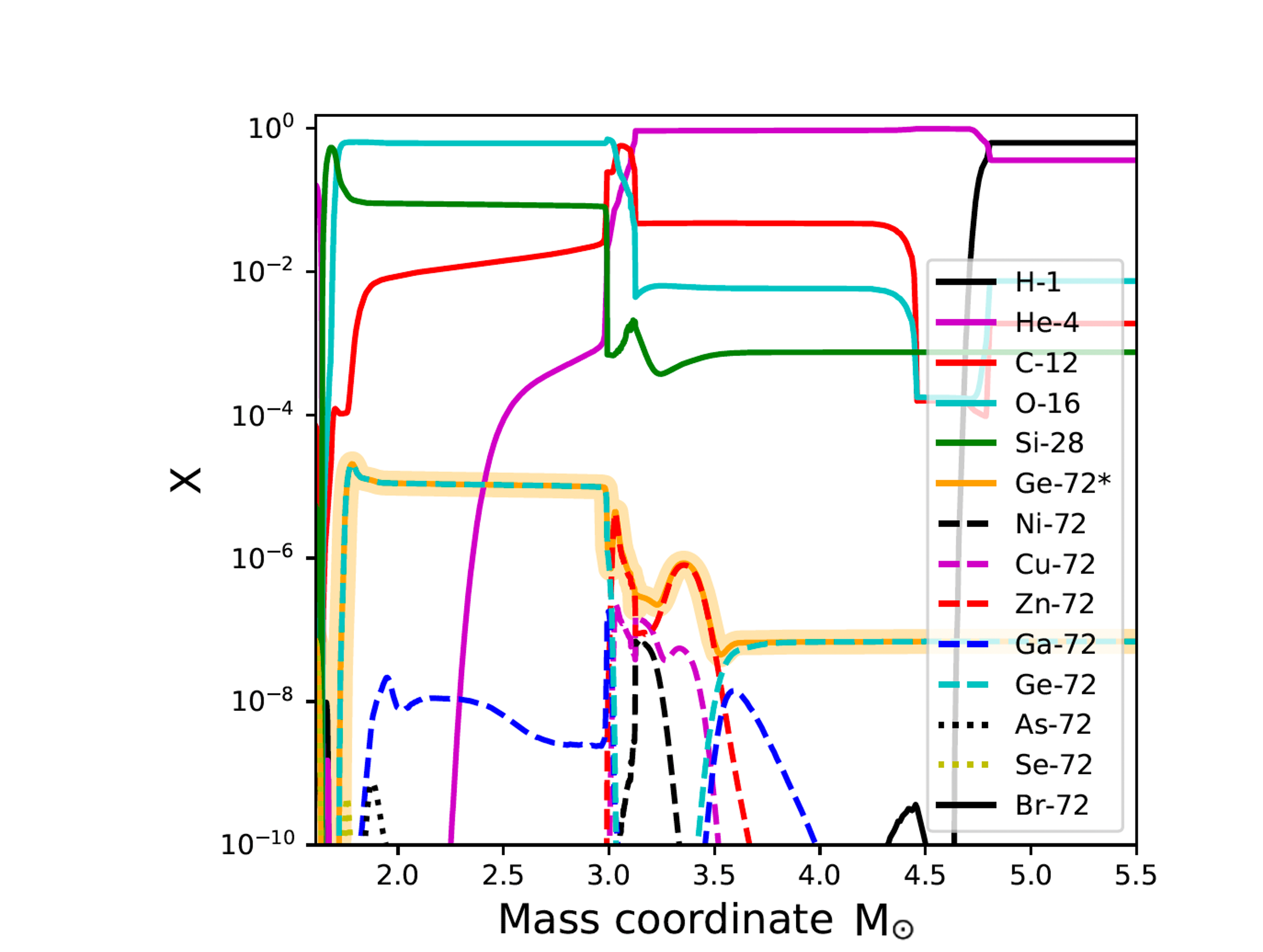}
    \includegraphics[scale=0.41]
    {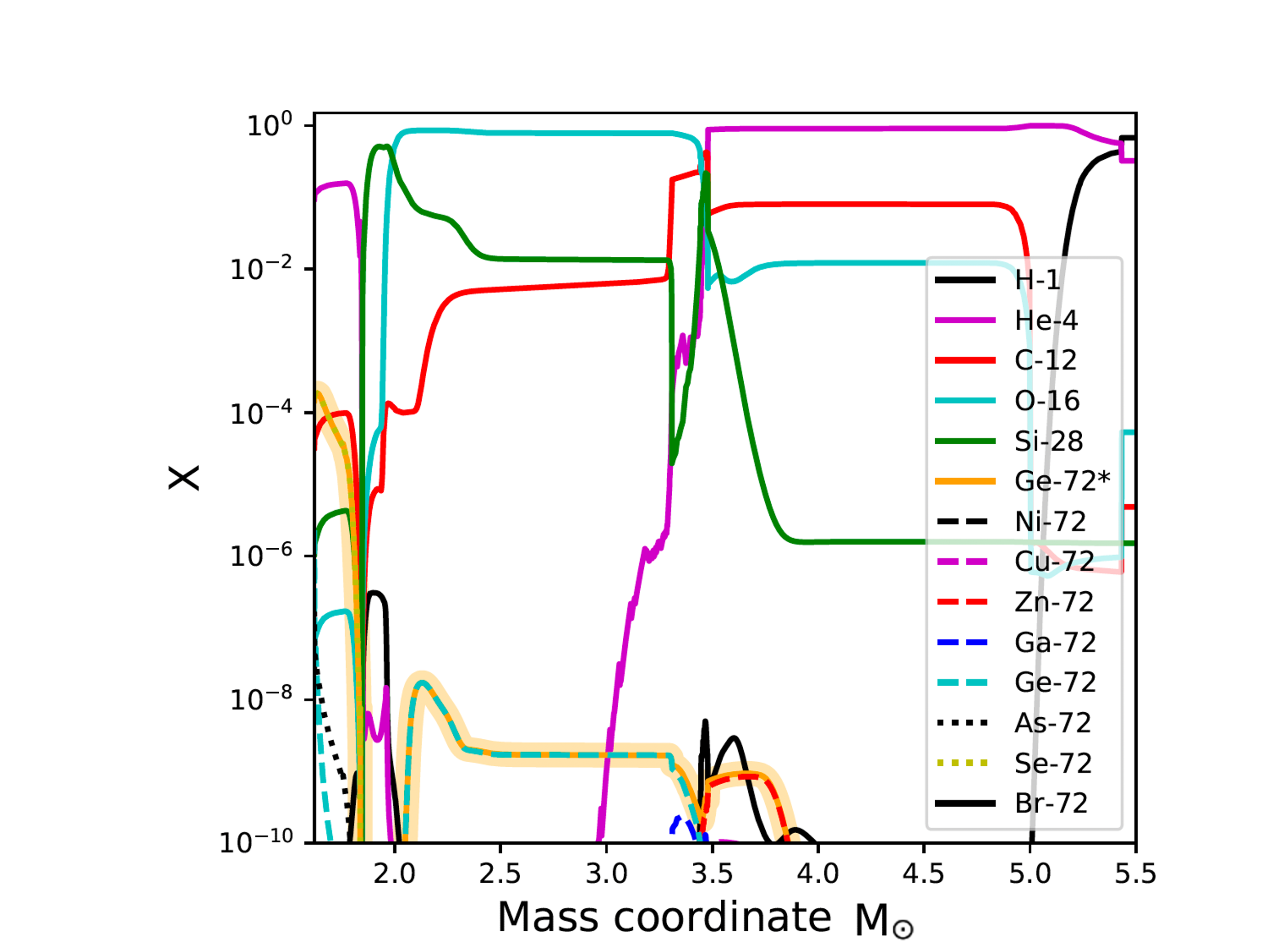}
    \caption{The same as in Figure \ref{fig:15_cu63}, but for $^{72}$Ge. }
    \label{fig:15_ge72}
\end{figure*}
\begin{figure*}
    \centering
    \includegraphics[scale=0.41]
    {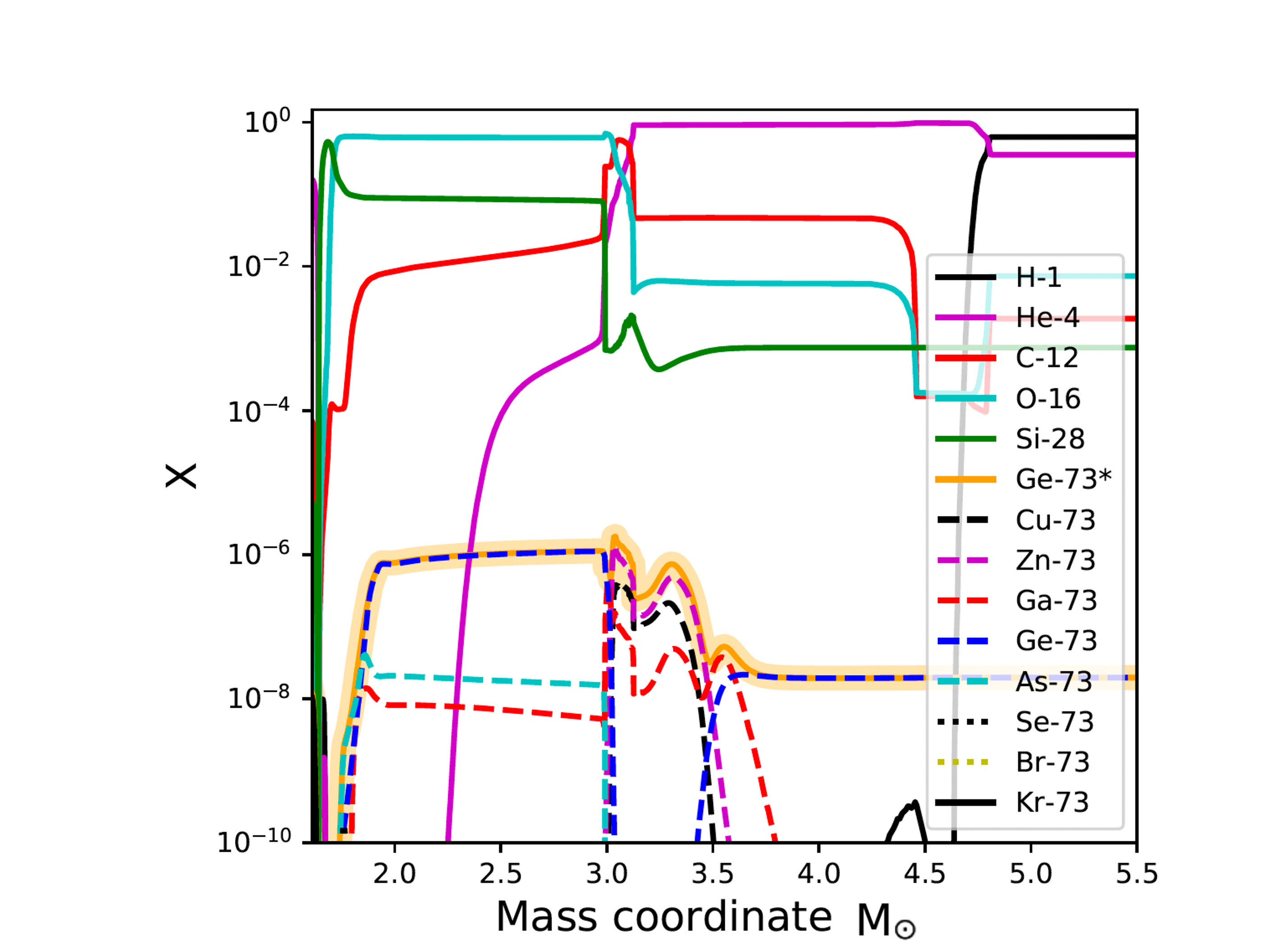}
    \includegraphics[scale=0.41]
    {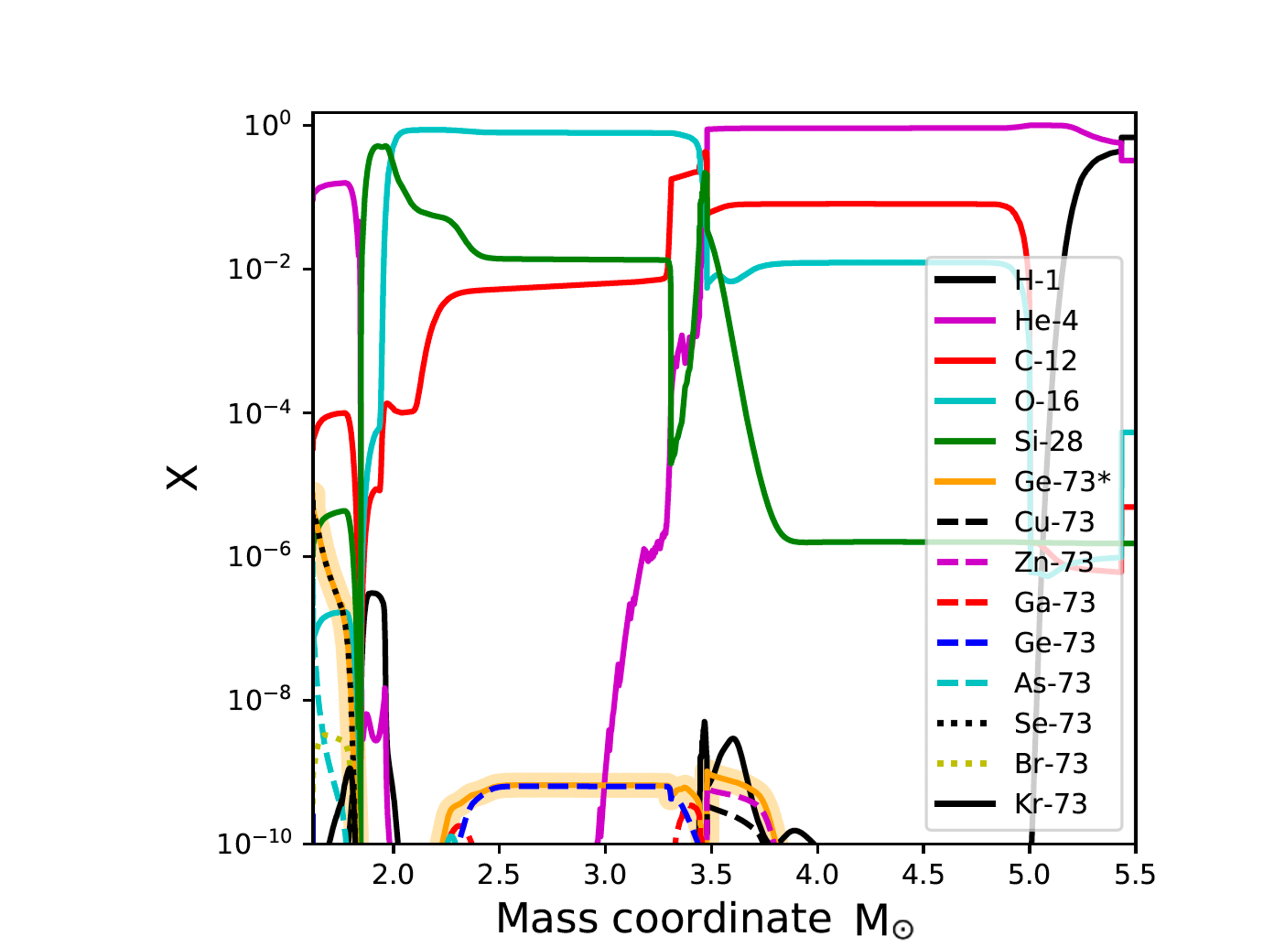}
    \caption{The same as in Figure \ref{fig:15_cu63}, but for $^{73}$Ge. }
    \label{fig:15_ge73}
\end{figure*}
\begin{figure*}
    \centering
    \includegraphics[scale=0.41]
    {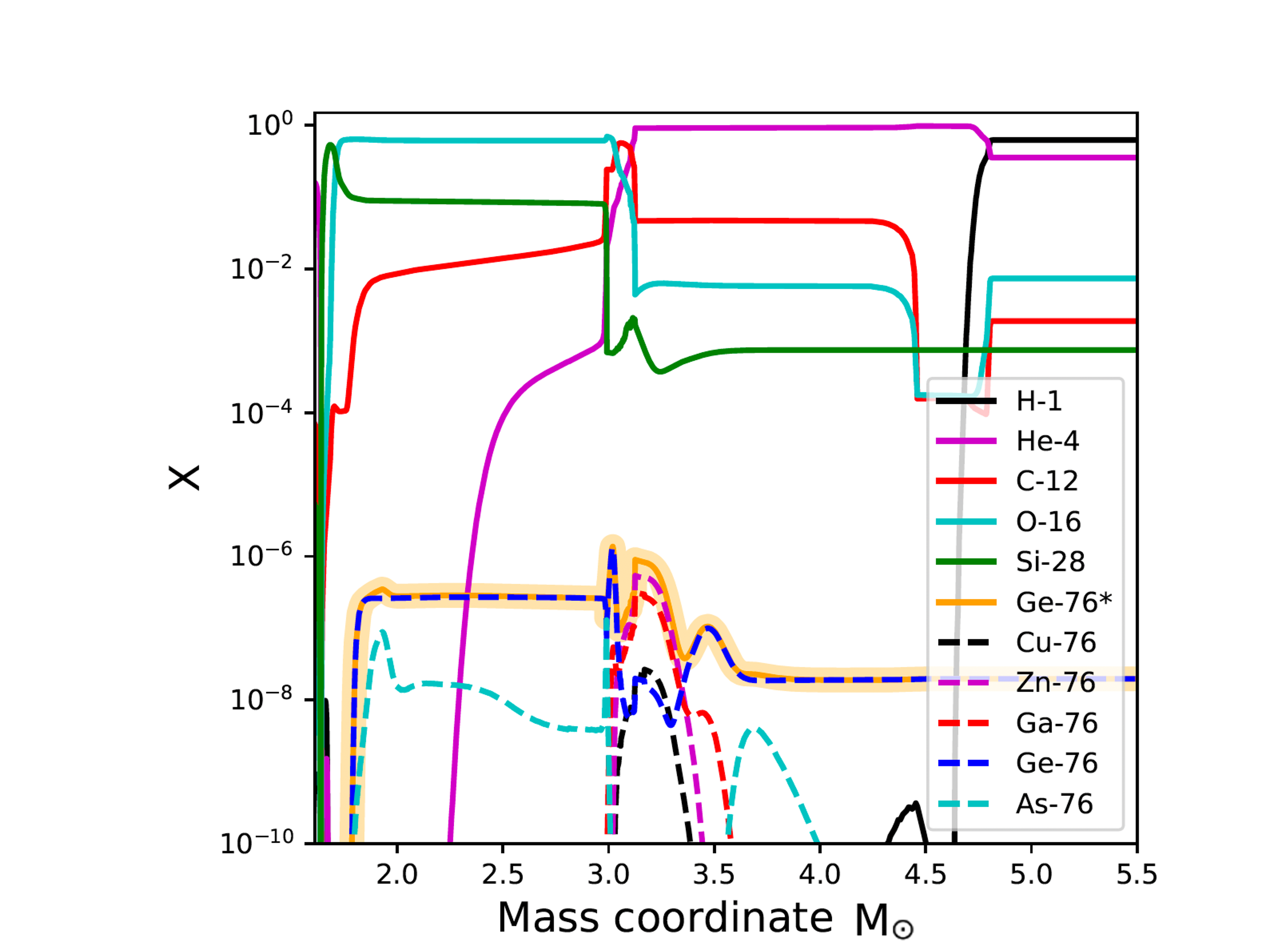}
    \includegraphics[scale=0.41]
    {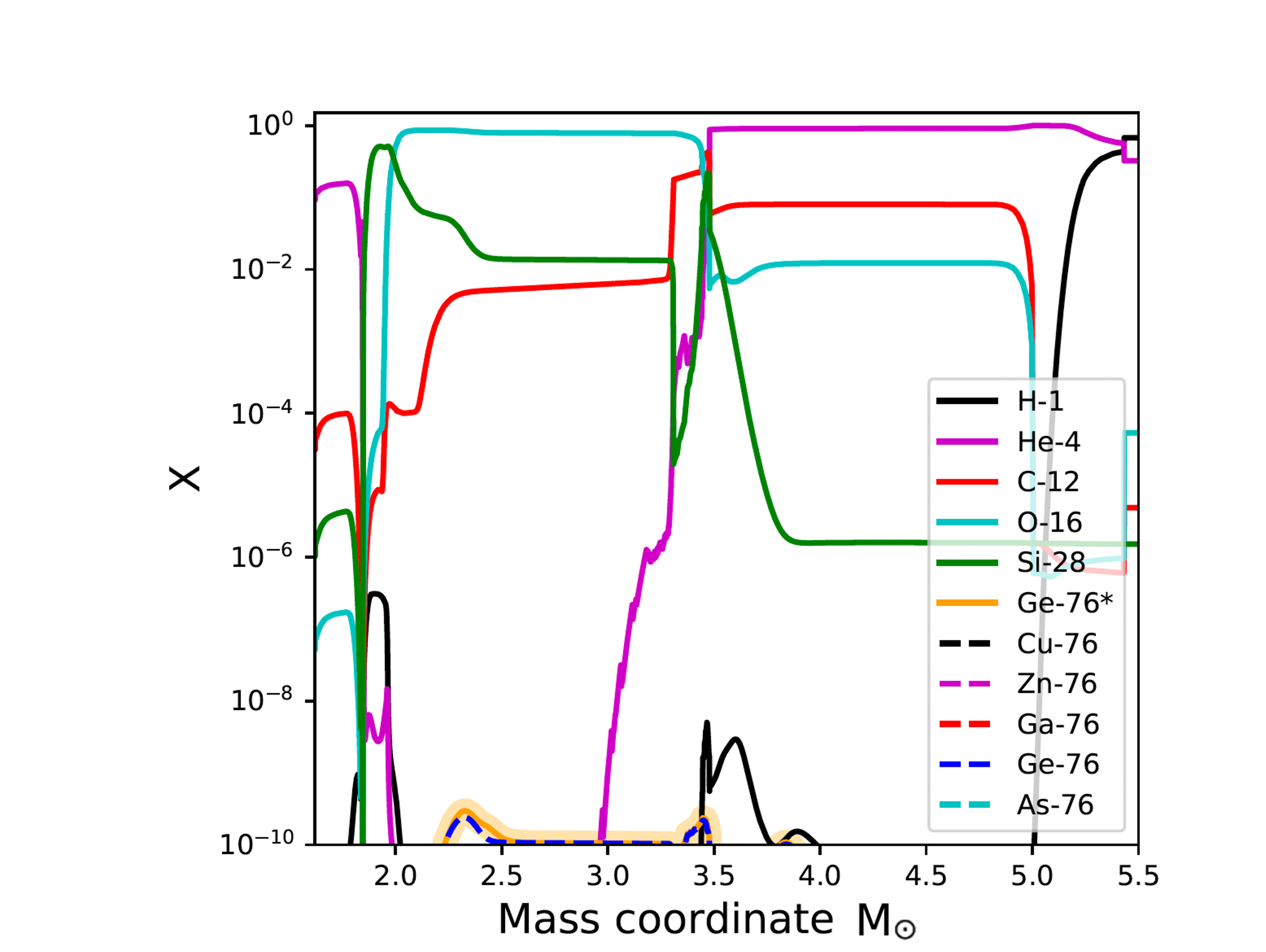}
    \caption{The same as in Figure \ref{fig:15_cu63}, but for $^{76}$Ge. }
    \label{fig:15_ge76}
\end{figure*}

\bibliography{sample631}{}

@ARTICLE{2018APJ_Cote,
       author = {{Côté}, Benoit and {Silvia}, Devin W. and {O'Shea}, Brian W. and {Smith}, Britton and {Wise}, John H.},
        title = "{Validating Semi-analytic Models of High-redshift Galaxy Formation Using Radiation Hydrodynamical Simulations}",
      journal = {\apj},
         year = "2018",
        month = "May",
       volume = {859},
       number = {67},
        pages = {20},
          doi = {10.3847/1538-4357/aabe8f},
 primaryClass = {astro-ph.GA},
       adsurl = {https://iopscience.iop.org/article/10.3847/1538-4357/aabe8f}
}

@ARTICLE{2017APJ_Cote,
       author = {{Côté}, Benoit and {O'Shea}, Brian W. and {Ritter}, Christian and {Herwig}, Falk and {Venn}, Kim A.},
        title = "{The Impact of Modeling Assumptions in Galactic Chemical Evolution Models}",
      journal = {\apj},
         year = "2017",
        month = "January",
       volume = {835},
       number = {128},
        pages = {15},
          doi = {10.3847/1538-4357/835/2/128},
 primaryClass = {astro-ph.GA},
       adsurl = {https://iopscience.iop.org/article/10.3847/1538-4357/aabe8f}
}

@ARTICLE{1997ApJ...477..765C,
       author = {{Chiappini}, C. and {Matteucci}, F. and {Gratton}, R.},
        title = "{The Chemical Evolution of the Galaxy: The Two-Infall Model}",
      journal = {\apj},
         year = 1997,
        month = mar,
       volume = {477},
       number = {2},
        pages = {765-780},
          doi = {10.1086/303726},
archivePrefix = {arXiv},
       eprint = {astro-ph/9609199},
 primaryClass = {astro-ph},
       adsurl = {https://ui.adsabs.harvard.edu/abs/1997ApJ...477..765C}
}

@article{10.1093/mnras/stac3180,
    author = {Womack, Kate A and Vincenzo, Fiorenzo and Gibson, Brad K and Côté, Benoit and Pignatari, Marco and Brinkman, Hannah E and Ventura, Paolo and Karakas, Amanda},
    title = "{Chemical evolution of fluorine in the Milky Way}",
    journal = {Monthly Notices of the Royal Astronomical Society},
    volume = {518},
    number = {1},
    pages = {1543-1556},
    year = {2022},
    month = {11},
    issn = {0035-8711},
    doi = {10.1093/mnras/stac3180},
    url = {https://doi.org/10.1093/mnras/stac3180},
    eprint = {https://academic.oup.com/mnras/article-pdf/518/1/1543/47195666/stac3180.pdf}
}

@ARTICLE{rauscher:13,
       author = {{Rauscher}, T. and {Dauphas}, N. and {Dillmann}, I. and {Fr{\"o}hlich}, C. and {F{\"u}l{\"o}p}, Zs and {Gy{\"u}rky}, Gy},
        title = "{Constraining the astrophysical origin of the p-nuclei through nuclear physics and meteoritic data}",
      journal = {Reports on Progress in Physics},
         year = 2013,
        month = jun,
       volume = {76},
       number = {6},
          eid = {066201},
        pages = {066201},
          doi = {10.1088/0034-4885/76/6/066201},
archivePrefix = {arXiv},
       eprint = {1303.2666},
 primaryClass = {astro-ph.SR},
       adsurl = {https://ui.adsabs.harvard.edu/abs/2013RPPh...76f6201R}
}

@ARTICLE{cowan:21,
       author = {{Cowan}, John J. and {Sneden}, Christopher and {Lawler}, James E. and {Aprahamian}, Ani and {Wiescher}, Michael and {Langanke}, Karlheinz and {Mart{\'\i}nez-Pinedo}, Gabriel and {Thielemann}, Friedrich-Karl},
        title = "{Origin of the heaviest elements: The rapid neutron-capture process}",
      journal = {Reviews of Modern Physics},
         year = 2021,
        month = jan,
       volume = {93},
       number = {1},
          eid = {015002},
        pages = {015002},
          doi = {10.1103/RevModPhys.93.015002},
archivePrefix = {arXiv},
       eprint = {1901.01410},
 primaryClass = {astro-ph.HE},
       adsurl = {https://ui.adsabs.harvard.edu/abs/2021RvMP...93a5002C}
}

@ARTICLE{pignatari:16,
       author = {{Pignatari}, Marco and {G{\"o}bel}, Kathrin and {Reifarth}, Ren{\'e} and {Travaglio}, Claudia},
        title = "{The production of proton-rich isotopes beyond iron: The {\ensuremath{\gamma}}-process in stars}",
      journal = {International Journal of Modern Physics E},
         year = 2016,
        month = apr,
       volume = {25},
       number = {4},
          eid = {1630003-232},
        pages = {1630003-232},
          doi = {10.1142/S0218301316300034},
archivePrefix = {arXiv},
       eprint = {1605.03690},
 primaryClass = {astro-ph.SR},
       adsurl = {https://ui.adsabs.harvard.edu/abs/2016IJMPE..2530003P}
}

@ARTICLE{psaltis:22,
       author = {{Psaltis}, A. and {Arcones}, A. and {Montes}, F. and {Mohr}, P. and {Hansen}, C.~J. and {Jacobi}, M. and {Schatz}, H.},
        title = "{Constraining Nucleosynthesis in Neutrino-driven Winds: Observations, Simulations, and Nuclear Physics}",
      journal = {\apj},
         year = 2022,
        month = aug,
       volume = {935},
       number = {1},
          eid = {27},
        pages = {27},
          doi = {10.3847/1538-4357/ac7da7},
archivePrefix = {arXiv},
       eprint = {2204.07136},
 primaryClass = {astro-ph.HE},
       adsurl = {https://ui.adsabs.harvard.edu/abs/2022ApJ...935...27P}
}

@ARTICLE{raiteri:91,
       author = {{Raiteri}, C.~M. and {Busso}, M. and {Gallino}, R. and {Picchio}, G. and {Pulone}, L.},
        title = "{S-Process Nucleosynthesis in Massive Stars and the Weak Component. I. Evolution and Neutron Captures in a 25 M$_{sun}$ Star}",
      journal = {\apj},
         year = 1991,
        month = jan,
       volume = {367},
        pages = {228},
          doi = {10.1086/169622},
       adsurl = {https://ui.adsabs.harvard.edu/abs/1991ApJ...367..228R}
}

@ARTICLE{raiteri:91a,
       author = {{Raiteri}, C.~M. and {Busso}, M. and {Gallino}, R. and {Picchio}, G.},
        title = "{S-Process Nucleosynthesis in Massive Stars and the Weak Component. II. Carbon Burning and Galactic Enrichment}",
      journal = {\apj},
         year = 1991,
        month = apr,
       volume = {371},
        pages = {665},
          doi = {10.1086/169932},
       adsurl = {https://ui.adsabs.harvard.edu/abs/1991ApJ...371..665R}
}

@ARTICLE{baraffe:92,
       author = {{Baraffe}, I. and {El Eid}, M.~F. and {Prantzos}, N.},
        title = "{The s-process in massive stars of variable composition}",
      journal = {\aap},
         year = 1992,
        month = may,
       volume = {258},
       number = {2},
        pages = {357-367},
       adsurl = {https://ui.adsabs.harvard.edu/abs/1992A&A...258..357B}
}

@ARTICLE{the:07,
       author = {{The}, Lih-Sin and {El Eid}, Mounib F. and {Meyer}, Bradley S.},
        title = "{s-Process Nucleosynthesis in Advanced Burning Phases of Massive Stars}",
      journal = {\apj},
         year = 2007,
        month = feb,
       volume = {655},
       number = {2},
        pages = {1058-1078},
          doi = {10.1086/509753},
archivePrefix = {arXiv},
       eprint = {astro-ph/0609788},
 primaryClass = {astro-ph},
       adsurl = {https://ui.adsabs.harvard.edu/abs/2007ApJ...655.1058T}
}

@ARTICLE{rauscher:02,
       author = {{Rauscher}, T. and {Heger}, A. and {Hoffman}, R.~D. and {Woosley}, S.~E.},
        title = "{Nucleosynthesis in Massive Stars with Improved Nuclear and Stellar Physics}",
      journal = {\apj},
         year = 2002,
        month = sep,
       volume = {576},
       number = {1},
        pages = {323-348},
          doi = {10.1086/341728},
archivePrefix = {arXiv},
       eprint = {astro-ph/0112478},
 primaryClass = {astro-ph},
       adsurl = {https://ui.adsabs.harvard.edu/abs/2002ApJ...576..323R}
}

@ARTICLE{chieffi:98,
       author = {{Chieffi}, Alessandro and {Limongi}, Marco and {Straniero}, Oscar},
        title = "{The Evolution of a 25 M$_{{\ensuremath{\odot}}}$ Star from the Main Sequence up to the Onset of the Iron Core Collapse}",
      journal = {\apj},
         year = 1998,
        month = aug,
       volume = {502},
       number = {2},
        pages = {737-762},
          doi = {10.1086/305921},
       adsurl = {https://ui.adsabs.harvard.edu/abs/1998ApJ...502..737C}
}

@ARTICLE{sukhbold:16,
       author = {{Sukhbold}, Tuguldur and {Ertl}, T. and {Woosley}, S.~E. and {Brown}, Justin M. and {Janka}, H. -T.},
        title = "{Core-collapse Supernovae from 9 to 120 Solar Masses Based on Neutrino-powered Explosions}",
      journal = {\apj},
         year = 2016,
        month = apr,
       volume = {821},
       number = {1},
          eid = {38},
        pages = {38},
          doi = {10.3847/0004-637X/821/1/38},
archivePrefix = {arXiv},
       eprint = {1510.04643},
 primaryClass = {astro-ph.HE},
       adsurl = {https://ui.adsabs.harvard.edu/abs/2016ApJ...821...38S}
}

@ARTICLE{nishimura:17,
       author = {{Nishimura}, N. and {Hirschi}, R. and {Rauscher}, T. and {St. J. Murphy}, A. and {Cescutti}, G.},
        title = "{Uncertainties in s-process nucleosynthesis in massive stars determined by Monte Carlo variations}",
      journal = {\mnras},
         year = 2017,
        month = aug,
       volume = {469},
       number = {2},
        pages = {1752-1767},
          doi = {10.1093/mnras/stx696},
archivePrefix = {arXiv},
       eprint = {1701.00489},
 primaryClass = {astro-ph.SR},
       adsurl = {https://ui.adsabs.harvard.edu/abs/2017MNRAS.469.1752N}
}

@ARTICLE{pignatari:23,
       author = {{Pignatari}, Marco and {Gallino}, Roberto and {Reifarth}, Rene},
        title = "{The s process in massive stars, a benchmark for neutron capture reaction rates}",
      journal = {European Physical Journal A},
         year = 2023,
        month = dec,
       volume = {59},
       number = {12},
          eid = {302},
        pages = {302},
          doi = {10.1140/epja/s10050-023-01206-1},
       adsurl = {https://ui.adsabs.harvard.edu/abs/2023EPJA...59..302P}
}

@ARTICLE{lugaro:23,
       author = {{Lugaro}, Maria and {Pignatari}, Marco and {Reifarth}, Ren{\'e} and {Wiescher}, Michael},
        title = "{The s Process and Beyond}",
      journal = {Annual Review of Nuclear and Particle Science},
         year = 2023,
        month = sep,
       volume = {73},
        pages = {315-340},
          doi = {10.1146/annurev-nucl-102422-080857},
       adsurl = {https://ui.adsabs.harvard.edu/abs/2023ARNPS..73..315L}
}

@ARTICLE{kobayashi:20,
       author = {{Kobayashi}, Chiaki and {Karakas}, Amanda I. and {Lugaro}, Maria},
        title = "{The Origin of Elements from Carbon to Uranium}",
      journal = {\apj},
         year = 2020,
        month = sep,
       volume = {900},
       number = {2},
          eid = {179},
        pages = {179},
          doi = {10.3847/1538-4357/abae65},
archivePrefix = {arXiv},
       eprint = {2008.04660},
 primaryClass = {astro-ph.GA},
       adsurl = {https://ui.adsabs.harvard.edu/abs/2020ApJ...900..179K}
}

@ARTICLE{limongi:18,
       author = {{Limongi}, Marco and {Chieffi}, Alessandro},
        title = "{Presupernova Evolution and Explosive Nucleosynthesis of Rotating Massive Stars in the Metallicity Range -3 {\ensuremath{\leq}} [Fe/H] {\ensuremath{\leq}} 0}",
      journal = {\apjs},
         year = 2018,
        month = jul,
       volume = {237},
       number = {1},
          eid = {13},
        pages = {13},
          doi = {10.3847/1538-4365/aacb24},
archivePrefix = {arXiv},
       eprint = {1805.09640},
 primaryClass = {astro-ph.SR},
       adsurl = {https://ui.adsabs.harvard.edu/abs/2018ApJS..237...13L}
}

@ARTICLE{bisterzo:14,
       author = {{Bisterzo}, S. and {Travaglio}, C. and {Gallino}, R. and {Wiescher}, M. and {K{\"a}ppeler}, F.},
        title = "{Galactic Chemical Evolution and Solar s-process Abundances: Dependence on the $^{13}$C-pocket Structure}",
      journal = {\apj},
         year = 2014,
        month = may,
       volume = {787},
       number = {1},
          eid = {10},
        pages = {10},
          doi = {10.1088/0004-637X/787/1/10},
archivePrefix = {arXiv},
       eprint = {1403.1764},
 primaryClass = {astro-ph.SR},
       adsurl = {https://ui.adsabs.harvard.edu/abs/2014ApJ...787...10B}
}

@article{Ritter2018,
    author = {Ritter, C and Herwig, F and Jones, S and Pignatari, M and Fryer, C and Hirschi, R},
    title = "{NuGrid stellar data set – II. Stellar yields from H to Bi for stellar models with MZAMS = 1–25 M⊙ and Z = 0.0001–0.02}",
    journal = {Monthly Notices of the Royal Astronomical Society},
    volume = {480},
    number = {1},
    pages = {538-571},
    year = {2018},
    month = {06},
    issn = {0035-8711},
    doi = {10.1093/mnras/sty1729},
    url = {https://doi.org/10.1093/mnras/sty1729},
    eprint = {https://academic.oup.com/mnras/article-pdf/480/1/538/25244629/sty1729.pdf}
}

@article{10.1046/j.1365-8711.2001.04022.x,
    author = {Kroupa, Pavel},
    title = "{On the variation of the initial mass function}",
    journal = {Monthly Notices of the Royal Astronomical Society},
    volume = {322},
    number = {2},
    pages = {231-246},
    year = {2001},
    month = {04},
    issn = {0035-8711},
    doi = {10.1046/j.1365-8711.2001.04022.x},
    url = {https://doi.org/10.1046/j.1365-8711.2001.04022.x},
    eprint = {https://academic.oup.com/mnras/article-pdf/322/2/231/2852412/322-2-231.pdf}
}

@ARTICLE{2016ApJS..225...24P,
       author = {{Pignatari}, M. and {Herwig}, F. and {Hirschi}, R. and {Bennett}, M. and {Rockefeller}, G. and {Fryer}, C. and {Timmes}, F.~X. and {Ritter}, C. and {Heger}, A. and {Jones}, S. and {Battino}, U. and {Dotter}, A. and {Trappitsch}, R. and {Diehl}, S. and {Frischknecht}, U. and {Hungerford}, A. and {Magkotsios}, G. and {Travaglio}, C. and {Young}, P.},
        title = "{NuGrid Stellar Data Set. I.Stellar Yields from H to Bi for Stars with Metallicities Z = 0.02 and Z = 0.01}",
      journal = {\apjs},
         year = 2016,
        month = aug,
       volume = {225},
       number = {2},
          eid = {24},
        pages = {24},
          doi = {10.3847/0067-0049/225/2/24},
archivePrefix = {arXiv},
       eprint = {1307.6961},
 primaryClass = {astro-ph.SR},
       adsurl = {https://ui.adsabs.harvard.edu/abs/2016ApJS..225...24P}
}

@ARTICLE{2011ApJS..192....3P,
       author = {{Paxton}, Bill and {Bildsten}, Lars and {Dotter}, Aaron and {Herwig}, Falk and {Lesaffre}, Pierre and {Timmes}, Frank},
        title = "{Modules for Experiments in Stellar Astrophysics (MESA)}",
      journal = {\apjs},
         year = 2011,
        month = jan,
       volume = {192},
       number = {1},
          eid = {3},
        pages = {3},
          doi = {10.1088/0067-0049/192/1/3},
archivePrefix = {arXiv},
       eprint = {1009.1622},
 primaryClass = {astro-ph.SR},
       adsurl = {https://ui.adsabs.harvard.edu/abs/2011ApJS..192....3P}
}

@article{OTA2020135256,
title = {Decay properties of 22Ne + α resonances and their impact on s-process nucleosynthesis},
journal = {Physics Letters B},
volume = {802},
pages = {135256},
year = {2020},
issn = {0370-2693},
doi = {https://doi.org/10.1016/j.physletb.2020.135256},
url = {https://www.sciencedirect.com/science/article/pii/S0370269320300605},
author = {S. Ota and G. Christian and G. Lotay and W.N. Catford and E.A. Bennett and S. Dede and D.T. Doherty and S. Hallam and J. Hooker and C. Hunt and H. Jayatissa and A. Matta and M. Moukaddam and G.V. Rogachev and A. Saastamoinen and J.A. Tostevin and S. Upadhyayula and R. Wilkinson}
}

@ARTICLE{2018ApJS..238...36A,
       author = {{Abohalima}, Abdu and {Frebel}, Anna},
        title = "{JINAbase{\textemdash}A Database for Chemical Abundances of Metal-poor Stars}",
      journal = {\apjs},
         year = 2018,
        month = oct,
       volume = {238},
       number = {2},
          eid = {36},
        pages = {36},
          doi = {10.3847/1538-4365/aadfe9},
archivePrefix = {arXiv},
       eprint = {1711.04410},
 primaryClass = {astro-ph.SR},
       adsurl = {https://ui.adsabs.harvard.edu/abs/2018ApJS..238...36A}
}

@ARTICLE{2010ApJ...710.1557P,
       author = {{Pignatari}, M. and {Gallino}, R. and {Heil}, M. and {Wiescher}, M. and {K{\"a}ppeler}, F. and {Herwig}, F. and {Bisterzo}, S.},
        title = "{The Weak s-Process in Massive Stars and its Dependence on the Neutron Capture Cross Sections}",
      journal = {\apj},
         year = 2010,
        month = feb,
       volume = {710},
       number = {2},
        pages = {1557-1577},
          doi = {10.1088/0004-637X/710/2/1557},
       adsurl = {https://ui.adsabs.harvard.edu/abs/2010ApJ...710.1557P}
}

@ARTICLE{Kobayashi2006,
       author = {{Kobayashi}, C. and {Umeda}, H. and {Nomoto}, K. and {Tominaga}, N. and {Ohkubo}, T.},
        title = "{GALACTIC CHEMICAL EVOLUTION: CARBON THROUGH ZINC}",
      journal = {\apj},
         year = {2006},
       volume = {653},
       number = {2},
        pages = {1145-1171},
          doi = {10.1086/508914},
       adsurl = {https://iopscience.iop.org/article/10.1086/508914}
}

@ARTICLE{Nomoto2013,
       author = {{Nomoto}, K. and {Kobayashi}, C. and {Nomoto}, K. and {Tominaga}},
        title = "{Nucleosynthesis in Stars and the Chemical Enrichment of Galaxies}",
      journal = {Ann. Rev. Astro. Astrophys.},
         year = {2013},
       volume = {51},
       number = {},
        pages = {457-509},
          doi = {10.1146/annurev-astro-082812-140956},
       adsurl = {https://www.annualreviews.org/content/journals/10.1146/annurev-astro-082812-140956}
}

@ARTICLE{Kappeler2011,
       author = {{Kappeler}, F. and {Gallino}, R. and {Bisterzo}, S. and {Aoki}, W.},
        title = "The s process: Nuclear physics, stellar models, and observations",
      journal = {Rev. Mod. Phys.},
         year = {2011},
        month = {apr},
       volume = {83},
       number = {157},
        pages = {},
          doi = {10.1103/RevModPhys.83.157},
       adsurl = {https://journals.aps.org/rmp/abstract/10.1103/RevModPhys.83.157}
}

@ARTICLE{Siegel2022,
       author = {{Siegel}, D. M.},
        title = "r-Process nucleosynthesis in gravitational-wave and other explosive astrophysical events",
      journal = {Nature Reviews Physics},
         year = {2022},
        month = {apr},
       volume = {4},
       number = {},
        pages = {306-318},
          doi = {10.1038/s42254-022-00439-1},
       adsurl = {https://www.nature.com/articles/s42254-022-00439-1}
}

@ARTICLE{Denissenkov2019,
       author = {{Denissenkov}, P. A. and {Herwig}, F. and {Woodward}, P. and {Andrassy}, R. and {Pignatari}, M. and {Jones}, S.},
        title = "The i-process yields of rapidly accreting white dwarfs from multicycle He-shell flash stellar evolution models with mixing parametrizations from 3D hydrodynamics simulations",
      journal = {Monthly Notices of the Royal Astronomical Society},
         year = {2019},
        month = {sep},
       volume = {488},
       number = {3},
        pages = {4258-4270},
          doi = {10.1093/mnras/stz1921},
       adsurl = {https://academic.oup.com/mnras/article/488/3/4258/5531775}
}

@ARTICLE{Choplin2019,
       author = {{Choplin}, A. and {Goriely}, S. and {Hirschi}, R. and {Tominaga}, N. and {Meynet}, G.},
        title = "The p-process in exploding rotating massive stars",
      journal = {A\&A},
         year = {2019},
        month = {sep},
       volume = {661},
       number = {A86},
        pages = {},
          doi = {10.1051/0004-6361/202243331},
       adsurl = {https://www.aanda.org/articles/aa/full_html/2022/05/aa43331-22/aa43331-22.html}
}

@ARTICLE{Wiescher2023,
       author = {{Wiescher}, M. and {deBoer}, R. J. and {Gorres}, J.},
        title = "The resonances in the $^{22}$Ne+$\alpha$ fusion reactions",
      journal = {Eur. Phys. J. A},
         year = {2023},
        month = {},
       volume = {59},
       number = {11},
        pages = {},
          doi = {10.1140/epja/s10050-023-00917-9},
       adsurl = {https://link.springer.com/article/10.1140/epja/s10050-023-00917-9}
}

@ARTICLE{mishenina:17,
       author = {{Mishenina}, T. and {Pignatari}, M. and {C{\^o}t{\'e}}, B. and {Thielemann}, F. -K. and {Soubiran}, C. and {Basak}, N. and {Gorbaneva}, T. and {Korotin}, S.~A. and {Kovtyukh}, V.~V. and {Wehmeyer}, B. and {Bisterzo}, S. and {Travaglio}, C. and {Gibson}, B.~K. and {Jordan}, C. and {Paul}, A. and {Ritter}, C. and {Herwig}, F. and {NuGrid Collaboration}},
        title = "{Observing the metal-poor solar neighbourhood: a comparison of galactic chemical evolution predictions*{\textdagger}}",
      journal = {\mnras},
         year = 2017,
        month = aug,
       volume = {469},
       number = {4},
        pages = {4378-4399},
          doi = {10.1093/mnras/stx1145},
archivePrefix = {arXiv},
       eprint = {1705.03642},
 primaryClass = {astro-ph.SR},
       adsurl = {https://ui.adsabs.harvard.edu/abs/2017MNRAS.469.4378M}
}

@ARTICLE{lach:20,
       author = {{Lach}, F. and {R{\"o}pke}, F.~K. and {Seitenzahl}, I.~R. and {Cot{\'e}}, B. and {Gronow}, S. and {Ruiter}, A.~J.},
        title = "{Nucleosynthesis imprints from different Type Ia supernova explosion scenarios and implications for galactic chemical evolution}",
      journal = {\aap},
         year = 2020,
        month = dec,
       volume = {644},
          eid = {A118},
        pages = {A118},
          doi = {10.1051/0004-6361/202038721},
archivePrefix = {arXiv},
       eprint = {2010.14084},
 primaryClass = {astro-ph.SR},
       adsurl = {https://ui.adsabs.harvard.edu/abs/2020A&A...644A.118L}
}

@ARTICLE{Best2016,
       author = {{Best}, A. and {Gorres}, J. and {Junker}, M. and {Kratz}, K. -L. and {Laubenstein}, M. and {Long}, A. and {Nisi}, S. and {Smith}, K. and {Wiescher}, M.},
        title = "Low energy neutron background in deep underground laboratories",
      journal = {Nucl. Instr. Meth. A},
         year = {2016},
        month = {},
       volume = {812},
       number = {11},
        pages = {1-6},
          doi = {10.1016/j.nima.2015.12.034},
       adsurl = {https://www.sciencedirect.com/science/article/pii/S0168900215016058?via%3Dihub}
}

@ARTICLE{Gao2022,
       author = {{Gao}, B. and {JUNA Collaboration}},
        title = "Deep Underground Laboratory Measurement of $^{13}$C⁡($\alpha,n$)⁢$^{16}$O in the Gamow Windows of the s and i Processes",
      journal = {Phys. Rev. Lett.},
         year = {2022},
        month = {sep},
       volume = {129},
       number = {132701},
        pages = {},
          doi = {10.1103/PhysRevLett.129.132701},
       adsurl = {https://journals.aps.org/prl/abstract/10.1103/PhysRevLett.129.132701}
}

@ARTICLE{Shahina2022,
       author = {{Shahina}, S. and {Gorres}, J. and {Robertson}, D. and {Couder}, M. and {Gomez}, O. and {Gula}, A. and {Hanhardt}, M. and {Kadlecek}, T. and {Kelmar}, R. and {Scholz}, P. and {Simon}, A. and {Stech}, E. and {Strieder}, F. and {Wiescher}, M.},
        title = "Direct measurement of the low-energy resonances in $^{22}$Ne($\alpha,\gamma$)$^{26}$Mg reaction",
      journal = {Phys. Rev. C},
         year = {2022},
        month = {sep},
       volume = {106},
       number = {025805},
        pages = {},
          doi = {10.1103/PhysRevC.106.025805},
       adsurl = {https://journals.aps.org/prl/abstract/10.1103/PhysRevLett.129.132701}
}

@ARTICLE{Wolke1989,
       author = {{Wolke}, K. and {Harms}, V. and {Becket}, H. W. and {Couder}, M. and {Hammer}, J. W. and {Kratz}, K. L. and {Rolfs}, C. and {Schroder}, U. and {Trautvetter}, H. P. and {Wiescher}, M. and {Wohr}, A.},
        title = "Helium burning of $^{22}Ne",
      journal = {Z. Phys. A Atomic Nuclei},
         year = {1989},
        month = {},
       volume = {334},
       number = {},
        pages = {491-510},
          doi = {10.1007/BF01294757},
       adsurl = {https://link.springer.com/article/10.1007/BF01294757}
}

@ARTICLE{Drotleff1991,
       author = {{Drotleff}, W. and {Denker}, A. and {Hammer}, J. W. and {Knee}, H. and {Kuchler}, S. and {Streit}, D. and {Rolfs}, C. and {Trautvetter}, H. P.},
        title = "New $^{22}Ne($\alpha,n$)$^{25}Mg-resonances at very low energies relevant for the astrophysical s-process",
      journal = {Z. Phys. A Atomic Nuclei},
         year = {1991},
        month = {},
       volume = {338},
       number = {},
        pages = {367-368},
          doi = {10.1007/BF01288203},
       adsurl = {https://link.springer.com/article/10.1007/BF01288203}
}

@ARTICLE{Jaeger2001,
       author = {{Jaeger}, M. and {Kunz}, R. and {Mayer}, A. and {Hammer}, J. W. and {Staudt}, G. and {Kratz}, K. L. and {Pfeiffer}, B.},
        title = "$^{22}$Ne($\alpha,n$)$^{25}Mg: The Key Neutron Source in Massive Stars",
      journal = {Phys. Rev. Lett.},
         year = {2001},
        month = {},
       volume = {87},
       number = {202501},
        pages = {},
          doi = {10.1103/PhysRevLett.87.202501},
       adsurl = {https://journals.aps.org/prl/abstract/10.1103/PhysRevLett.87.202501}
}

@ARTICLE{Shahina2024,
       author = {{Shahina}, S. and {deBoer}, R. J. and {Gorres}, J. and {Fang}, R. and {Febbraro}, M. and {Kelmar}, R. and {Matney}, M. and {Manukyan}, K. and {Nattress}, T. and {Robles}, E. and {Ruland}, T. J. and {King}, T. T. and {Sanchez}, A. and {Sidhu}, R. S. and {Stech}, E. and {Wiescher}, M.},
        title = "Strength measurement of the $E_{\alpha}^{lab}=830 keV resonance in the $^{22}$Ne($\alpha,n$)$^{25}$Mg reaction using a stilbene detector",
      journal = {Phys. Rev. C},
         year = {2024},
        month = {},
       volume = {110},
       number = {015801},
        pages = {},
          doi = {10.1103/PhysRevC.110.015801},
       adsurl = {https://journals.aps.org/prc/abstract/10.1103/PhysRevC.110.015801}
}

@ARTICLE{Hunt2019,
       author = {{Hunt}, C. and {Iliadis}, C. and {Champagne}, A. and {Downen}, L. and {Cooper}, A.},
        title = "New measurement of the $^E_{lab}$ $\alpha$ = 0.83 MeV resonance in $^{22}$Ne($\alpha,\gamma$)$^{26}$Mg",
      journal = {Phys. Rev. C},
         year = {2019},
        month = {},
       volume = {99},
       number = {045804},
        pages = {},
          doi = {10.1103/PhysRevC.99.045804},
       adsurl = {https://journals.aps.org/prc/abstract/10.1103/PhysRevC.99.045804}
}

@ARTICLE{Giesen1993,
       author = {{Giesen}, U. and {Browne}, C. P. and {Gorres}, J. and {Graff}, S. and {Iliadis}, C. and {Trautvetter}, H. P. and {Wiescher}, M. and {Harms}, W. and {Kratz}, K. L. and {Pfeiffer}, B. and {Azuma}, R. R. and {Buckby}, M. and {King}, J. D.},
        title = "The astrophysical implications of low-energy resonances in $^{22}$Ne + $\alpha$",
      journal = {Nucl. Phys. A},
         year = {1993},
        month = {},
       volume = {561},
       number = {1},
        pages = {95-111},
          doi = {10.1016/0375-9474(93)90167-V},
       adsurl = {https://www.sciencedirect.com/science/article/pii/037594749390167V?via%3Dihub}
}

@ARTICLE{Longland2009,
       author = {{Longland}, R. and {Iliadis}, G. and {Rusev}, G. and {Tonchev}, A. P. and {Iliadis}, C. and {Trautvetter}, H. P. and {Wiescher}, M.},
        title = "Photoexcitation of astrophysically important states in $^{26}$Mg",
      journal = {Phys. Rev. C},
         year = {2009},
        month = {},
       volume = {80},
       number = {055803},
        pages = {},
          doi = {10.1103/PhysRevC.80.055803},
       adsurl = {https://journals.aps.org/prc/abstract/10.1103/PhysRevC.80.055803}
}

@ARTICLE{deBoer2010,
       author = {{deBoer}, R. J. and {Wiescher}, M. and {Gorres}, J. and {Longland}, R. and {Iliadis}, C. and {Rusev}, G. and {Tonchev}, A. P.},
        title = "Photoexcitation of astrophysically important states in $^{26}$Mg. II. Ground-state-transition partial widths",
      journal = {Phys. Rev. C},
         year = {2010},
        month = {},
       volume = {82},
       number = {025802},
        pages = {},
          doi = {10.1103/PhysRevC.82.025802},
       adsurl = {https://journals.aps.org/prc/abstract/10.1103/PhysRevC.82.025802}
}

@ARTICLE{Massimi2017,
       author = {{Massimi}, C. and {Altstadt}, S. and {Andrzejewski}, J. and {et al.}},
        title = "Neutron spectroscopy of $^{26}$Mg states: Constraining the stellar neutron source $^{22}$Ne($\alpha,n$)$^{25}$Mg",
      journal = {Phys. Lett. B},
         year = {2017},
        month = {},
       volume = {768},
       number = {},
        pages = {1-6},
          doi = {10.1016/j.physletb.2017.02.025},
       adsurl = {https://www.sciencedirect.com/science/article/pii/S0370269317301260?via%3Dihub}
}

@ARTICLE{Talwar2016,
       author = {{Talwar}, R. and {Adachi}, T. and {Berg}, G. P. A. and {Bin}, L. and {Bisterzo}, S. and {Couder}, M. and {deBoer}, R. J. and {Fang}, X. and {Fujita}, H. and {Fujita}, Y. and {Gorres}, J. and {Hatanaka}, K. and {Itoh}, T. and {Kadoya}, T. and {Long}, A. and {Miki}, K. and {Patel}, D. and {Pignatari}, M. and {Shimbara}, Y. and {Tamii}, A. and {Wiescher}, M. and {Yamamoto}, T. and {Yosoi}, M.},
        title = "Probing astrophysically important states in the $^{26}Mg nucleus to study neutron sources for the s process",
      journal = {Phys. Rev. C},
         year = {2016},
        month = {},
       volume = {93},
       number = {055803},
        pages = {},
          doi = {10.1103/PhysRevC.93.055803},
       adsurl = {https://journals.aps.org/prc/abstract/10.1103/PhysRevC.93.055803}
}

@ARTICLE{Adsley2017,
       author = {{Adsley}, P. and {Brummer}, J. W. and {Li}, K. C. W. and {Marin-Lambarri}, D. J. and {Kheswa}, N. Y. and {Donaldson}, L. M. and {Neveling}, R. and {Papka}, P. and {Pellegri}, L. and {Pesudo}, V. and {Pool}, L. C. and {Smit}, F. D. and {van Zyl}, J. J.},
        title = "Re-examining the $^{26}$Mg($\alpha,\alpha\prime$)$^{26}$Mg reaction: Probing astrophysically important states in $^{26}Mg",
      journal = {Phys. Rev. C},
         year = {2017},
        month = {},
       volume = {96},
       number = {055802},
        pages = {},
          doi = {10.1103/PhysRevC.96.055802},
       adsurl = {https://journals.aps.org/prc/abstract/10.1103/PhysRevC.96.055802}
}

@ARTICLE{Adsley2018,
       author = {{Adsley}, P. and {Brummer}, J. W. and {Faestermann}, T. and {Fox}, S. P. {Hammache}, F. and {Hertenberger}, R. and {Meyer}, A. and {Neveling}, R. and {Seiler}, D. and {de Sereville}, N. and {Wirth}, H. -F.},
        title = "High-resolution study of levels in the astrophysically important nucleus $^{26}Mg and resulting updated level assignments",
      journal = {Phys. Rev. C},
         year = {2018},
        month = {},
       volume = {97},
       number = {045807},
        pages = {},
          doi = {10.1103/PhysRevC.97.045807},
       adsurl = {https://journals.aps.org/prc/abstract/10.1103/PhysRevC.97.045807}
}

@ARTICLE{Jayatissa2020,
       author = {{Jayatissa}, H. and {Rogachev}, G. V. and {Goldberg}, V. Z. and {Koshchiy}, E. {Christian}, G. and {Hooker}, J. and {Ota}, S. and {Roeder}, B. T. and {Saastamoinen}, A. and {Trippella}, O. and {Upadhyayula}, S. and {Uberseder}, E.},
        title = "Constraining the $^{22}$Ne($\alpha,\gamma$)$^{26}$Mg and $^{22}$Ne($\alpha,n$)$&{25}$Mg reaction rates using sub-Coulomb $\alpha$-transfer reactions",
      journal = {Phys. Lett. B},
         year = {2020},
        month = {},
       volume = {802},
       number = {135267},
        pages = {},
          doi = {10.1016/j.physletb.2020.135267},
       adsurl = {https://www.sciencedirect.com/science/article/pii/S037026932030071X?via%3Dihub}
}

@ARTICLE{Ota2021,
       author = {{Ota}, S. and {Christian}, G. and {Catford}, W. N. and {Lotay}, G. {Pignatari}, M. and {Battino}, U. and {Bennett}, E. A. and {Dede}, S. and {Doherty}, D. T. and {Hallam}, S. and {Herwig}, F. and {Hooker}, J. and {Hunt}, C. and {Jayatissa}, H. and {Matta}, A. and {Moukaddam}, M. and {Rao}, E. and {Rogachev}, G. V. and {Saastamoinen}, A. and {Scriven}, D. and {Tostevin}, J. A. and {Upadhyayula}, S. and {Wilkinson}, R.},
        title = "($^6$Li,$d$) and ($^6$Li,$t$) reactions on $^{22}$Ne and implications for s-process nucleosynthesis",
      journal = {Phys. Rev. C},
         year = {2021},
        month = {},
       volume = {104},
       number = {055806},
        pages = {},
          doi = {10.1103/PhysRevC.104.055806},
       adsurl = {https://journals.aps.org/prc/abstract/10.1103/PhysRevC.104.055806}
}

@ARTICLE{Chen2021,
       author = {{Chen}, Y. and {Berg}, G. P. A. and {deBoer}, R. J. and {Gorres}, J. {Jung}, H. and {Long}, A. and {Seetedohnia}, K. and {Talwar}, R. and {Wiescher}, M. and {Adachi}, S. and {Fujita}, H. and {Fujita}, Y. and {Hatanaka}, K. and {Iwamoto}, C. and {Liu}, B. and {Noji}, S. and {Ong}, H. -J. and {Tamii}, A.},
        title = "Neutron transfer studies on $^{25}$Mg and its correlation to neutron radiative capture processes",
      journal = {Phys. Rev. C},
         year = {2021},
        month = {},
       volume = {103},
       number = {035809},
        pages = {},
          doi = {10.1103/PhysRevC.103.035809},
       adsurl = {https://journals.aps.org/prc/abstract/10.1103/PhysRevC.103.035809}
}

@ARTICLE{Adsley2021,
       author = {{Adsley}, P. and {Battino}, U. and {Best}, A. and {Caciolli}, A. {Guglielmetti}, A. and {Imbriani}, G. and {Jayatissa}, H. and {La Cognata}, M. and {Lamia}, L. and {Masha}, E. and {Massimi}, C. and {Palmerini}, S. and {Tattersall}, A. and Hirschi, R},
        title = "Reevaluation of the $^{22}$Ne($\alpha,\gamma$)$^{26}$Mg and $^{22}Ne($\alpha,n$)$^{25}$Mg reaction rates",
      journal = {Phys. Rev. C},
         year = {2021},
        month = {},
       volume = {103},
       number = {015805},
        pages = {},
          doi = {10.1103/PhysRevC.103.015805},
       adsurl = {https://journals.aps.org/prc/abstract/10.1103/PhysRevC.103.015805}
}

@ARTICLE{Longland2012,
       author = {{Longland}, R. and {Iliadis}, C. and {Karakas}, A. I.},
        title = "Reaction rates for the s-process neutron source $^{22}$Ne+$\alpha$",
      journal = {Phys. Rev. C},
         year = {2012},
        month = {},
       volume = {85},
       number = {065809},
        pages = {},
          doi = {10.1103/PhysRevC.85.065809},
       adsurl = {https://journals.aps.org/prc/abstract/10.1103/PhysRevC.85.065809}
}

@ARTICLE{Harms1991,
       author = {{Harms}, V. and {Kratz}, K. -L. and {Wiescher}, M.},
        title = "Properties of $^{22}$Ne($\alpha,n$)$^{25}$Mg resonances",
      journal = {Phys. Rev. C},
         year = {1991},
        month = {},
       volume = {43},
       number = {},
        pages = {2849-2861},
          doi = {10.1103/PhysRevC.43.2849},
       adsurl = {https://
doi.org/10.1103/PhysRevC.43.2849}
}

@ARTICLE{Fryer2012,
       author = {{Fryer}, C. L. and {Belczynski}, K. and {Wiktorowicz}, G.},
        title = "Properties of $^{22}$Ne($\alpha,n$)$^{25}$Mg resonances",
      journal = {\apj},
         year = {2012},
        month = {},
       volume = {749},
       number = {91},
        pages = {},
          doi = {10.1088/0004-637X/749/1/91},
       adsurl = {https://iopscience.iop.org/article/10.1088/0004-637X/749/1/91}
}

@ARTICLE{Bisterzo2005,
       author = {{Bisterzo}, S. and {Pompeia}, L. and {Gallino}, R. and {Pignatari}, M. and {Cunha}, K. and {Heger}, A. and {Smith}, V.},
        title = "Cu and Zn in different stellar conditions: Infering their astrophysical origin",
      journal = {Nucl. Phys. A},
         year = {2005},
        month = {},
       volume = {758},
       number = {},
        pages = {284-287},
          doi = {10.1016/j.nuclphysa.2005.05.049},
       adsurl = {https://www.sciencedirect.com/science/article/pii/S0375947405006913}
}

@ARTICLE{woosley:92,
       author = {{Woosley}, S.~E. and {Hoffman}, Robert D.},
        title = "{The alpha -Process and the r-Process}",
      journal = {\apj},
         year = 1992,
        month = aug,
       volume = {395},
        pages = {202},
          doi = {10.1086/171644},
       adsurl = {https://ui.adsabs.harvard.edu/abs/1992ApJ...395..202W}
}

@ARTICLE{raiteri:92,
       author = {{Raiteri}, C.~M. and {Gallino}, R. and {Busso}, M.},
        title = "{S-Processing in Massive Stars as a Function of Metallicity and Interpretation of Observational Trends}",
      journal = {\apj},
         year = 1992,
        month = mar,
       volume = {387},
        pages = {263},
          doi = {10.1086/171078},
       adsurl = {https://ui.adsabs.harvard.edu/abs/1992ApJ...387..263R}
}

@ARTICLE{pignatari:18,
       author = {{Pignatari}, Marco and {Hoppe}, Peter and {Trappitsch}, Reto and {Fryer}, Chris and {Timmes}, F.~X. and {Herwig}, Falk and {Hirschi}, Raphael},
        title = "{The neutron capture process in the He shell in core-collapse supernovae: Presolar silicon carbide grains as a diagnostic tool for nuclear astrophysics}",
      journal = {\gca},
         year = 2018,
        month = jan,
       volume = {221},
        pages = {37-46},
          doi = {10.1016/j.gca.2017.06.005},
       adsurl = {https://ui.adsabs.harvard.edu/abs/2018GeCoA.221...37P}
}

@ARTICLE{cowan:05,
       author = {{Cowan}, John J. and {Sneden}, Christopher and {Beers}, Timothy C. and {Lawler}, James E. and {Simmerer}, Jennifer and {Truran}, James W. and {Primas}, Francesca and {Collier}, Jason and {Burles}, Scott},
        title = "{Hubble Space Telescope Observations of Heavy Elements in Metal-Poor Galactic Halo Stars}",
      journal = {\apj},
         year = 2005,
        month = jul,
       volume = {627},
       number = {1},
        pages = {238-250},
          doi = {10.1086/429952},
archivePrefix = {arXiv},
       eprint = {astro-ph/0502591},
 primaryClass = {astro-ph},
       adsurl = {https://ui.adsabs.harvard.edu/abs/2005ApJ...627..238C}
}

@ARTICLE{meyer:00,
       author = {{Meyer}, B.~S. and {Clayton}, D.~D. and {The}, L. -S.},
        title = "{Molybdenum and Zirconium Isotopes from a Supernova Neutron Burst}",
      journal = {\apjl},
         year = 2000,
        month = sep,
       volume = {540},
       number = {1},
        pages = {L49-L52},
          doi = {10.1086/312865},
       adsurl = {https://ui.adsabs.harvard.edu/abs/2000ApJ...540L..49M}
}

@ARTICLE{pignatari:08,
       author = {{Pignatari}, M. and {Gallino}, R. and {Meynet}, G. and {Hirschi}, R. and {Herwig}, F. and {Wiescher}, M.},
        title = "{The s-Process in Massive Stars at Low Metallicity: The Effect of Primary $^{14}$N from Fast Rotating Stars}",
      journal = {\apjl},
         year = 2008,
        month = nov,
       volume = {687},
       number = {2},
        pages = {L95},
          doi = {10.1086/593350},
archivePrefix = {arXiv},
       eprint = {0810.0182},
 primaryClass = {astro-ph},
       adsurl = {https://ui.adsabs.harvard.edu/abs/2008ApJ...687L..95P}
}

@ARTICLE{Prantzos2020,
       author = {{Prantzos}, N. and {Abia}, C. and {Cristallo}, S. and {Limongi}, M. and {Chieffi}, A.},
        title = "Chemical evolution with rotating massive star yields II. A new assessment of the solar s- and r-process components",
      journal = {Monthly Notices of the Royal Astronomical Society},
         year = {2020},
        month = {},
       volume = {491},
       number = {1832},
        pages = {},
          doi = {doi.org/10.1093/mnras/stz3154},
       adsurl = {https://academic.oup.com/mnras/article/491/2/1832/5621509}
}

@ARTICLE{Massimi2012,
       author = {{Massimi}, C. and {Koehler}, P. and {Bisterzo}, S. and {et al.}},
        title = "Resonance neutron-capture cross sections of stable magnesium isotopes and their astrophysical implications",
      journal = {Phys. Rev. C},
         year = {2012},
        month = {},
       volume = {85},
       number = {044615},
        pages = {},
          doi = {10.1103/PhysRevC.85.044615},
       adsurl = {https://journals.aps.org/prc/abstract/10.1103/PhysRevC.85.044615}
}

@ARTICLE{Grevesse1993,
       author = {{Grevesse}, N. and {Noels}, A.},
        title = "Origin and Evolution of the Elements",
      journal = {Cambridge Univ. Press},
         year = {1993},
        month = {},
       volume = {},
       number = {},
        pages = {15-25},
          doi = {},
       adsurl = {}
}

@ARTICLE{Asplund2009,
       author = {{Asplund}, M. and {Grevesse}, N. and {Sauval}, J. and {Scott}, P.},
        title = "The Chemical Composition of the Sun",
      journal = {Annu. Rev. Astron. Astrophys.},
         year = {2009},
        month = {},
       volume = {47},
       number = {},
        pages = {481-522},
          doi = {10.1146/annurev.astro.46.060407.145222},
       adsurl = {https://www.annualreviews.org/content/journals/10.1146/annurev.astro.46.060407.145222}
}

@ARTICLE{Wiescher2012,
       author = {{Wiescher}, M. and {Kappeler}, F. and {Langanke}, K.},
        title = "Critical Reactions in Contemporary Nuclear Astrophysics",
      journal = {Annu. Rev. Astron. Astrophys.},
         year = {2012},
        month = {},
       volume = {50},
       number = {},
        pages = {165-210},
          doi = {10.1146/annurev-astro-081811–125543},
       adsurl = {https://www.annualreviews.org/docserver/fulltext/astro/50/1/annurev-astro-081811-125543.pdf?expires=1718530847&id=id&accname=ar-366613&checksum=CDAF90C3467C50FFC09C3543FDB38531}
}

@ARTICLE{Wang2021,
       author = {{Wang}, M. and {Huang}, W. J. and {Kondev}, F. G. and {Audi}, G. and {Naimi}, S.},
        title = "The AME 2020 atomic mass evaluation (II). Tables, graphs and references",
      journal = {Chinese Phys.},
         year = {2021},
        month = {},
       volume = {45},
       number = {3},
        pages = {},
          doi = {10.1088/1674-1137/abddaf},
       adsurl = {https://iopscience.iop.org/article/10.1088/1674-1137/abddaf}
}

@ARTICLE{Straniero2006,
       author = {{Straniero}, O. and {Gallino}, R. and {Cristallo}, S. },
        title = "s process in low-mass asymptotic giant branch stars",
      journal = {Nucl. Phys. A},
         year = {2006},
        month = {},
       volume = {777},
       number = {},
        pages = {311-339},
          doi = {10.1016/j.nuclphysa.2005.01.011},
       adsurl = {https://www.sciencedirect.com/science/article/pii/S037594740500028X?via%3Dihub}
}

@ARTICLE{Ritter2018syg,
       author = {Ritter, C. and {Cote}, B. and {Herwig}, F. and {Navarro}, J. F. and {Fryer}, C. L.},
        title = "SYGMA: Stellar Yields for Galactic Modeling Applications",
      journal = {ApJ Suppl.},
         year = {2018},
        month = {},
       volume = {237},
       number = {},
        pages = {42},
          doi = {10.3847/1538-4365/aad691},
       adsurl = {https://iopscience.iop.org/article/10.3847/1538-4365/aad691}
}

@ARTICLE{Masha2022,
       author = {{Masha}, E. and {LUNA Collaboration}.},
        title = "First direct limit on the 395 keV resonance of the
$^{22}$Ne($\alpha,\gamma$)$^{26}$Mg reaction",
      journal = {EPJ Web of Conferences},
         year = {2022},
        month = {},
       volume = {260},
       number = {11017},
        pages = {42},
          doi = {10.1051/epjconf/202226011017},
       adsurl = {https://www.epj-conferences.org/articles/epjconf/abs/2022/04/epjconf_nic16th2022_11017/epjconf_nic16th2022_11017.html}
}

@ARTICLE{Pignatari2013,
       author = {{Pignatari}, M. and {Hirschi}, R. and {Wiescher}, M. and {Gallino}, R. and {Bennett}, M. and {Beard}, M. and {Fryer}, C. and {Herwig}, F. and {Rockefeller}, G. and {Timmes}, F. X.},
        title = "The $^{12}$C + $^{12}$C reaction and the impact on nucleosynthesis in massive stars",
      journal = {Astrophys. J. },
         year = {2013},
        month = {},
       volume = {762},
       number = {1},
        pages = {31},
          doi = {10.1088/0004-637X/762/1/31},
       adsurl = {https://iopscience.iop.org/article/10.1088/0004-637X/762/1/31}
}

@ARTICLE{Iwamoto1999,
       author = {{Iwamoto}, K. and {Brachwitz}, F. and {Nomoto}, K. and {Kishimoto}, N. and {Umeda}, H. and {Hix}, W. R. and {Thielemann}, F.},
        title = "Nucleosynthesis in Chandrasekhar mass models for Type Ia supernovae AND
constraints on progenitor systems and burning-front propagation",
      journal = {Astrophys. J. Suppl.},
         year = {1999},
        month = {},
       volume = {125},
       number = {2},
        pages = {439},
          doi = {10.1086/313278},
       adsurl = {https://iopscience.iop.org/article/10.1086/313278/meta}
}

@ARTICLE{Matteucci1993,
       author = {{Matteucci}, F. and {Raiteri}, C. M. and {Busso}, M. and {Gallino}, R. and {Gratton}, R.},
        title = "Constraints on the nucleosynthesis of Cu and Zn from models of chemical evolution of the Galaxy",
      journal = {A\&A},
         year = {1993},
        month = {},
       volume = {272},
       number = {},
        pages = {421-429},
          doi = {},
       adsurl = {https://ui.adsabs.harvard.edu/abs/1993A%26A...272..421M/abstract}
}

@ARTICLE{Mishenia2002,
       author = {{Mishenia}, T. V. and {Kovtyukh}, V. V. and {Soubiran}, C. and {Travaglio}, C. and {Busso}, M.},
        title = "Abundances of Cu and Zn in metal-poor stars: Clues for Galaxy evolution",
      journal = {A\&A},
         year = {2002},
        month = {},
       volume = {396},
       number = {},
        pages = {189-201},
          doi = {10.1051/0004-6361:20021399},
       adsurl = {https://www.aanda.org/articles/aa/pdf/2002/46/aa2651.pdf}
}

@ARTICLE{Woosley1994,
       author = {{Woosley}, S. E.. and {Wilson}, J. R. and {Mathews}, G. J. and {Hoffman}, R. D. and {Meyer}, B. S.},
        title = "The r-Process and Neutrino-heated Supernova Ejecta",
      journal = {Astrophys. J.},
         year = {1994},
        month = {},
       volume = {},
       number = {433},
        pages = {229-246},
          doi = {10.1086/174638},
       adsurl = {}
}

@ARTICLE{Kappeler1994,
       author = {{Kappeler}, S. E. and {Weischer}, M. and {Giesen}, U. and {Gorres}, J. and {et al.}},
        title = "Reaction rates for $^{18}$O($\alpha,\gamma$)$^{22}$Ne, $^{22}$Ne($\alpha,\gamma$)$^{26}$Mg, and $^{22}$Ne($\alpha,n$)$^{25}$Mg in stellar helium burning and s-process nucleosynthesis in massive stars",
      journal = {Astrophys. J.},
         year = {1994},
        month = {},
       volume = {},
       number = {437},
        pages = {396-409},
          doi = {},
       adsurl = {}
}

@ARTICLE{battino:16,
       author = {{Battino}, U. and {Pignatari}, M. and {Ritter}, C. and {Herwig}, F. and {Denisenkov}, P. and {Den Hartogh}, J. W. and {Trappitsch}, R. and {Hirschi}, R. and {Freytag}, B. and {Thielemann}, F. and {Paxton}, B.},
        title = "APPLICATION OF A THEORY AND SIMULATION-BASED CONVECTIVE BOUNDARY
MIXING MODEL FOR AGB STAR EVOLUTION AND NUCLEOSYNTHESIS",
      journal = {Astrophys. J.},
         year = {2016},
        month = {},
       volume = {},
       number = {827},
        pages = {1-30},
          doi = {10.3847/0004-637X/827/1/30},
       adsurl = {}
}

@ARTICLE{Timmes1995,
       author = {{Timmes}, F. X. and {Woosley}, S. E. and {Weaver}, T. A.},
        title = "GALACTIC CHEMICAL EVOLUTION: HYDROGEN THROUGH ZINC",
      journal = {Astrophys. J. Suppl.},
         year = {1995},
        month = {},
       volume = {},
       number = {98},
        pages = {617-658},
          doi = {10.1086/192172},
       adsurl = {}
}

@ARTICLE{Cote2016,
       author = {{Côté}, B. and {Ritter}, C. and {O'Shea}, B. W. and {Herwig}, F. and {Pignatari}, M. and {Jones}, S. and {Fryer}, C. L.},
        title = "UNCERTAINTIES IN GALACTIC CHEMICAL EVOLUTION MODELS",
      journal = {Astrophys. J.},
         year = {2016},
        month = {},
       volume = {824},
       number = {82},
        pages = {},
          doi = {10.3847/0004-637X/824/2/82},
       adsurl = {}
}

@ARTICLE{Best2025,
       author = {{Best}, A. and {Adsley}, P. and {Amberger}, R. and {Battino}, U. and {Chillery}, T. and {La Cognata}, M. and {deBoer}, R. J. and {Mercogliano}, D. and {Ota}, S. and {Rapagnani}, D. and {Sidhu}, R. S. and {Sparta}, R. and {Tumino}, A. and {Wiescher}, M.},
        title = "The $^{22}$Ne($\alpha$,n)$^{25}$Mg reaction - state of the art, astrophysics, and perspectives",
      journal = {Eur. Phys. J. A},
         year = {2025},
        month = {},
       volume = {},
       number = {},
        pages = {(under review)},
          doi = {},
       adsurl = {}
}

@ARTICLE{Angulo1999,
       author = {{Angulo}, C. and {Arnould}, M. and {Rayet}, M. and {Descouvemont}, P. and {Baye}, D. and {Leclercq-Willain}, C. and {Coc}, A. and {Barhoumi}, S. and {Aguer}, P. and {Rolfs}, C. and {Kunz}, R. and {Hammer}, J. W. and {Mayer}, A. and {Paradellis}, T. and {Kossionides}, S. and {Chronidou}, C. and {Spyrou}, K. and {Degl'Innocenti}, S. and {Fiorentini}, G. and {Ricci}, B. and {Zavatarelli}, S. and {Providencia}, C. and {Wolters}, H. and {Soares}, J. and {Grama}, C. and {Rahighi}, J. and {Shotter}, A. and {Lamehi Rachti}, M.},
        title = "A compilation of charged-particle induced
thermonuclear reaction rates",
      journal = {Nucl. Phys. A},
         year = {1999},
        month = {},
       volume = {},
       number = {656},
        pages = {3-183},
          doi = {},
       adsurl = {}
}

@ARTICLE{Longland2010,
       author = {{Longland}, R. and {Iliadis}, C. and {Champagne}, A. E. and {Newton}, J. R. and {Ugalde}, C. and {Coc}, A. and {Fitzgerald}, R.},
        title = "Charged-particle thermonuclear reaction rates: I. Monte Carlo method and statistical distributions",
      journal = {Nucl. Phys. A},
         year = {2010},
        month = {},
       volume = {841},
       number = {1},
        pages = {},
          doi = {},
       adsurl = {}
}

@ARTICLE{Magg2022,
       author = {{Magg}, E. and {Bergemann}, M. and {Serenelli}, A. and {Bautista}, M. and {Plez}, B. and {Heiter}, U. and {Gerber}, J. M. and {Ludwig}, H. G. and {Basu}, S. and {Ferguson}, J. W. and {Gallego1}, H. C. and {Gamrathl}, S. and {Palmeri}, P. and {Quinet1}, P.},
        title = "Observational constraints on the origin of the elements
IV. Standard composition of the Sun",
      journal = {A\&A},
         year = {2022},
        month = {},
       volume = {661},
       number = {140},
        pages = {},
          doi = {},
       adsurl = {}
}

@ARTICLE{Sedov1946,
       author = {{Sedov}, L. I.},
        title = "",
      journal = {ApMM},
         year = {1946},
        month = {},
       volume = {10},
       number = {241},
        pages = {},
          doi = {},
       adsurl ={}
}

@ARTICLE{Roberti2024,
       author = {{Roberti}, L. and {Pignatari}, M. and {Fryer}, C. and {Lugaro}, M.},
        title = "",
      journal = {A\&A},
         year = {2024},
        month = {},
       volume = {686},
       number = {L8},
        pages = {},
          doi = {},
       adsurl ={}
}

@ARTICLE{Dillmann2006,
       author = {{Dillmann}, I. and {Heil}, M. and {Kappeler}, F. and {et al.}},
        title = "",
      journal = {AIP Conf. Ser},
         year = {2006},
        month = {},
       volume = {819},
       number = {123},
        pages = {},
          doi = {},
       adsurl ={}
}

@ARTICLE{Berman1969,
       author = {{Berman}, B. L. and {Hemert}, R. L. and {Bowman}, C. D.},
        title = "",
      journal = {Phys. Rev. Lett.},
         year = {1969},
        month = {},
       volume = {23},
       number = {386},
        pages = {},
          doi = {https://doi.org/10.1103/PhysRevLett.23.386},
       adsurl ={}
}

@ARTICLE{Rapagnani2022,
       author = {{Rapagnani}, D. and {Ananna}, C. and {Di Leva}, A. and {Imbriani1}, G. and {Junker}, M. and {Pignatari}, M. and {Best}, A.},
        title = "",
      journal = {EPJ Web of Conferences},
         year = {2022},
        month = {},
       volume = {260},
       number = {11031},
        pages = {},
          doi = {},
       adsurl ={}
}

@ARTICLE{Lai2008,
       author = {{Lai}, D. K. and {Bolte}, M. and {Johnson}, J. A. and {Lucatello}, S. and {Heger}, A. and {Wooseley}, S. E.},
        title = "",
      journal = {Astrophys. J.},
         year = {2008},
        month = {},
       volume = {681},
       number = {1524},
        pages = {},
          doi = {},
       adsurl ={}
}

@ARTICLE{Bihain2004,
       author = {{Bihain}, G. and {Israelian}, G. and {Rebolo}, R. and {Bonifacio}, P. and {Molaro}, P.},
        title = "",
      journal = {A\&A},
         year = {2004},
        month = {},
       volume = {423},
       number = {777},
        pages = {},
          doi = {},
       adsurl ={}
}

@ARTICLE{Ivans2003,
       author = {{Ivans}, I. I. and {Sneden}, C. and {James}, C. R. and {Preston}, G. W. and {Fulbright}, J. P. and {Hoflich}, P. A. and {Carney}, B. W. and {Wheeler}, J. C.},
        title = "",
      journal = {Astrophys. J.},
         year = {2003},
        month = {},
       volume = {592},
       number = {906},
        pages = {},
          doi = {},
       adsurl ={}
}

@ARTICLE{Reddy2006,
       author = {{Reddy}, B. E. and {Lambert}, D. L. and {Prieto}, C. A. },
        title = "",
      journal = {MNRAS},
         year = {2006},
        month = {},
       volume = {367},
       number = {1329},
        pages = {},
          doi = {},
       adsurl ={}
}

@ARTICLE{Reddy2003,
       author = {{Reddy}, B. E. and {Tomkin}, J. and {Lambert}, D. L. and {Prieto}, C. A. },
        title = "",
      journal = {MNRAS},
         year = {2003},
        month = {},
       volume = {340},
       number = {},
        pages = {304},
          doi = {},
       adsurl ={}
}

@ARTICLE{Ishigaki2013,
       author = {{Ishigaki}, M. N. and {Wako}, A. and {Chiba}, M.},
        title = "",
      journal = {Astrophys. J.},
         year = {2013},
        month = {},
       volume = {771},
       number = {},
        pages = {67},
          doi = {},
       adsurl ={}
}

@ARTICLE{Roederer2014,
       author = {{Roederer}, I. U. and  {Preston}, G. W. and {Thompson}, I. B. and {Shectman}, S. A. and {Sneden}, C. and {Burley}, G. S. and {Kelson}, D. D.},
        title = "",
      journal = {Astrophys. J.},
         year = {2014},
        month = {},
       volume = {147},
       number = {},
        pages = {136},
          doi = {},
       adsurl ={}
}

@ARTICLE{Frebel2010,
       author = {{Frebel}, A. and {Simon}, J. D. and {Geha}, M. and {Willman}, B.},
        title = "",
      journal = {Astrophys. J.},
         year = {2010},
        month = {},
       volume = {708},
       number = {560},
        pages = {},
          doi = {},
       adsurl ={}
}

@ARTICLE{Bensby2005,
       author = {{Bensby}, T. and {Feltzing}, S. and {Lundstrom}, I. and {Ilyin}, I.},
        title = "",
      journal = {A\&A},
         year = {2005},
        month = {},
       volume = {433},
       number = {},
        pages = {185},
          doi = {},
       adsurl ={}
}

@ARTICLE{Bonifacio2009,
       author = {{Bonifacio}, P. and {Spite}, M. and {Cayrel}, R. and {Hill}, V. and {Spite}, F. and {Francois}, P. and {Plez}, B. and {Ludwig}, H. G. and {Caffau}, E. and {Molaro}, P. and {Depagne}, E. and {Andersen}, J. and {Barbuy}, B. and {Beers}, C. and {Nordstrom}, B. and {Primas}, F.},
        title = "",
      journal = {A\&A},
         year = {2000},
        month = {},
       volume = {501},
       number = {519},
        pages = {},
          doi = {},
       adsurl ={}
}

@ARTICLE{Cayrel2004,
       author = {{Cayrel}, R. and {Depagne}, E. and {Spite}, M. and {Hill}, V. and {Spite}, F. and {Francois}, P. and {Plez}, B. and {Beers}, C. and {Primas}, F. and {Andersen}, J. and {Barbuy}. B. and {Bonifacio}, P. and {Molaro}, P. and {Nordstrom}, B.},
        title = "",
      journal = {A\&A},
         year = {2004},
        month = {},
       volume = {416},
       number = {},
        pages = {1117},
          doi = {},
       adsurl ={}
}

@ARTICLE{Gratton2003,
       author = {{Gratton}, R. G. and {Carretta}, E. and {Claudi}, R. and {Lucatello}, S. and {Barbieri}, M.},
        title = "",
      journal = {A\&A},
         year = {2003},
        month = {},
       volume = {404},
       number = {},
        pages = {187},
          doi = {},
       adsurl ={}
}

@ARTICLE{Bensby2014,
       author = {{Bensby}, T. and {Feltzing}, S. and {Oey}, M. S.},
        title = "",
      journal = {A\&A},
         year = {2014},
        month = {},
       volume = {562},
       number = {},
        pages = {71},
          doi = {},
       adsurl ={}
}

@ARTICLE{Jacobson2015,
       author = {{Jacobson}, H. R. and {Keller}, S. and {Frebel}, A. and {Casey}, A. R. and {Asplund}, M. and {Bessell}, M. S. and {Da Costa}, G. S. and {Lind}, K. and {Marino}, A. F. and {Norris}, J. E. and {Penal}, J. M. and {Schmidt}, B. P. and {Tisserand}, P. and {Walsh}, J. M. and {Yong}, D. and {Yu}, Q.},
        title = "",
      journal = {Astrophys. J.},
         year = {2015},
        month = {},
       volume = {807},
       number = {},
        pages = {171},
          doi = {},
       adsurl ={}
}

@ARTICLE{Placco2015,
       author = {{Placco}, V. M. and {Beers}, T. C. and {Ivans}, I. I. and {Filler}, D. and {Imig}, J. A. and {Roederer}, U. U. },
        title = "",
      journal = {Astrophys. J.},
         year = {2015},
        month = {},
       volume = {812},
       number = {},
        pages = {109},
          doi = {},
       adsurl ={}
}

@ARTICLE{Sneden1998,
       author = {{Sneden}, C. and {Cowan}, J. J. and {Burris}, D. L. and {Truran}, J. W.},
        title = "",
      journal = {Astrophys. J.},
         year = {1998},
        month = {},
       volume = {496},
       number = {},
        pages = {235},
          doi = {},
       adsurl ={}
}

@ARTICLE{roberti:24,
       author = {{Roberti}, Lorenzo and {Limongi}, Marco and {Chieffi}, Alessandro},
        title = "{Zero and Extremely Low-metallicity Rotating Massive Stars: Evolution, Explosion, and Nucleosynthesis Up to the Heaviest Nuclei}",
      journal = {\apjs},
         year = 2024,
        month = feb,
       volume = {270},
       number = {2},
          eid = {28},
        pages = {28},
          doi = {10.3847/1538-4365/ad1686},
archivePrefix = {arXiv},
       eprint = {2312.02942},
 primaryClass = {astro-ph.SR},
       adsurl = {https://ui.adsabs.harvard.edu/abs/2024ApJS..270...28R}
}

@article{boccioli:23,
doi = {10.3847/1538-4357/acc06a},
url = {https://dx.doi.org/10.3847/1538-4357/acc06a},
year = {2023},
month = {may},
publisher = {The American Astronomical Society},
volume = {949},
number = {1},
pages = {17},
author = {Luca Boccioli and Lorenzo Roberti and Marco Limongi and Grant J. Mathews and Alessandro Chieffi},
title = {Explosion Mechanism of Core-collapse Supernovae: Role of the Si/Si–O Interface},
journal = {The Astrophysical Journal}
}

@ARTICLE{curtis:19,
       author = {{Curtis}, Sanjana and {Ebinger}, Kevin and {Fr{\"o}hlich}, Carla and {Hempel}, Matthias and {Perego}, Albino and {Liebend{\"o}rfer}, Matthias and {Thielemann}, Friedrich-Karl},
        title = "{PUSHing Core-collapse Supernovae to Explosions in Spherical Symmetry. III. Nucleosynthesis Yields}",
      journal = {\apj},
         year = 2019,
        month = jan,
       volume = {870},
       number = {1},
          eid = {2},
        pages = {2},
          doi = {10.3847/1538-4357/aae7d2},
archivePrefix = {arXiv},
       eprint = {1805.00498},
 primaryClass = {astro-ph.SR},
       adsurl = {https://ui.adsabs.harvard.edu/abs/2019ApJ...870....2C}
}

@ARTICLE{BR24,
       author = {{Boccioli}, Luca and {Roberti}, Lorenzo},
        title = "{The Physics of Core-Collapse Supernovae: Explosion Mechanism and Explosive Nucleosynthesis}",
      journal = {Universe},
         year = 2024,
        month = mar,
       volume = {10},
       number = {3},
          eid = {148},
        pages = {148},
          doi = {10.3390/universe10030148},
archivePrefix = {arXiv},
       eprint = {2403.12942},
 primaryClass = {astro-ph.SR},
       adsurl = {https://ui.adsabs.harvard.edu/abs/2024Univ...10..148B}
}

@ARTICLE{Cohen2013,
       author = {{Cohen}, J. G. and {Christlieb}, N. and {Thompson}, I. and {McWilliam}, A.},
        title = "",
      journal = {Astrophys. J.},
         year = {2013},
        month = {},
       volume = {778},
       number = {56},
        pages = {},
          doi = {},
       adsurl ={}
}

@ARTICLE{karakas:14,
       author = {{Karakas}, A. I. and {Marino}, A. F. and {Nataf}, D. M.},
        title = "",
      journal = {Astrophys. J.},
         year = {2014},
        month = {},
       volume = {784},
       number = {32},
        pages = {},
          doi = {},
       adsurl ={}
}

@ARTICLE{kajino:19,
       author = {{Kajino}, T. and {Aoki}, W. and {Balantekin}, A. B. and {Diehl}, R. and {Famiano}, M. A. and {Mathew}, G. J.},
        title = "Current status of r-process nucleosynthesis",
      journal = {Progress in Particle and Nuclear Physics},
         year = {2019},
        month = {},
       volume = {107},
       number = {},
        pages = {109-166},
          doi = {10.1016/j.ppnp.2019.02.008},
       adsurl ={https://www.sciencedirect.com/science/article/abs/pii/S0146641019300201?via%3Dihub}
}

@ARTICLE{kubryk:15,
       author = {{Kubryk}, M. and {Prantzos}, N. and {Athanassoula}, E.},
        title = "Evolution of the Milky Way with radial motions of stars and gas
II. The evolution of abundance profiles from H to Ni",
      journal = {A\&A},
         year = {2015},
        month = {},
       volume = {580},
       number = {A127},
        pages = {},
          doi = {10.1051/0004-6361/201424599},
       adsurl ={https://www.aanda.org/articles/aa/abs/2015/08/aa24599-14/aa24599-14.html}
}

@ARTICLE{pumo:10,
       author = {{Pumo}, M. L. and {Contino}, G. and {Bonanno}, A. and {Zappala}, R. A.},
        title = "Convective overshooting and production of s-nuclei in massive stars during their core He-burning phase",
      journal = {A\&A},
         year = {2010},
        month = {},
       volume = {524},
       number = {A45},
        pages = {},
          doi = {10.1051/0004-6361/201015518},
       adsurl ={https://www.aanda.org/articles/aa/full_html/2010/16/aa15518-10/aa15518-10.html}
}

@ARTICLE{2019APJ_Cote,
       author = {{Côté}, Benoit and {Lugaro}, M. and {Reifarth}, R. and {Pignatari}, M. and {Világos}, B. and {Yague}, A. and {Gibson}, B.},
        title = "{Galactic Chemical Evolution of Radioactive Isotopes}",
      journal = {\apj},
         year = "2019",
        month = "May",
       volume = {878},
       number = {156},
        pages = {},
          doi = {10.3847/1538-4357/ab21d1},
 primaryClass = {},
       adsurl = {https://iopscience.iop.org/article/10.3847/1538-4357/ab21d1}
}

@ARTICLE{best:13,
       author = {{Best}, A. and {Beard}, M. and {Gorres}, J. and {Couder}, M. and {deBoer}, R. and {Falahat}, S. },
        title = "Measurement of the $^{17}$O($\alpha,n$)$^{20}$O reaction ONe and its impact on the process in massive stars",
      journal = {Phys. Rev. C},
         year = {2013},
        month = {},
       volume = {87},
       number = {045805},
        pages = {},
          doi = {10.1103/PhysRevC.87.045805},
       adsurl ={https://journals.aps.org/prc/abstract/10.1103/PhysRevC.87.045805}
}

@ARTICLE{frischknecht:16,
       author = {{Frischknecht}, Urs and {Hirschi}, Raphael and {Pignatari}, Marco and {Maeder}, Andr{\'e} and {Meynet}, George and {Chiappini}, Cristina and {Thielemann}, Friedrich-Karl and {Rauscher}, Thomas and {Georgy}, Cyril and {Ekstr{\"o}m}, Sylvia},
        title = "{s-process production in rotating massive stars at solar and low metallicities}",
      journal = {\mnras},
         year = 2016,
        month = feb,
       volume = {456},
       number = {2},
        pages = {1803-1825},
          doi = {10.1093/mnras/stv2723},
archivePrefix = {arXiv},
       eprint = {1511.05730},
 primaryClass = {astro-ph.SR},
       adsurl = {https://ui.adsabs.harvard.edu/abs/2016MNRAS.456.1803F}
}
\bibliographystyle{aasjournal}



\end{document}